\documentclass[twocolumn]{aastex631}

\usepackage{amsmath}

\begin{document}

\title{Parallel Alignments between Magnetic Fields and Dense Structures in the Central Molecular Zone}

\author[0000-0003-1337-9059]{Xing Pan}
\affiliation{School of Astronomy and Space Science, Nanjing University, 163 Xianlin Avenue, Nanjing 210023, P.R.China}
\affiliation{Key Laboratory of Modern Astronomy and Astrophysics (Nanjing University), Ministry of Education, Nanjing 210023, P.R.China}
\affiliation{Center for Astrophysics $\vert$ Harvard \& Smithsonian, 60 Garden Street, Cambridge, MA, 02138, USA}

\author[0000-0003-2384-6589]{Qizhou Zhang}
\affiliation{Center for Astrophysics $\vert$ Harvard \& Smithsonian, 60 Garden Street, Cambridge, MA, 02138, USA}

\author[0000-0002-5093-5088]{Keping Qiu}
\affiliation{School of Astronomy and Space Science, Nanjing University, 163 Xianlin Avenue, Nanjing 210023, P.R.China}
\affiliation{Key Laboratory of Modern Astronomy and Astrophysics (Nanjing University), Ministry of Education, Nanjing 210023, P.R.China}

\author[0000-0002-5811-0136]{Dylan M. Par\'e}
\affiliation{Department of Physics, Villanova University, 800 E. Lancaster Ave., Villanova, PA 19085, USA}

\author[0000-0003-0016-0533]{David T. Chuss}
\affiliation{Department of Physics, Villanova University, 800 E. Lancaster Ave., Villanova, PA 19085, USA}

\author[0000-0002-4013-6469]{Natalie O. Butterfield}
\affiliation{National Radio Astronomy Observatory, 520 Edgemont Road, Charlottesville, VA 22903, USA}

\author[0000-0002-9483-7164]{Robin G. Tress}
\affiliation{Institute of Physics, Laboratory for Galaxy Evolution and Spectral Modelling, EPFL, Observatoire de Sauverny, Chemin Pegasi 51, 1290 Versoix, Switzerland}

\author[0000-0001-6113-6241]{Mattia C. Sormani}
\affiliation{Universit\`a dell’Insubria, via Valleggio 11, 22100 Como, Italy}

\author[0000-0002-6581-3307]{Yuping Tang}
\affiliation{Shanghai Key Lab for Astrophysics, Shanghai Normal University, Shanghai 200234, P.R.China}

\author[0000-0001-6353-0170]{Steven N. Longmore}
\affiliation{Astrophysics Research Institute, Liverpool John Moores University, 146 Brownlow Hill, Liverpool L3 5RF, UK}
\affiliation{Cosmic Origins Of Life (COOL) Research DAO, Germany}

\author[0000-0003-2133-4862]{Thushara Pillai}
\affiliation{Haystack Observatory, Massachusetts Institute of Technology, 99 Millstone Road, Westford, MA 01886, USA}



\begin{abstract}
The recent Far-Infrared Polarimetric Large-Area Central Molecular Zone Exploration (FIREPLACE) survey with SOFIA has mapped plane-of-the-sky magnetic field orientations within the Central Molecular Zone (CMZ) of the Milky Way. Applying the Histogram of Relative Orientation (HRO) analysis to the FIREPLACE data, we find that the relative orientation between magnetic fields and column density structures is random in low-density regions ($2\times10^{22} \lesssim N_\mathrm{H_2} \lesssim 10^{23}\mathrm{cm^{-2}}$), but becomes preferentially parallel in high-density regions ($\gtrsim 10^{23}\mathrm{cm^{-2}}$). This trend is in contrast with that of the nearby molecular clouds, where the relative orientation transitions from parallel to perpendicular with increasing column densities. However, the relative orientation varies between individual CMZ clouds. Comparisons with MHD simulations specific to the CMZ conditions suggest that the observed parallel alignment is intrinsic rather than artifacts caused by the projection effect. The origin of this parallel configuration may arise from the fact that most dense structures in the CMZ are not self-gravitating, as they are in super-virial states, except for the mini-starburst region Sgr B2. These findings are consistent with the low star formation efficiency observed in the CMZ compared to that in the Galactic disk.

\end{abstract}

\keywords{}


\section{Introduction} \label{sec:intro}
The Central Molecular Zone (CMZ) is the innermost 150 parsecs of our Galaxy, harboring a significant reservoir of molecular gas ($2-6\times10^7M_\odot$) \citep{Morris1996} with high surface densities \citep[$\sim10^{23}~\mathrm{cm^{-2}}$ from][]{Longmore2013, Battersby2024}. Several massive molecular clouds, with gas mass over $10^5M_\odot$, have been identified in the CMZ, including well-known regions such as Sgr B2, G0.253+0.016, Sgr C, and the Dust Ridge \citep{Goldsmith1990, Huettemeister1995, Longmore2012, Kendrew2013, Lu2019, Lu2019b, Walker2018}. Meanwhile, the CMZ contains an extreme star-forming environment, distinct from that of the Galactic disk. \cite{Heyer2015} investigated the size–linewidth relation in molecular gas and found that the turbulent energy in the CMZ is significantly higher than that in the Galactic disk, indicating that gas in the CMZ is more turbulent than in the disk. Additionally, the CMZ is characterized with strong magnetic fields, with mean field strengths ranging from 0.1 to 10 mG \citep{Pillai2015, Mangilli2019, Lu2024, Pan2024}.

Despite the large amount of molecular gas, the present star formation rates (SFRs) of $\simeq0.07~\mathrm{M_\odot~yr^{-1}}$ for the CMZ is an order of magnitude lower than that expected from the dense-gas star formation relations \citep[][]{Lada2010,Longmore2013,Barnes2017,Kauffmann2017,Lu2019}. Some studies proposed that the high turbulent pressure may increase the density threshold for star formation and hence decrease the SFR \citep[e.g.,][]{Kruijssen2014}. Another hypothesis suggests that star formation in the CMZ occurs episodically and is currently in an inactive phase \citep[][]{Kruijssen2014,Krumholz2015}. However, the strong magnetic field in the CMZ can also delay the collapse of molecular clouds, thereby suppressing star formation. For example, \cite{Pillai2015} found that G0.253+0.016 is strongly magnetized, with the magnetic field dominating over turbulence and gravity. The cloud exhibits only a single site of star formation \citep{Kauffmann2013,Lu2019,Walker2021}. In contrast, the mini-starburst region Sgr B2 \citep[$\mathrm{SFR}\sim0.01M_\odot~\mathrm{yr^{-1}}$,][]{Kauffmann2017,Ginsburg2018} exhibits a relatively weak magnetic field compared to its gravitational and turbulent kinetic energies \citep{Pan2024}. Therefore, it is necessary to take magnetic field into account to investigate the low star formation efficiency in the CMZ. Nevertheless, our understanding of magnetic fields in the CMZ remains incomplete, largely due to the limited number of detailed observations.

Recent advances have been made through several large-scale infrared and millimeter surveys targeting the magnetic fields in this region. For example, the PILOT survey \citep[$2.2\arcmin$ resolution, at 240 $\mu$m from][]{Mangilli2019} and the Atacama Cosmology Telescope observations \citep[$\sim1\arcmin$ resolution, at 98, 150, and 224 GHz from][]{Guan2021} have uncovered large-scale magnetic fields ($\gtrsim10$ pc) that are both ordered and tilted relative to the Galactic Plane. Additionally, higher-resolution surveys such as the Far-Infrared Polarimetric Large-Area CMZ Exploration \citep[FIREPLACE, $19.6\arcsec$ at 214 $\mu$m][]{Butterfield2024, Pare2024}, JCMT/POL2 observations of 11 CMZ molecular clouds from \cite{Lu2024} and the B-fields In STar-forming Region Observations \citep[BISTRO, $12\arcsec$ at 850 $\mu$m][]{Karoly2025,Yang2025} have revealed diverse magnetic field morphologies within individual molecular clouds. Additionally, \cite{Hu2022} used Gradient Technique (GT) and revealed the magnetic field in the CMZ globally consistent with the polarization measurements, indicating that the magnetic field and turbulence are dynamically important in the CMZ. These datasets provide an unprecedented opportunity to investigate the role of magnetic fields in the dynamics and star formation processes of CMZ clouds.

The role of magnetic fields in star formation is often assessed by measuring their strength using the Zeeman effect or the Davis–Chandrasekhar–Fermi (DCF) method \citep{Davis1951, Chandrasekhar1953}, and comparing the associated magnetic energy to the turbulent kinetic and gravitational energies. Alternatively, magnetohydrodynamic (MHD) simulations \citep{Soler2013, Soler2017b} have introduced the Histogram of Relative Orientations (HRO), a statistical method to assess the role of magnetic fields. This statistical approach characterizes the relative orientation between the magnetic field and column density structures to infer the energy balance between the magnetic field, gravity and turbulence. Observational studies \citep[e.g.,][]{PlanckXXXV2016, Soler2017, Soler2019, Fissel2019} have applied the HRO analysis to Galactic molecular clouds. These studies reveal a clear trend: as column density increases, the relative orientation between magnetic fields and density structures transitions from predominantly parallel in diffuse regions to perpendicular in denser regions. Areas where the magnetic fields are orthogonal to the density structures are typically gravitationally unstable, facilitating star formation. This behavior is consistent with predictions from MHD simulations with initial conditions of sub- to trans-Alf{\'v}enic turbulence.

Using the Histogram of Relative Orientations (HRO) analysis, \cite{Pare2025} found that magnetic fields in the CMZ are generally aligned parallel to column density structures in dense regions, which is different from that observed in the Galactic disk. However, \cite{PlanckXXXV2016} have shown that such parallel alignments in 2D projection do not necessarily imply a true 3D parallel configuration. The underlying distribution may still include structures with nearly perpendicular orientations. This highlights the need to test whether the observed preferentially parallel alignments in the CMZ is intrinsic or merely a result of projection effects. Moreover, MHD simulations of the Galactic disk have shown that a transition from parallel to perpendicular relative orientations is linked to the balance between magnetic, turbulent kinetic, and gravitational energies. Understanding the energy balance in the CMZ may therefore provide insight into the origin of the observed preferential alignment.

In this paper, we investigate the evolution of the relative orientation between magnetic fields and column density structures by applying the HRO analysis with a normalized parameter. To better resolve the relative orientation in dense regions, we increase the number of column density bins at high densities. We also derive a high-angular-resolution ($\sim$19\arcsec) column density map using spectral energy distribution (SED) fitting that combines Herschel and ATLASGAL data. Furthermore, we examine the impact of projection effects using MHD simulations specified to the CMZ environment and explore how energy balance influences the observed magnetic field–density alignment in this unique region. The paper is organized as follows: Section \ref{sec:obs} introduces the data used in our analysis. Section \ref{sec:results} presents the results of the relative orientation analysis for the entire CMZ and individual clouds within. We also compare these observational results with state-of-the-art MHD simulations to investigate the effects of line-of-sight projection. In Section \ref{sec:BfieldRole}, we explore the energy balance between magnetic, turbulent kinetic, and gravitational energy, as well as the possible origin of the observed relative orientation trends. Finally, Section \ref{sec:summary} summarizes the key findings of this study.

\section{Observations} \label{sec:obs}
\subsection{SOFIA observations}
We present the SOFIA \citep[][]{2018JAI.....740011T,2018JAI.....740008H}{}{} 214 $\mu$m polarization data from the FIREPLACE survey. The survey covered the entire CMZ, spanning from Sgr B2 to Sgr C (a roughly $1.5^\circ\times0.5^\circ$ region) and achieved an angular resolution of 19.6\arcsec. The resulting pixel size is 4.55\arcsec. We refer readers to \cite{Butterfield2024} and \cite{Pare2024} for an overview and observation details for the entire survey. These studies also demonstrate that the magnetic fields observed by FIREPLACE generally follow the morphologies of individual clouds, suggesting that the observed fields are predominantly local to the CMZ rather than arising from unrelated line-of-sight components.
\subsection{Column density map}\label{subsec:colden}
The column density distribution in the CMZ is derived by fitting a modified blackbody model to the dust spectral energy distribution (SED). To construct the dust SED, we used the \emph{Herschel} data at 160, 250, 350, and 500 $\mu$m from the Hi-GAL survey \citep{Molinari2010} and 870 $\mu$m dust emission obtained from the combination of APEX/LABOCA data from the APEX Telescope Large Area Survey of the Galaxy  \citep[ATLASGAL,][]{Schuller2009, Csengeri2014} and Planck/HFI \citep[][]{Planck2013I}. The ATLASGAL data did not account for line contamination. However, as discussed in \cite{Schuller2009}, the impact of line contamination is generally within the 15\% flux calibration uncertainty, except in extreme cases such as hot molecular cores, strong outflow sources, and bright photon-dominated regions. Thus, it is unlikely to significantly affect the majority of the CMZ. Assuming a single temperature modified blackbody model, the intensity at each wavelength is given by:
\begin{equation}
    I(\nu)=(1-e^{-\tau_\nu})B_\nu(T_d),
\end{equation}
where $B_\nu(T_d)$ is the Planck function at dust temperature $T_d$ and $\tau_\nu$ is the optical depth at frequency $\nu$, expressed as:
\begin{equation}
    \tau_\nu=\frac{\kappa_0\mu \mathrm{m_H} N_\mathrm{H_2}}{g} \left(\frac{\nu}{\nu_0}\right)^\beta
\end{equation}
where $\kappa_0$ is the dust opacity per unit mass at frequency $\nu_0$, $\mu=2.8$ is the mean molecular weight, $\mathrm{m_H}$ is mass of atomic hydrogen, $\beta$ is the dust emissivity index, $g=100$ is the gas-to-dust ratio. We adopt $\kappa_0=1.37~\mathrm{cm^2~g^{-1}}$ at $\nu_0=300$ GHz from \cite{Ossenkopf1994} for coagulated dust grains with thin ice mantles. However, it is important to note that fitting a single-temperature model can bias the temperature estimate toward higher values, particularly when short-wavelength data are missing. In such cases, warmer dust tends to dominate the SED, as it emits more strongly in the far-infrared.

\begin{figure*}[!ht]
    \centering
    \includegraphics[width=0.95\linewidth]{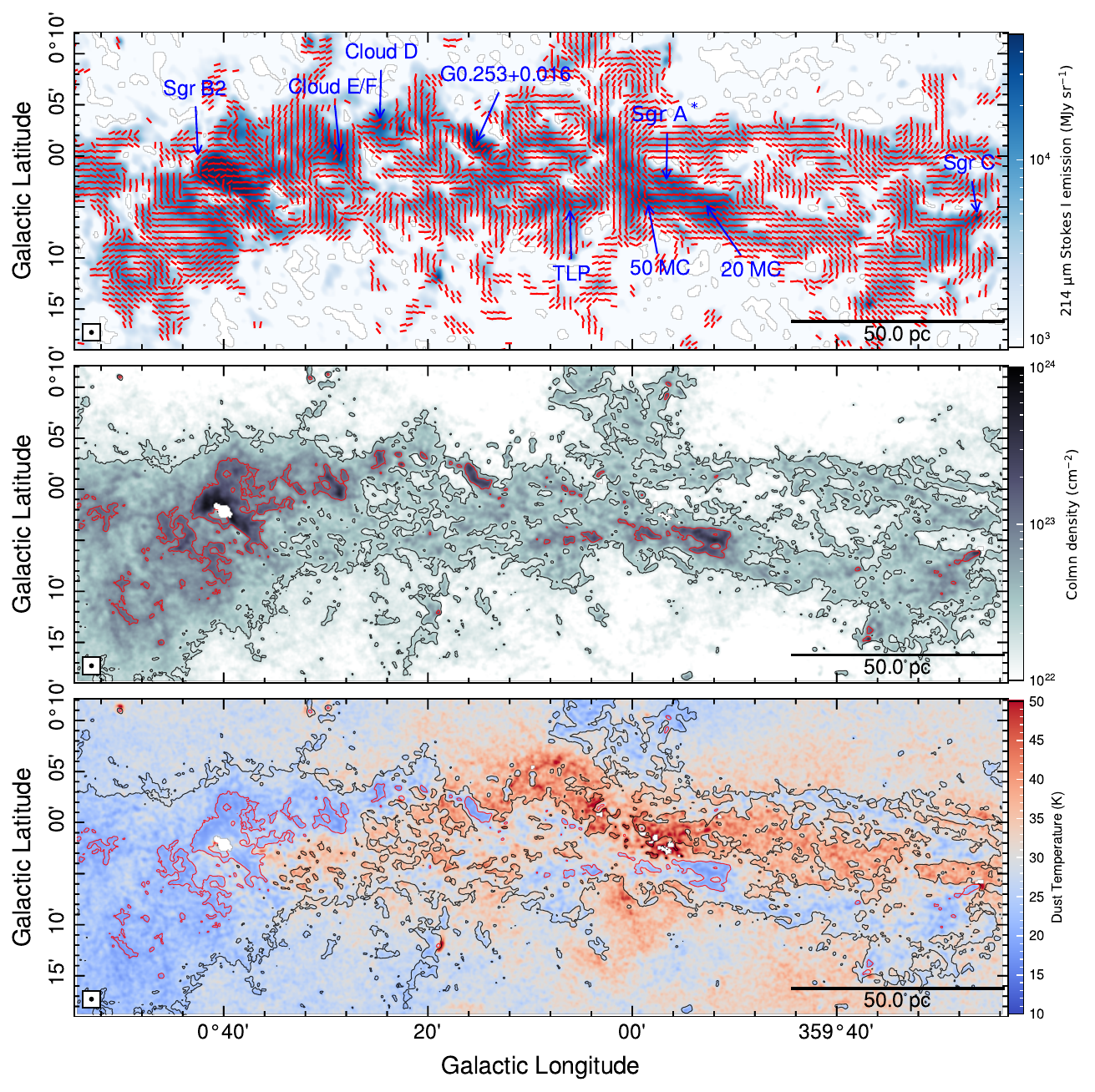}
    \caption{Column density and dust temperature map of the CMZ derived via SED fitting. Top panel shows 214 $\mu$m SOFIA/HAWC+ Stokes I emission overlaid with magnetic field orientations. Red segments indicate the plane-of-the-sky (POS) magnetic field orientations, and labeled blue arrows highlight prominent CMZ molecular clouds. The middle panel shows the distribution of column densities derived by fitting a modified blackbody function to the 160, 250, 350, 500 $\mu$m \emph{Herschel} data from \cite{Molinari2010} and 870 $\mu$m dust emission obtained from the combination of Planck and APEX data from \cite{Csengeri2014}. Column density values corresponding to dust temperature uncertainties over 6 K have been masked. The black contour marks a column density of $N_\mathrm{H_2}=2\times10^{22}~\mathrm{cm^{-2}}$. The red contour marks a column density of $N_\mathrm{H_2}=10^{23}~\mathrm{cm^{-2}}$, where the relative orientation between magnetic fields and column density structures for the CMZ becomes more parallel. Bottom panel shows the distribution of dust temperatures.}
    \label{fig:B_col_CMZ}
\end{figure*}

Using the SED fitting method described in \citet{Tang2021}, we derive the distributions of column density ($N_\mathrm{H_2}$), dust temperature ($T_d$), and dust emissivity index ($\beta$) across the CMZ. The resolution of the resulting column density map is primarily set by the longest-wavelength data from ATLASGAL, approximately 19$\arcsec$ \citep{Schuller2009}, which is comparable to the resolution of the SOFIA data. The column density map is sampled at 4.55$\arcsec$ per pixel, consistent with the FIREPLACE data, which allows us to make direct comparisons between magnetic fields and column density structures. Fig. \ref{fig:B_col_CMZ} shows the distribution of column densities and dust temperature in the CMZ overlaid with magnetic field orientations from SOFIA data. Due to saturation in the \emph{Herschel} data at 160 $\mu$m in the dense regions of Sgr B2, we masked out the saturated regions in the column density map. The distribution of column densities in the CMZ is similar to the previous work that derived the column density using different methods \citep[e.g.,][]{Molinari2011,Mills2017,Battersby2024,Pare2025}. We found pixels with unusually high temperatures ($\gtrsim$ 50 K) but large uncertainties ($\sigma_T> 6$ K, see Appendix \ref{app:tdust}). We attribute this to overfitting, likely due to the absence of short-wavelength data (e.g., $<100\ \mu$m), which are essential for accurately constraining high-temperature regions ($>30$ K). Therefore, we masked out the regions with large temperature uncertainties ($\sigma_T\geq6$ K) in the following analysis.

\section{Relative orientation analysis}\label{sec:results}

\begin{figure*}[ht!]
    \centering
    \includegraphics[width=0.95\linewidth]{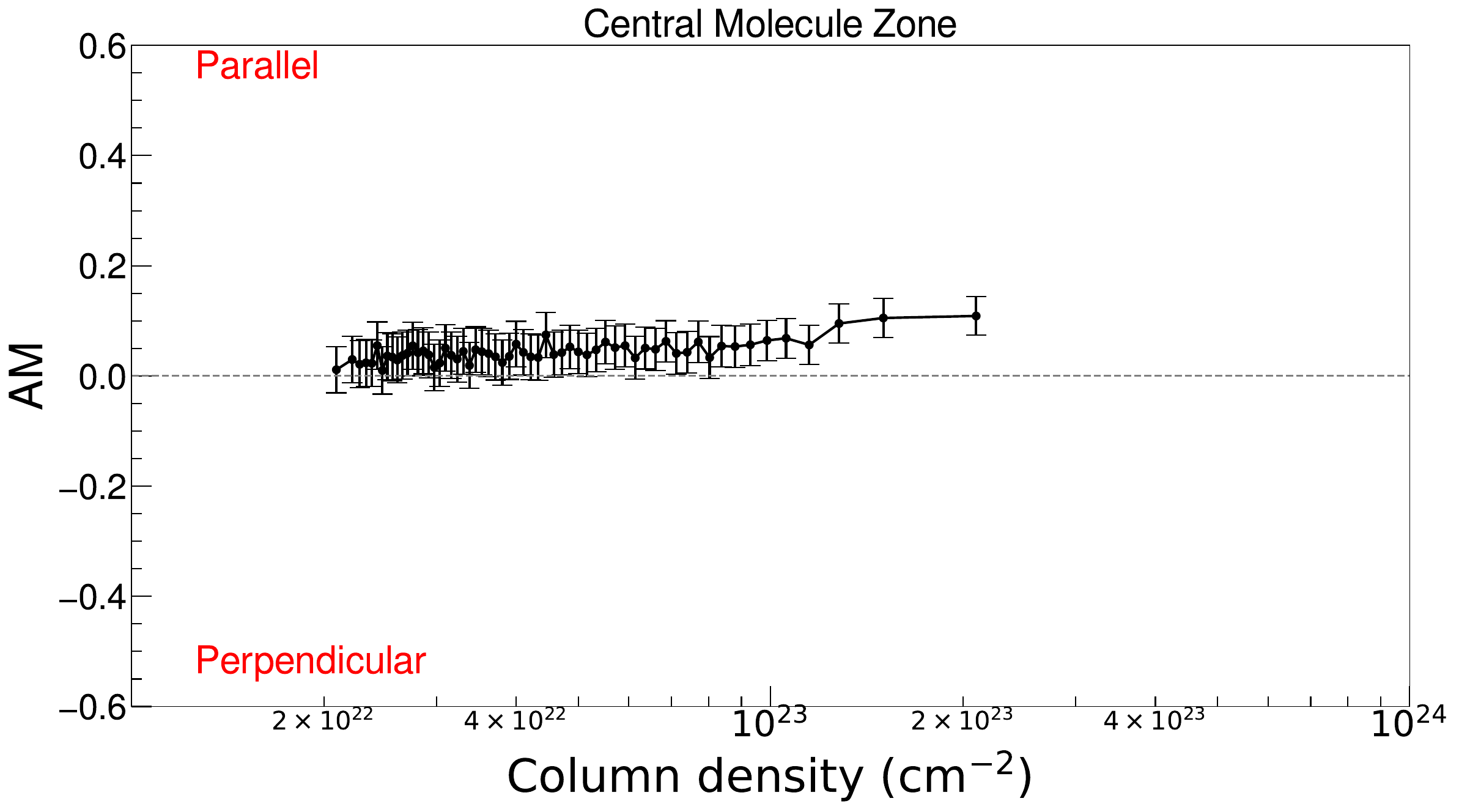}
    \caption{Relative orientation between magnetic field and column density structure as a function of column density for the entire CMZ, characterized by AM (alignment measure). AM $>0$ corresponds to magnetic field preferentially parallel to the column density structure, while AM $<0$ corresponds to magnetic field preferentially perpendicular to the column density structure. AM $\sim0$ indicates the relative orientation between magnetic field and column density structure is random.}
    \label{fig:HRO_CMZ}
\end{figure*}

Here, we use the HRO method proposed by \cite{Soler2013} to quantitatively analyze the relative orientation between the column density structures of the CMZ and the magnetic fields revealed by SOFIA. In this method, the column density structures are characterized by their gradient, which is perpendicular to the iso-density contours. We define the relative angle following \cite{Soler2017}:
\begin{equation}
    \phi = \tan^{-1}(|\nabla N_\mathrm{H_2}\times\hat{E}|,\nabla N_\mathrm{H_2}\cdot\hat{E}), \label{eq:phi}
\end{equation}
where $\nabla N_\mathrm{H_2}$ is the local gradient of column density map and $\hat{E}$ is the unit polarization pseudo-vector which is perpendicular to the magnetic field. The column density map, derived from SED fitting of \emph{Herschel} and ATLASGAL data, has a resolution of about 19\arcsec, matching that of the SOFIA observations. This consistency allows for a direct comparison between the magnetic field orientation and the column density gradient ($\nabla N_\mathrm{H_2}$). Meanwhile, the corresponding spatial resolution of the column density map is about 0.7 pc at the distance of 8.1 kpc. Therefore, our HRO analysis here focuses on the cloud scale, comparable to the scale investigated in MHD simulations \citep{Soler2013} and archival observations \citep[e.g.,][]{PlanckXXXV2016, Soler2017}.

To normalize the relative orientation, we use the normalized alignment measure (AM) parameter introduced in \cite{Lazarian2018}:
\begin{equation}
    \mathrm{AM}=\left\langle\cos 2\phi \right\rangle
\end{equation}

The uncertainty of AM is given by \cite{Liu2023} (see its Appendix B):
\begin{multline}
\delta\mathrm{AM}\\
=\sqrt{(\left\langle(\cos2\phi)^2\right\rangle-\mathrm{AM}^2+\sum(2\sin(2\phi)\delta\phi)^2)/n^\prime} 
\end{multline}
where $n^\prime$ is the number of independent data points within each density bin. In our study, $\mathrm{AM}>0$ indicates that the magnetic field is preferentially parallel to the density structure, $\mathrm{AM}<0$ indicates magnetic field is preferentially perpendicular to the density structure, and $\mathrm{AM}\sim0$ indicates no preferred relative orientation between magnetic field and density structure.

\subsection{Relative orientations in the entire CMZ}

Following \cite{Butterfield2024} and \cite{Pare2024}, we apply cuts in the Stokes I intensity threshold of $I/\sigma_I>200$, percentage polarization less than 50\%, and polarization intensity threshold of $P/\sigma_P>3$, to derive the magnetic field orientation. For the column density map, we apply a cut of $N_\mathrm{H_2}=2.0\times10^{22}~\mathrm{cm^{-2}}$, as indicated by the black contour in Fig. \ref{fig:B_col_CMZ}, which roughly corresponds to the 200$\sigma$ limit of the Stokes I emission in the SOFIA data. Then, we calculate AM over a set of relative angles within different column density bins.

Fig. \ref{fig:HRO_CMZ} shows the relative orientation between the magnetic field and density structures as a function of column densities. Following \cite{PlanckXXXV2016}, we use an equal number of data points in each bin to ensure comparable statistics across density bins. The number of data points in each $N_\mathrm{H_2}$ bin is selected to provide a sufficient number of independent measures per bin for reliable statistical analysis, while also maintaining enough $N_\mathrm{H_2}$ bins to adequately resolve high density regions. Therefore, we use 1800 pixels ($\sim$100 independent polarization measurements) per bin for the entire CMZ. We confirm that varying the number of independent measurements in each column density bin by a factor of two does not significantly affect the observed trend in the relative orientation–column density relation. While the relative orientations between column density structures and magnetic fields appear random in low-density regions ($2 \times 10^{22} < N_{\mathrm{H_2}} < 10^{23}\mathrm{cm^{-2}}$), they become more parallel in high-density regions ($N_{\mathrm{H_2}} \gtrsim 10^{23}\mathrm{cm^{-2}}$). \cite{Pare2025} also find a strong preference for parallel orientations to column density structures. This contrasts with previous HRO studies of Galactic disk molecular clouds \citep[e.g.,][]{PlanckXXXV2016,Soler2017,Malinen2016,Chen2024}, which consistently observe a transition from parallel to perpendicular alignment with increasing density. However, in those cases, the transition occurs at significantly lower densities $10^{21}-10^{22}~\mathrm{cm^{-2}}$ than the average density in the CMZ.

\subsection{Relative orientations in individual molecular clouds in the CMZ}

\begin{figure*}[!ht]
    \centering
    \includegraphics[width=0.48\linewidth]{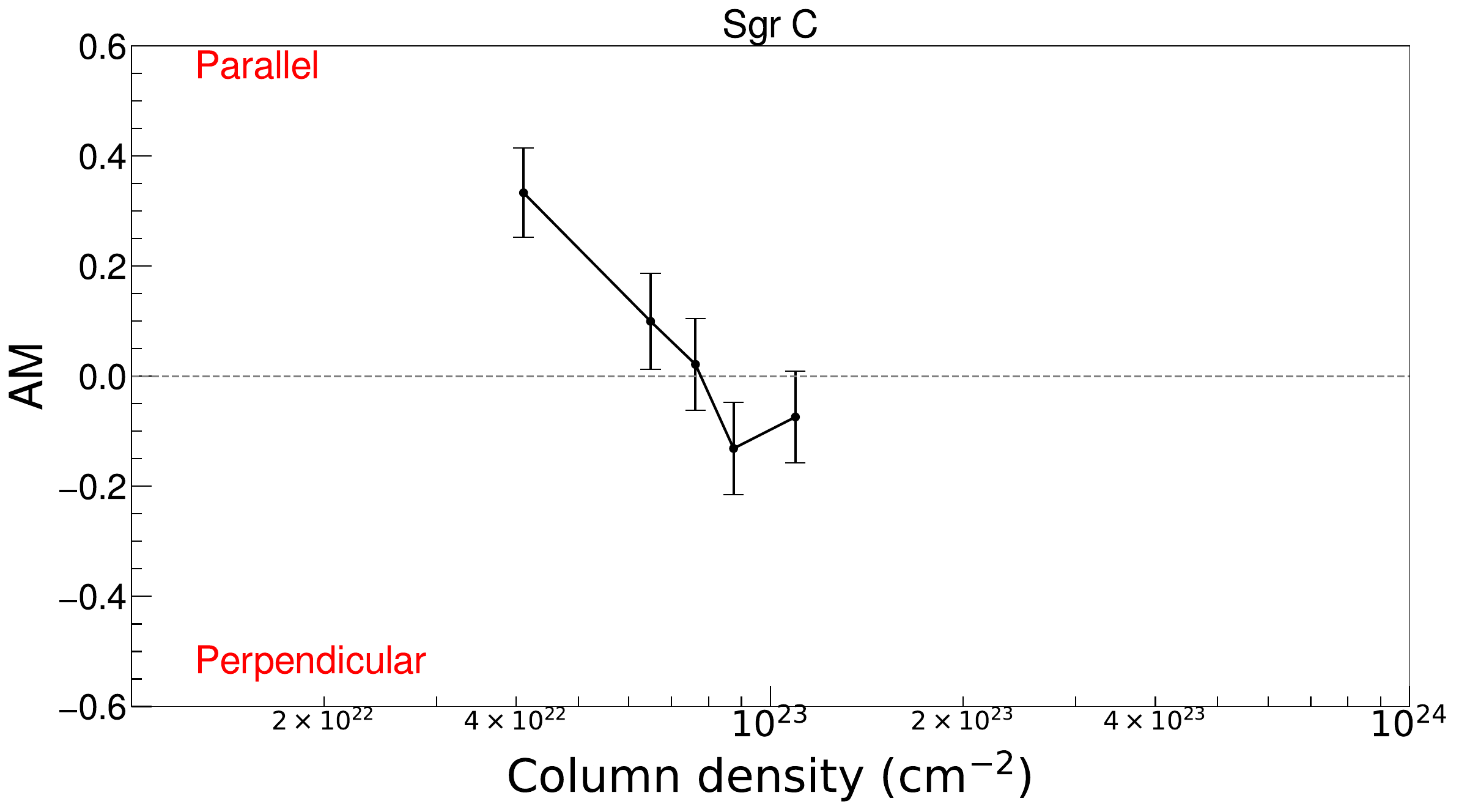}
    \includegraphics[width=0.48\linewidth]{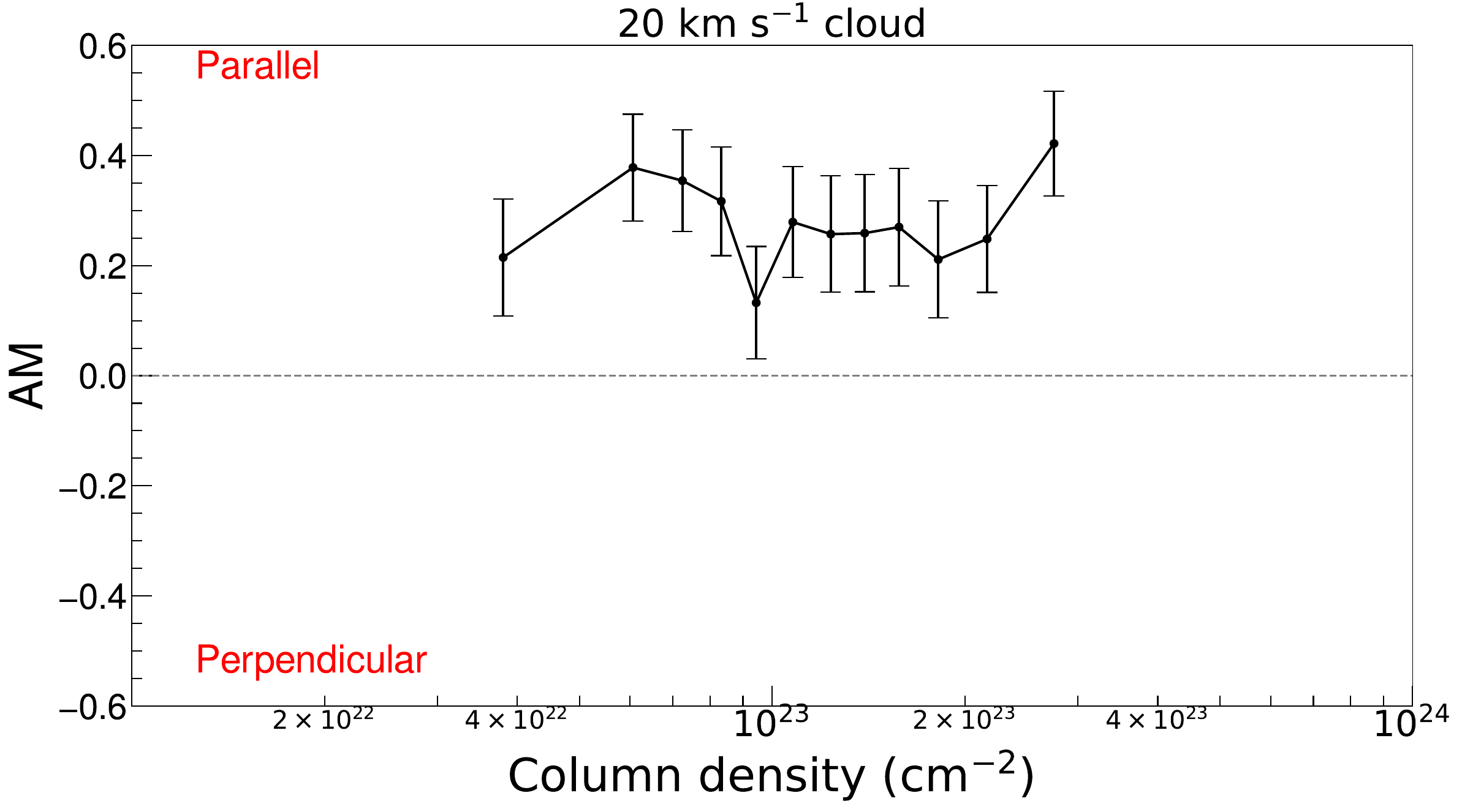}\\
    \includegraphics[width=0.48\linewidth]{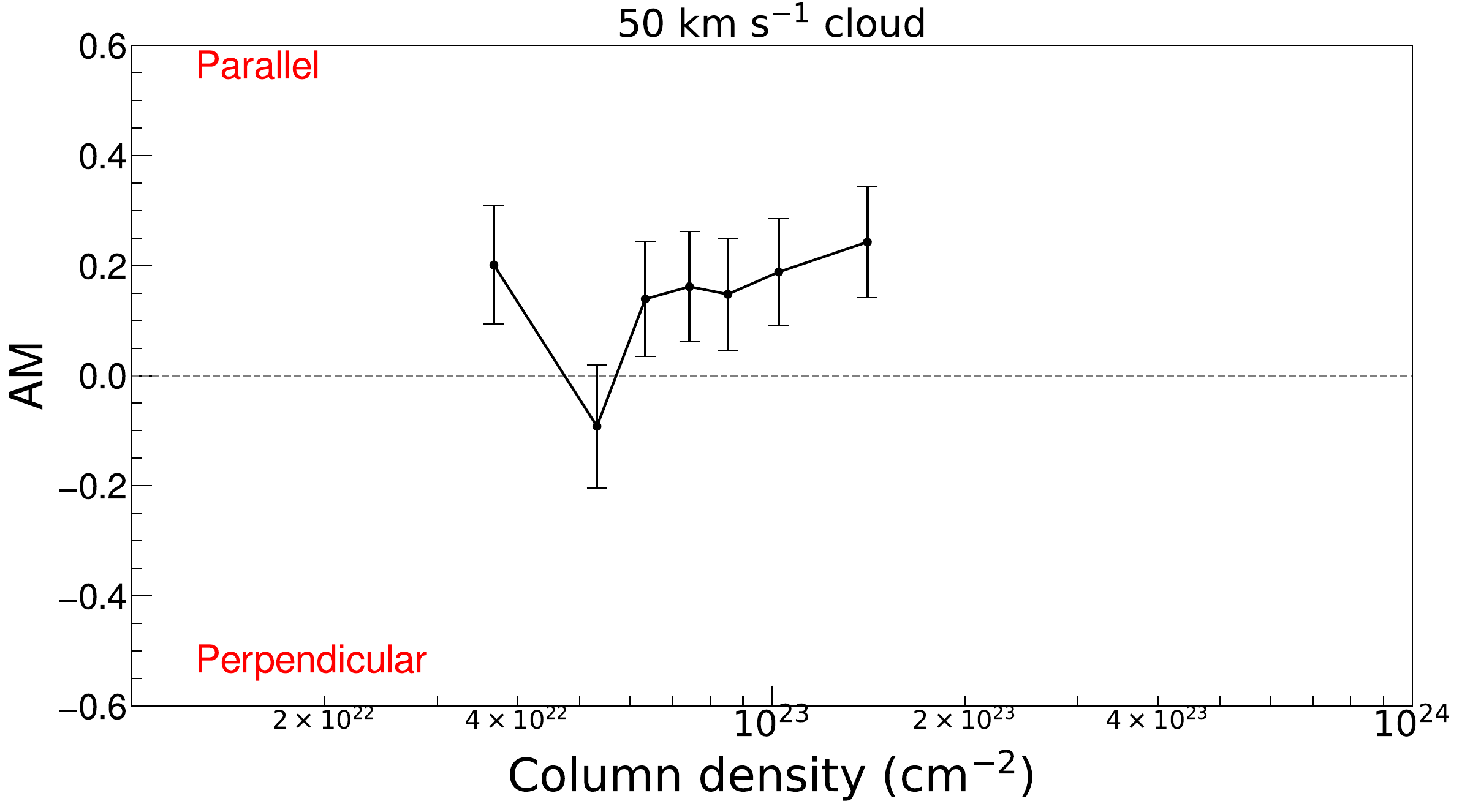}
    \includegraphics[width=0.48\linewidth]{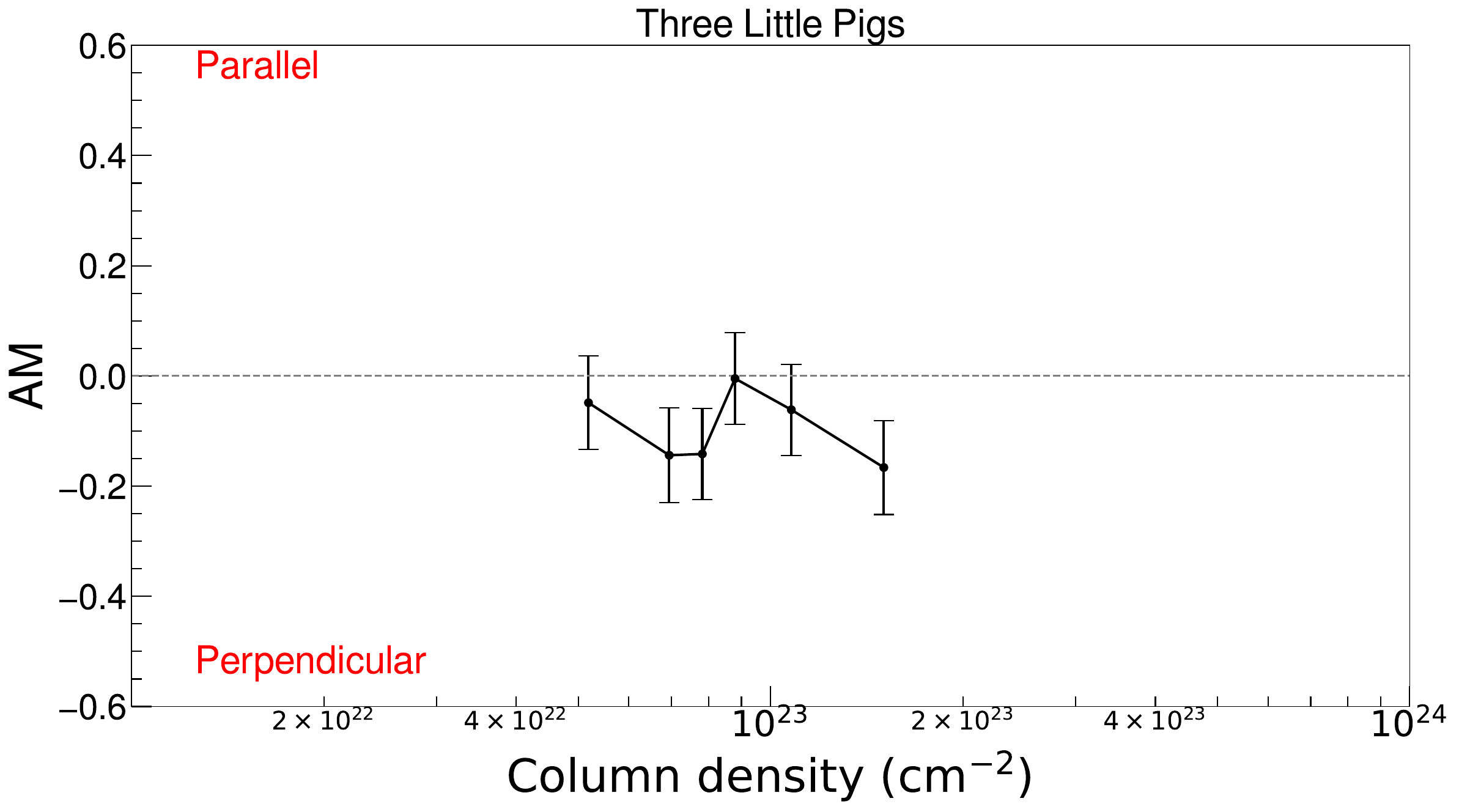}\\
    \includegraphics[width=0.48\linewidth]{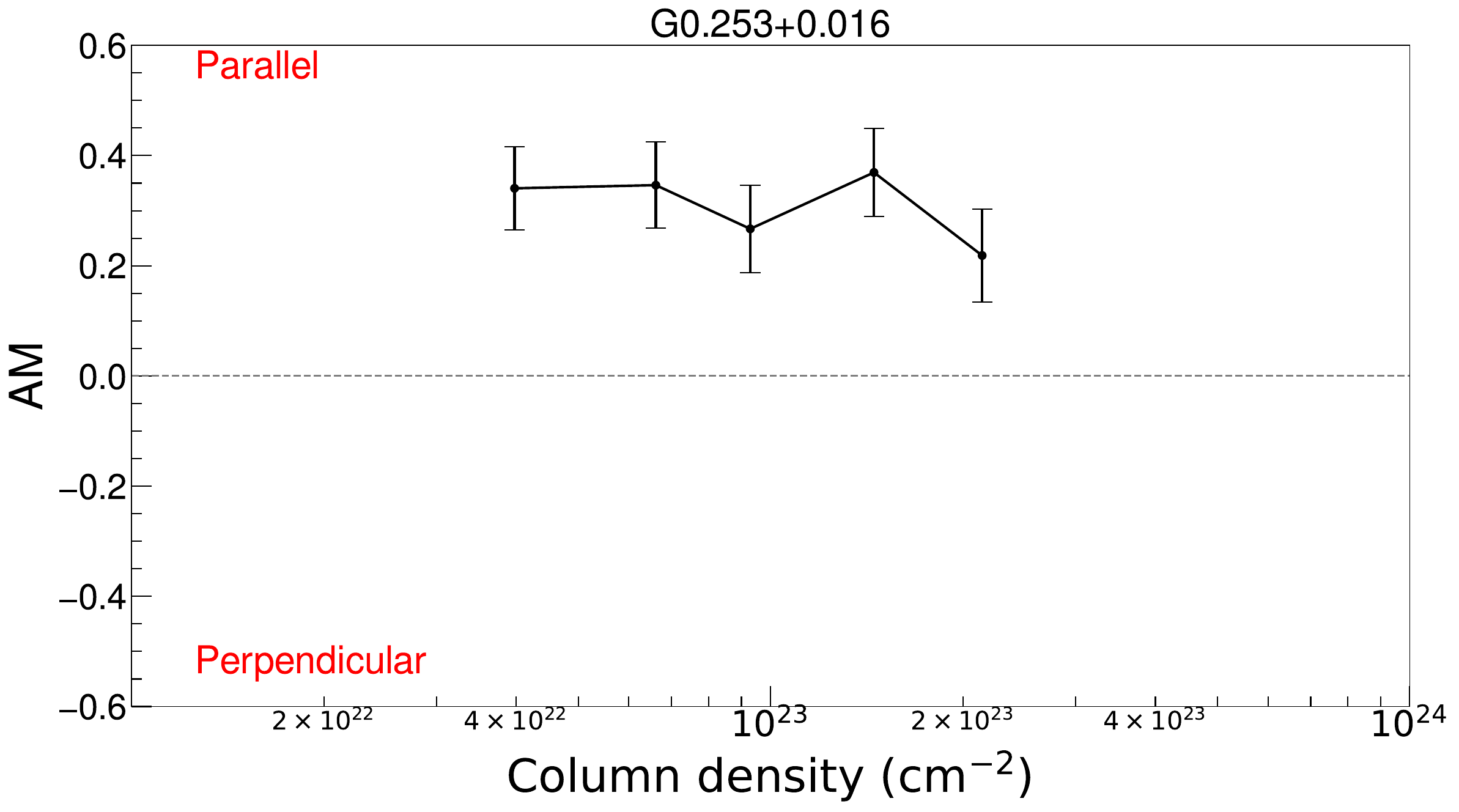}
    \includegraphics[width=0.48\linewidth]{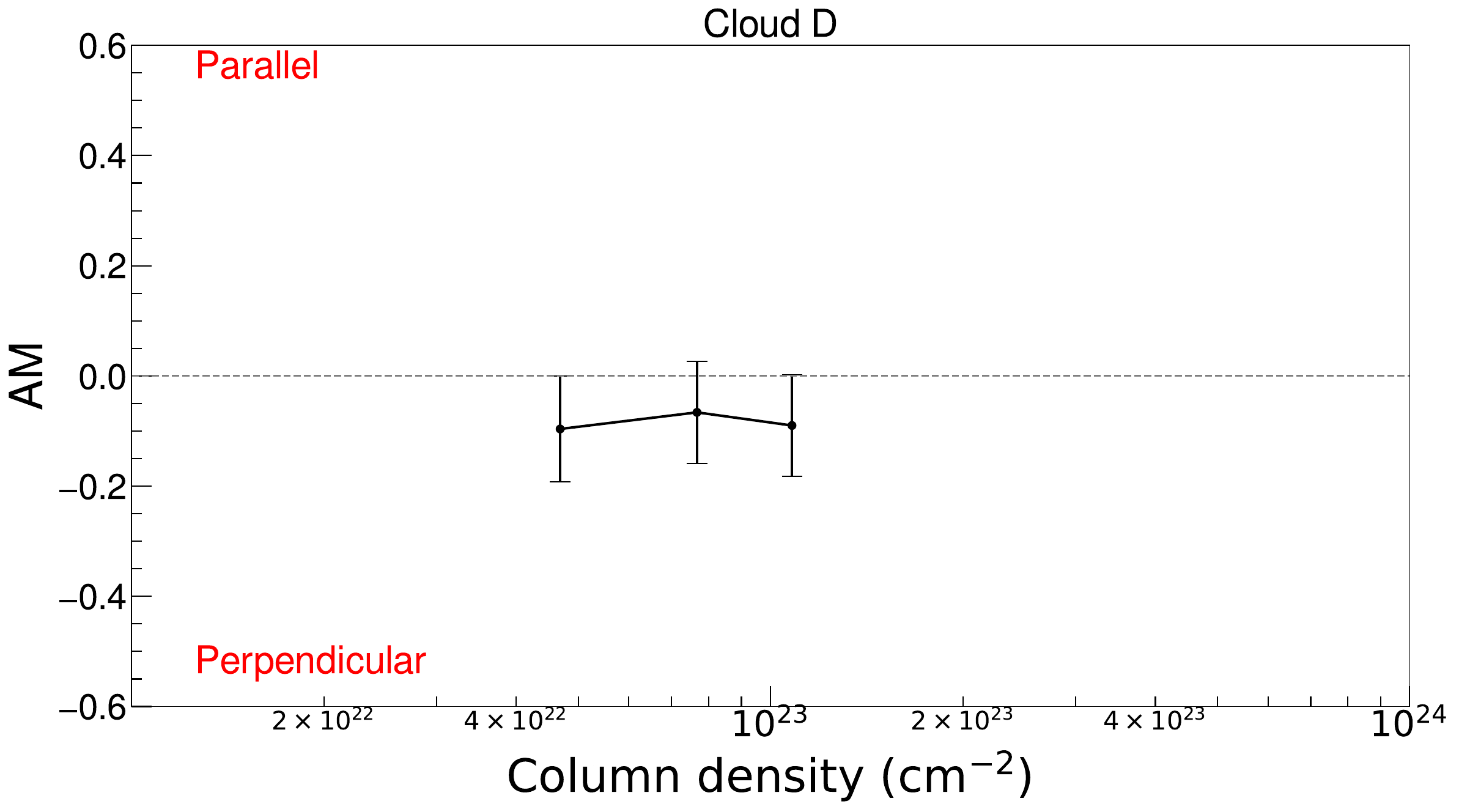}\\
    \includegraphics[width=0.48\linewidth]{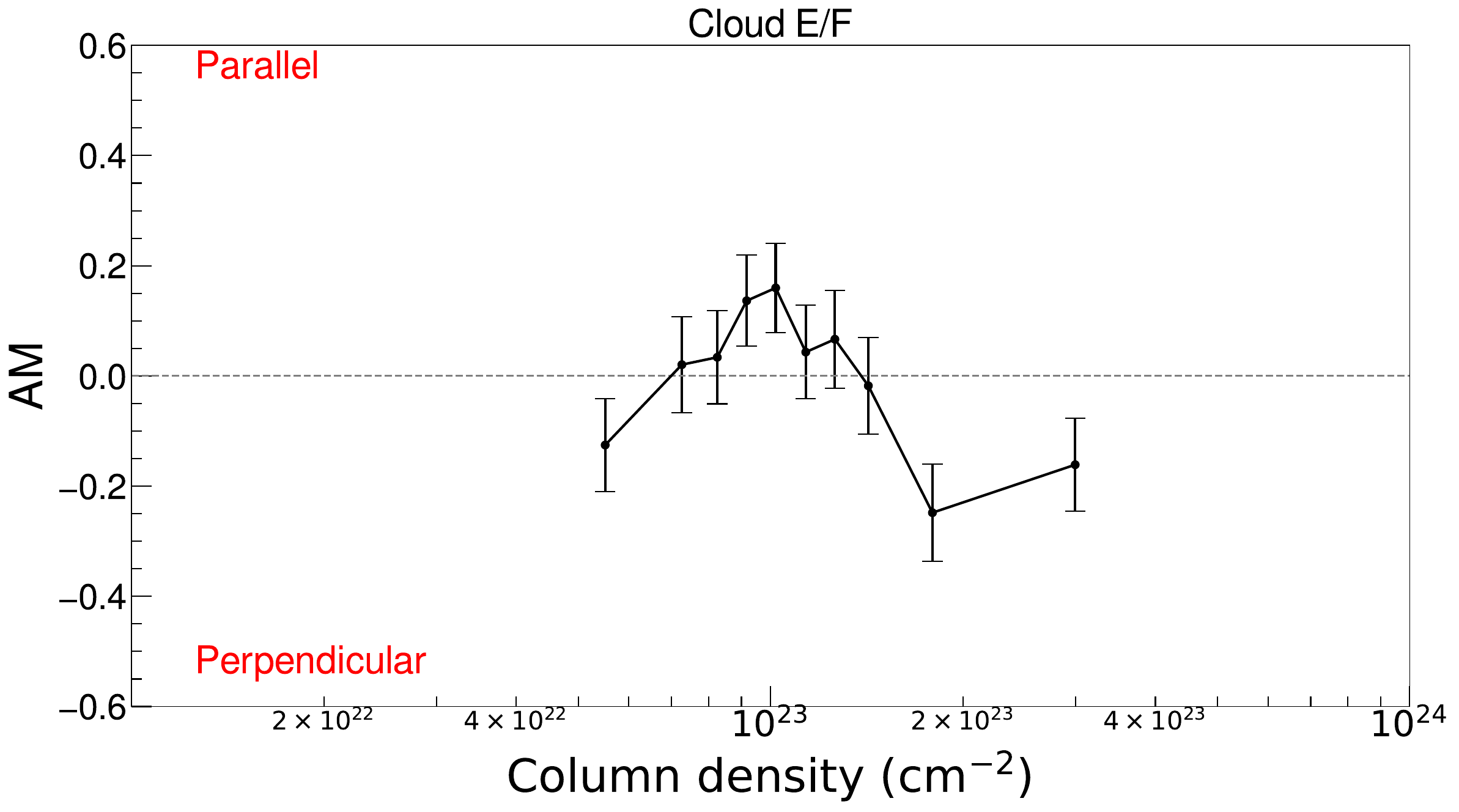}
    \includegraphics[width=0.48\linewidth]{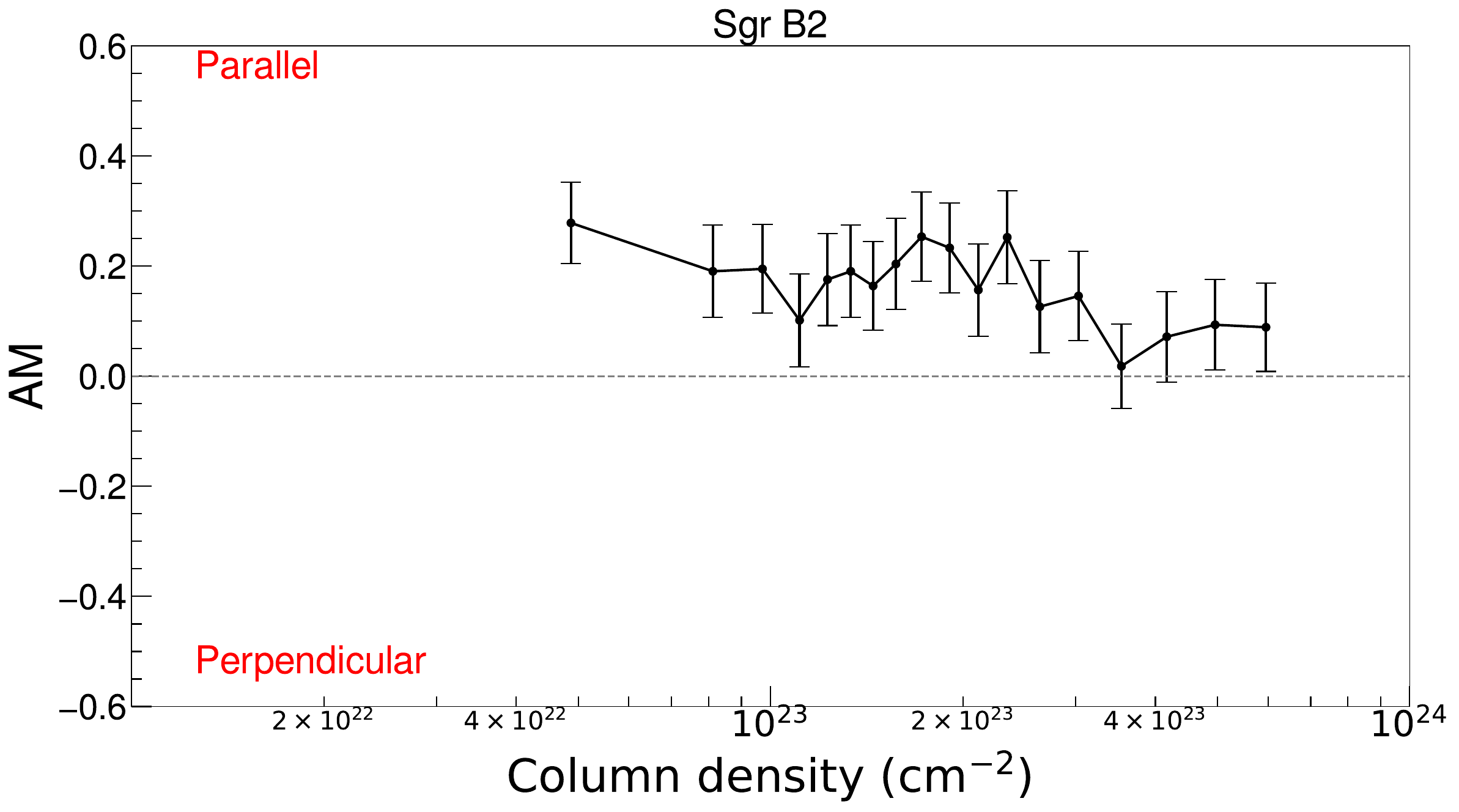}
    \caption{Relative orientation between magnetic field and column density structure as a function of column density for the individual clouds in the CMZ, characterized by AM (alignment measure). AM $>0$ corresponds to magnetic field preferentially parallel to the column density structure, while AM $<0$ corresponds to magnetic field preferentially perpendicular to the column density structure. AM $\sim0$ indicates the relative orientation between magnetic field and column density structure is random.}
    \label{fig:HRO_MCs}
\end{figure*}

\begin{figure*}[!ht]
    \centering
    \includegraphics[width=0.45\linewidth]{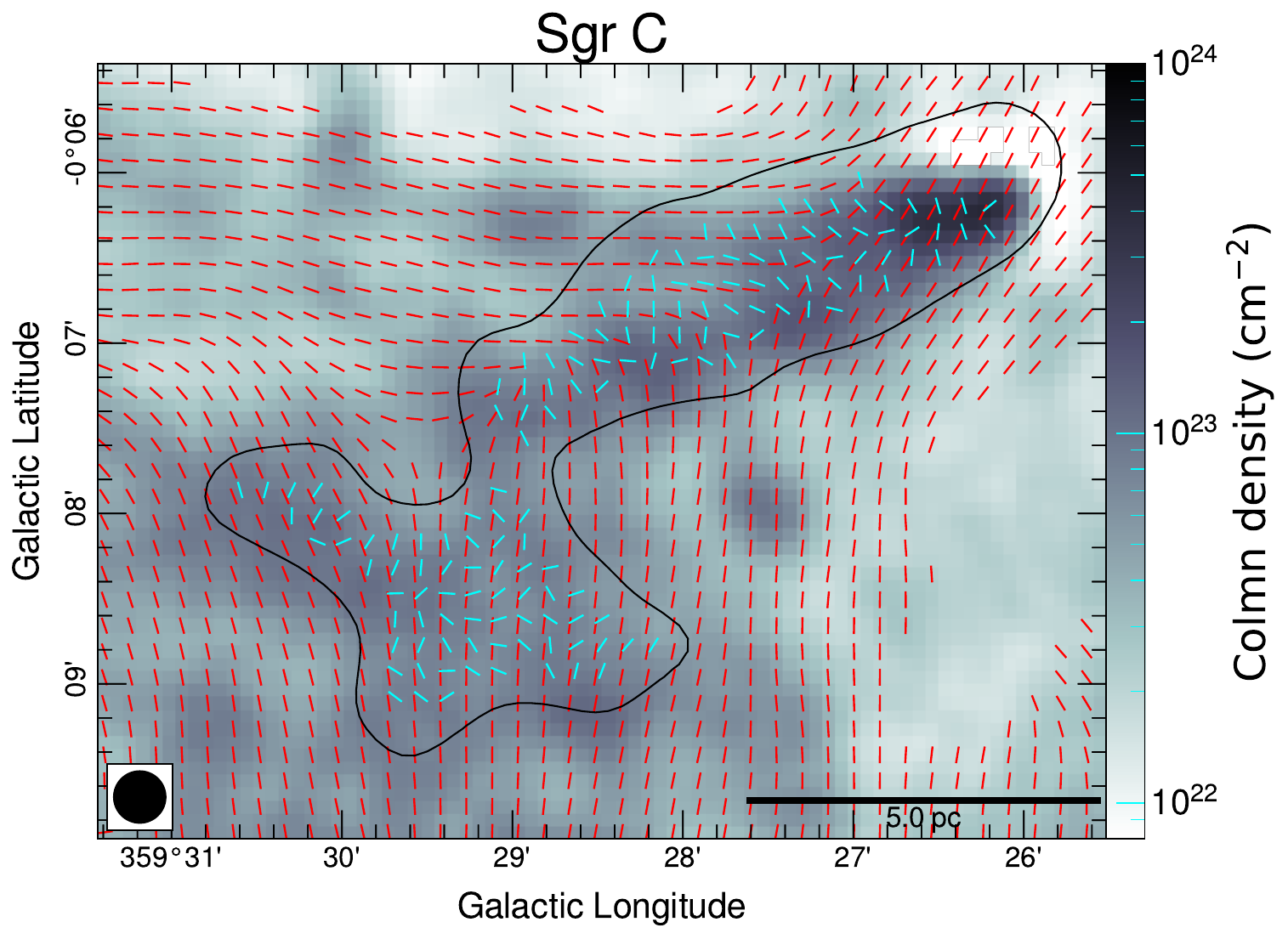}
    \includegraphics[width=0.54\linewidth]{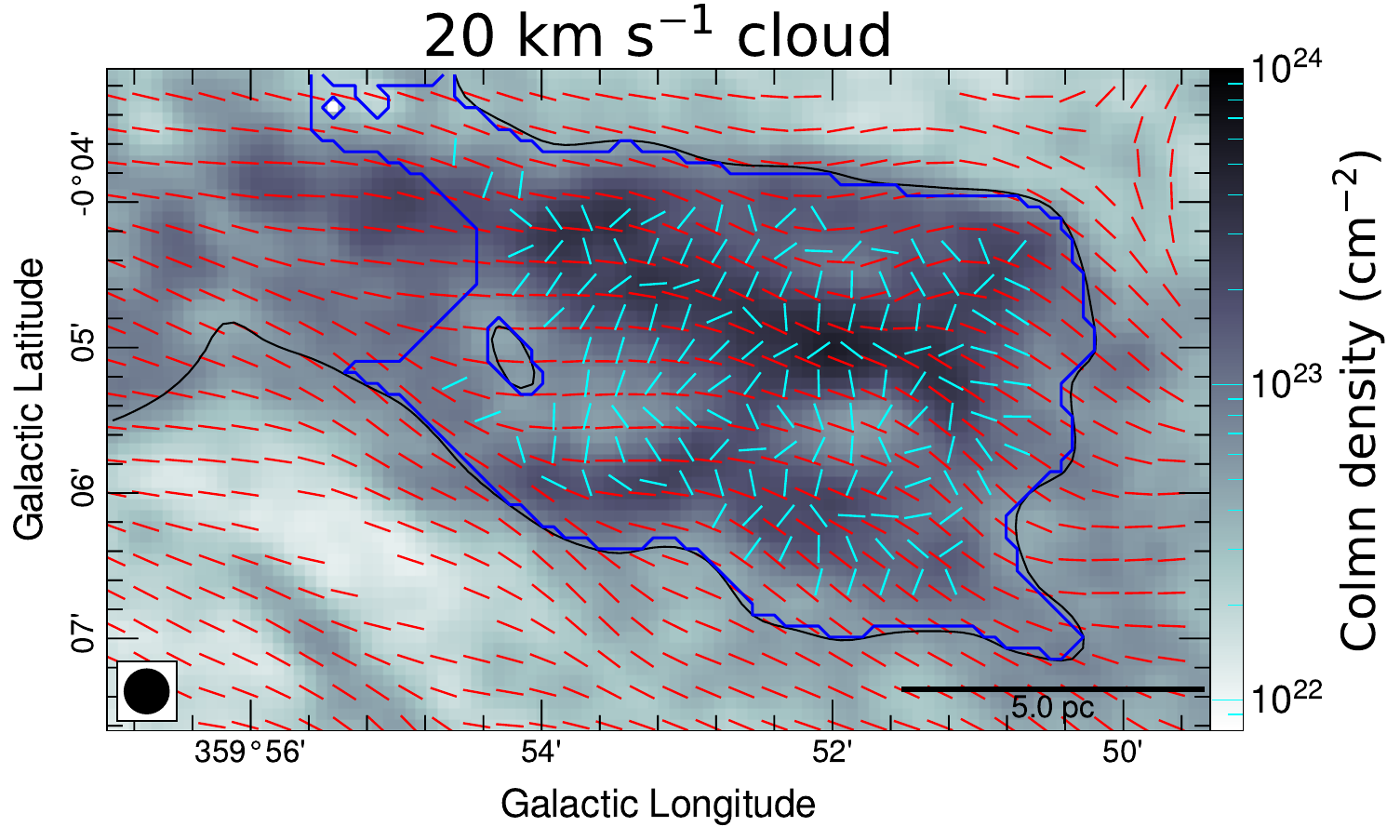}\\
    \includegraphics[width=0.44\linewidth]{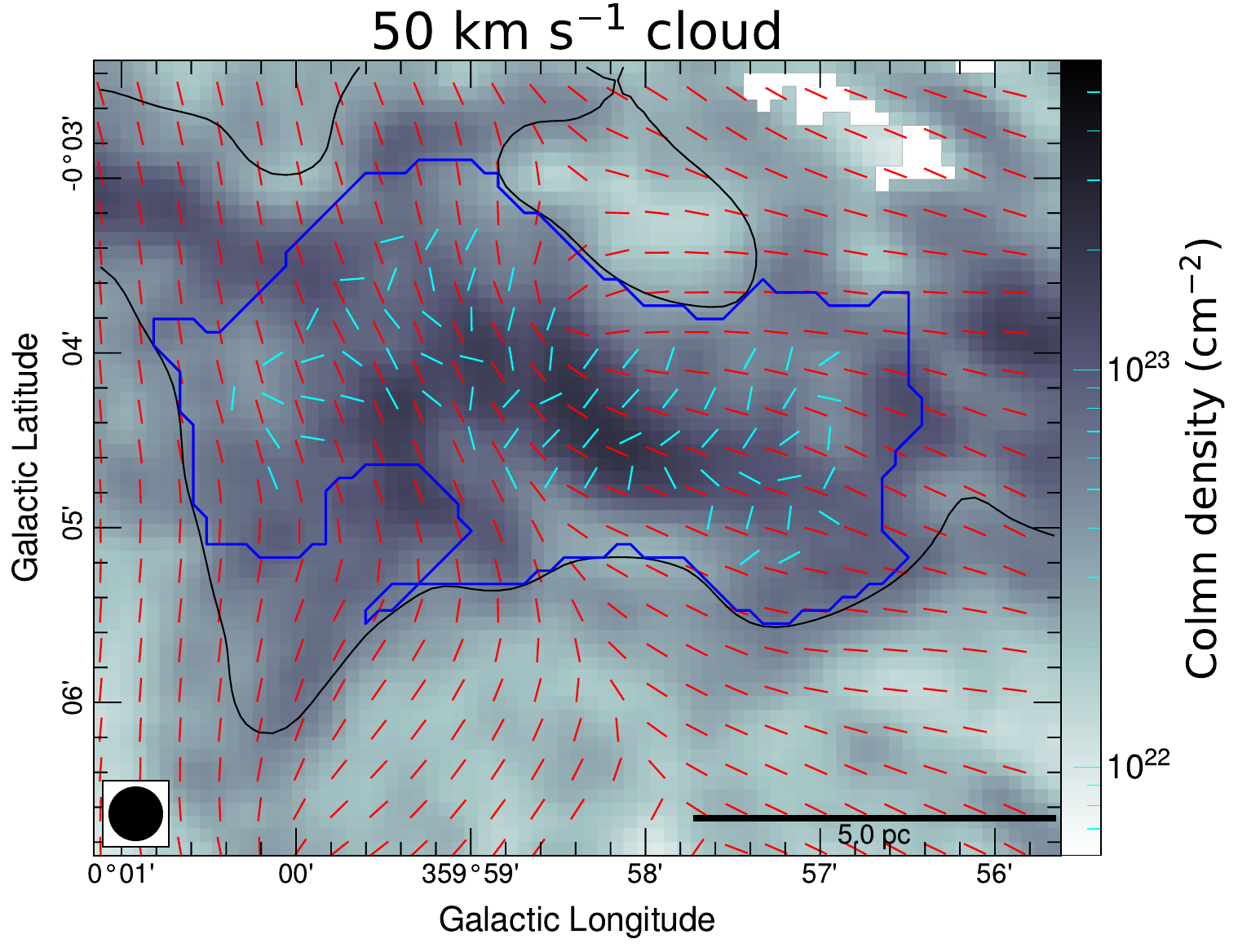}
    \includegraphics[width=0.55\linewidth]{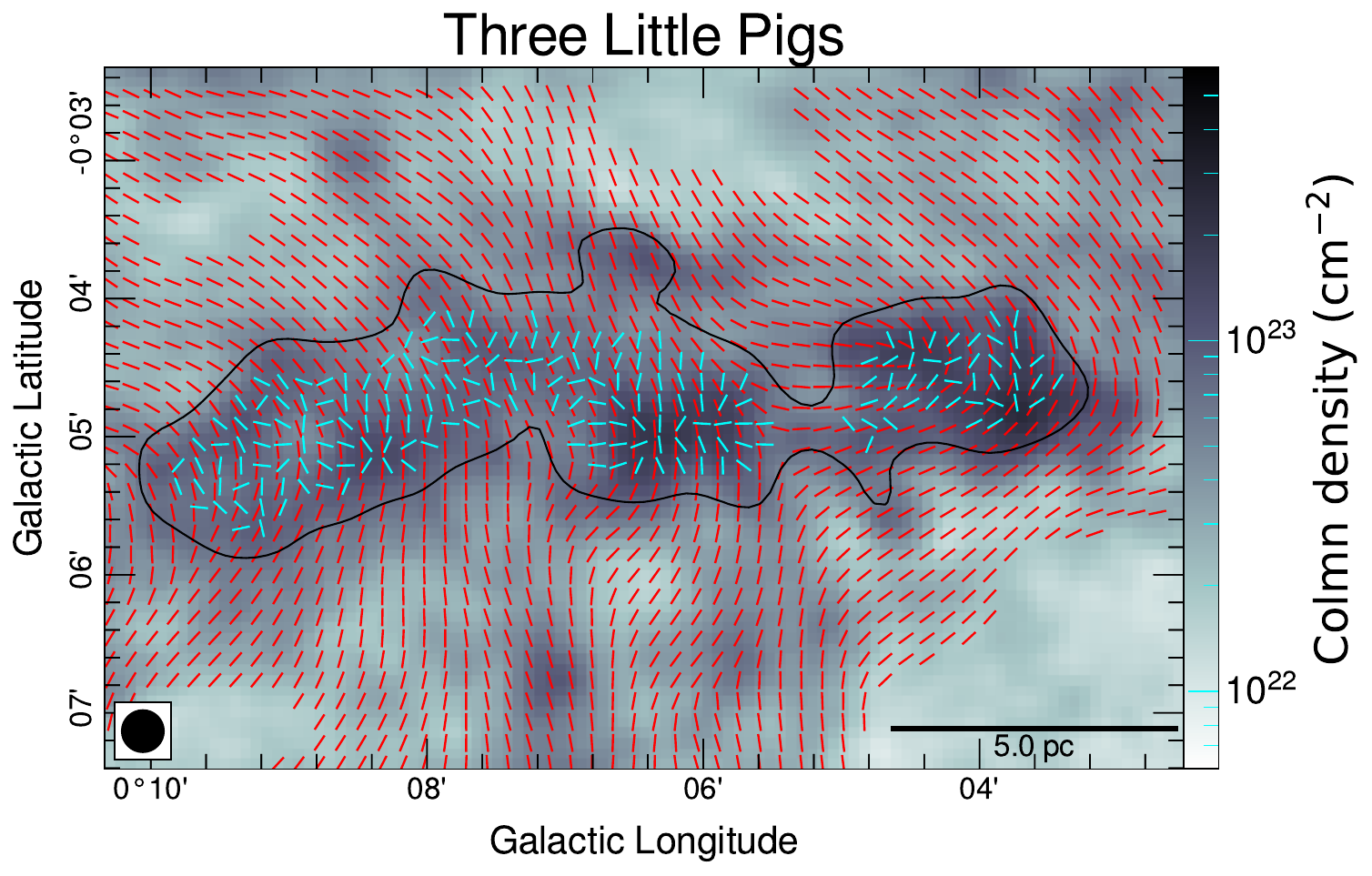}\\
    \includegraphics[width=0.59\linewidth]{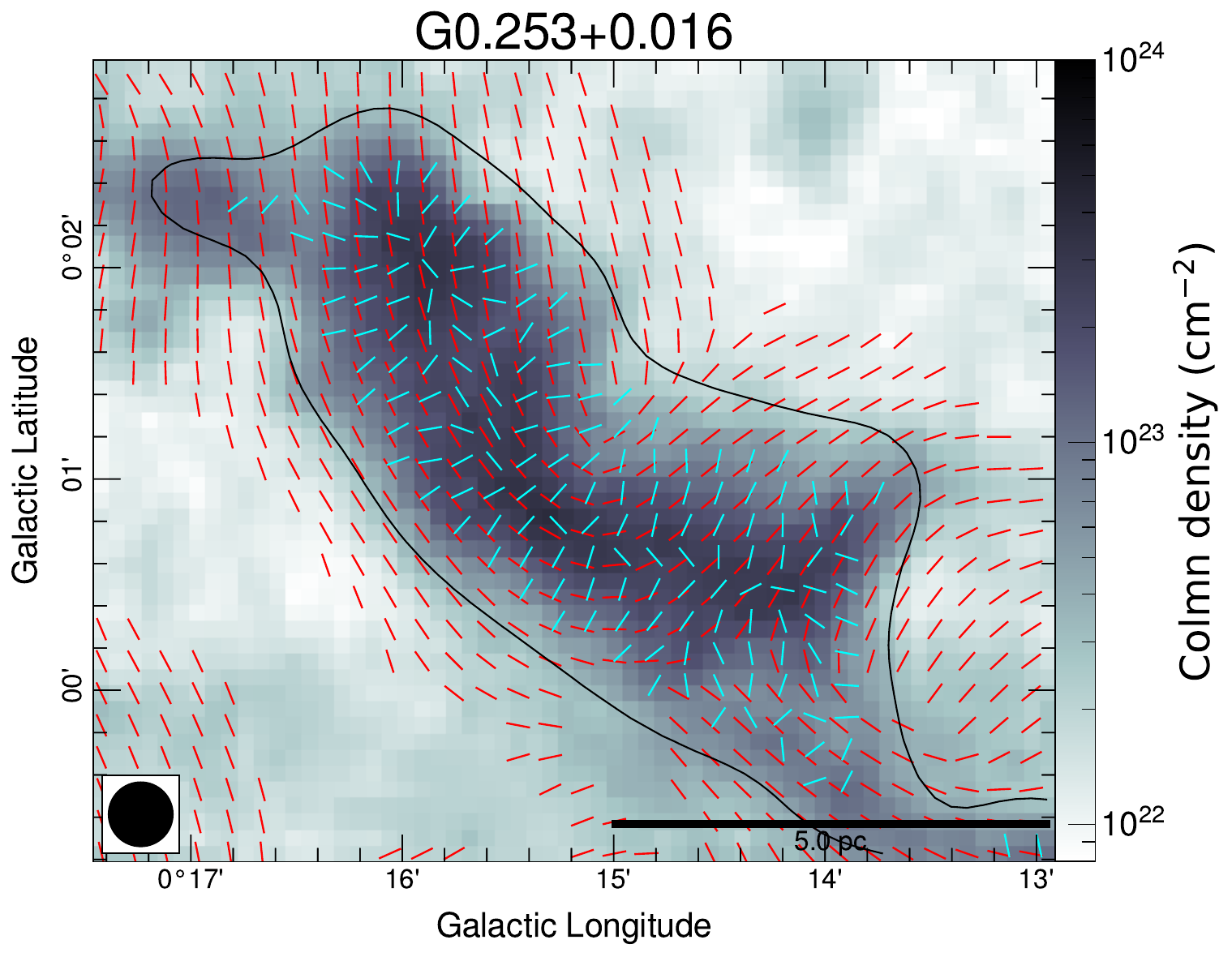}
    \includegraphics[width=0.4\linewidth]{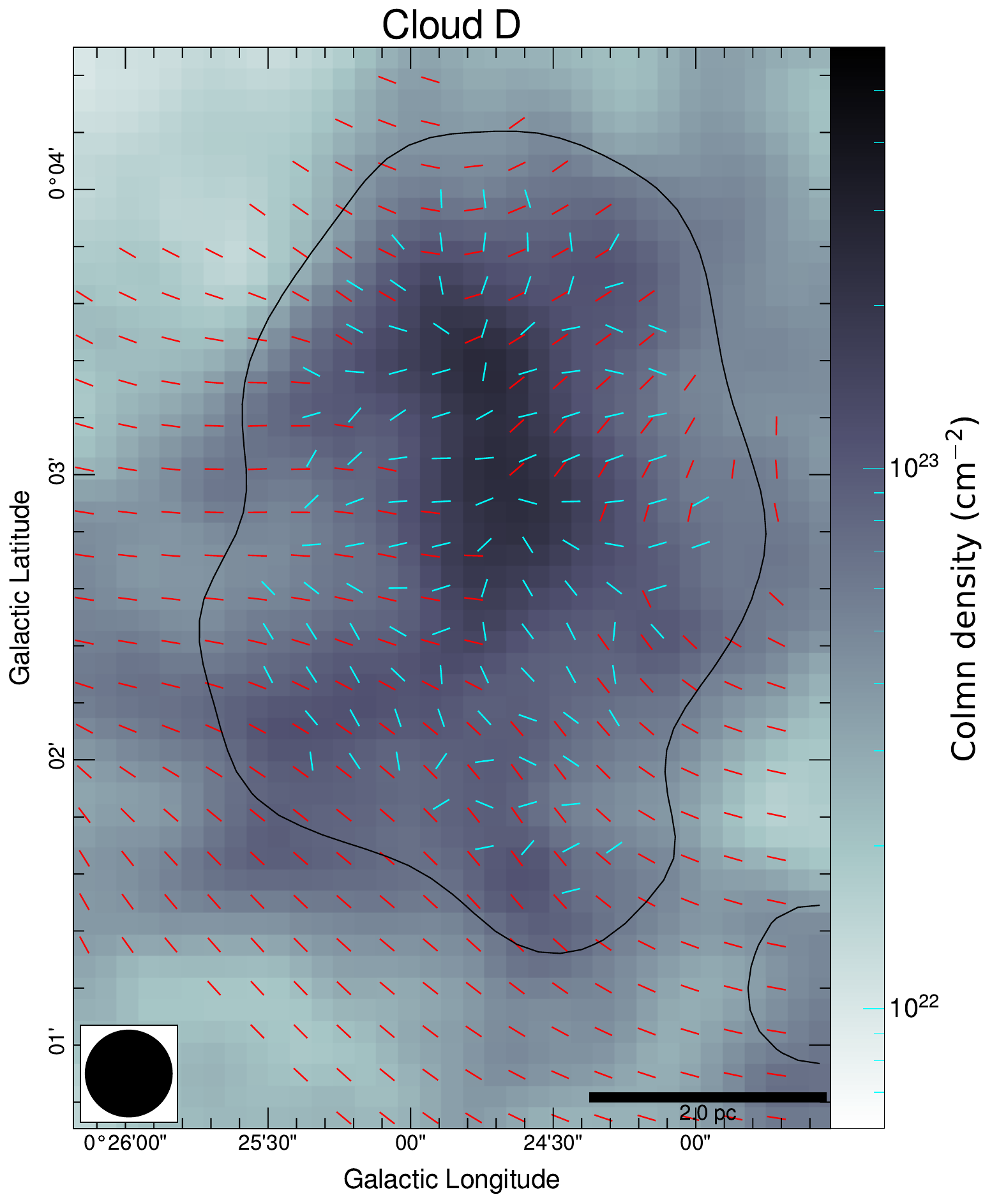}
    \caption{Comparisons between  orientations of magnetic fields and column density structures for individual molecular clouds. Red and cyan segments represent plane-of-the-sky magnetic field orientations and column density gradient orientations, respectively. The black contour marks an intensity threshold of 108 Jy/beam for 214 $\mu$m SOFIA/HAWC+ Stokes I emission, indicating the region used for HRO analysis. Blue contours in the 20 $\mathrm{km~s^{-1}}$ and 50 $\mathrm{km~s^{-1}}$ clouds highlight regions used for HRO analysis which are separated by kinematics.}
    \label{fig:ro_plots}
\end{figure*}

\begin{figure*}[!ht]
    \includegraphics[width=0.48\linewidth]{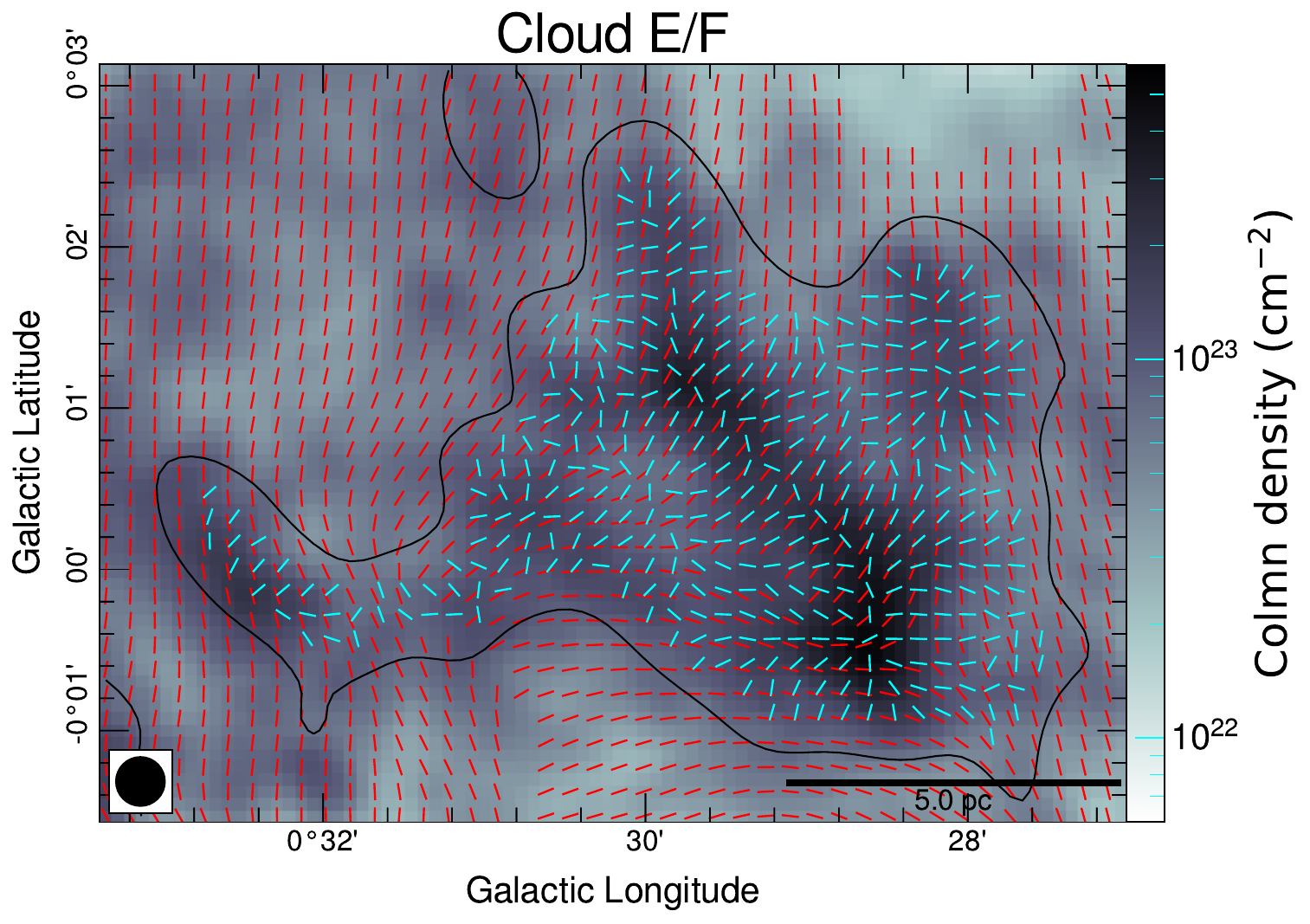}
    \includegraphics[width=0.48\linewidth]{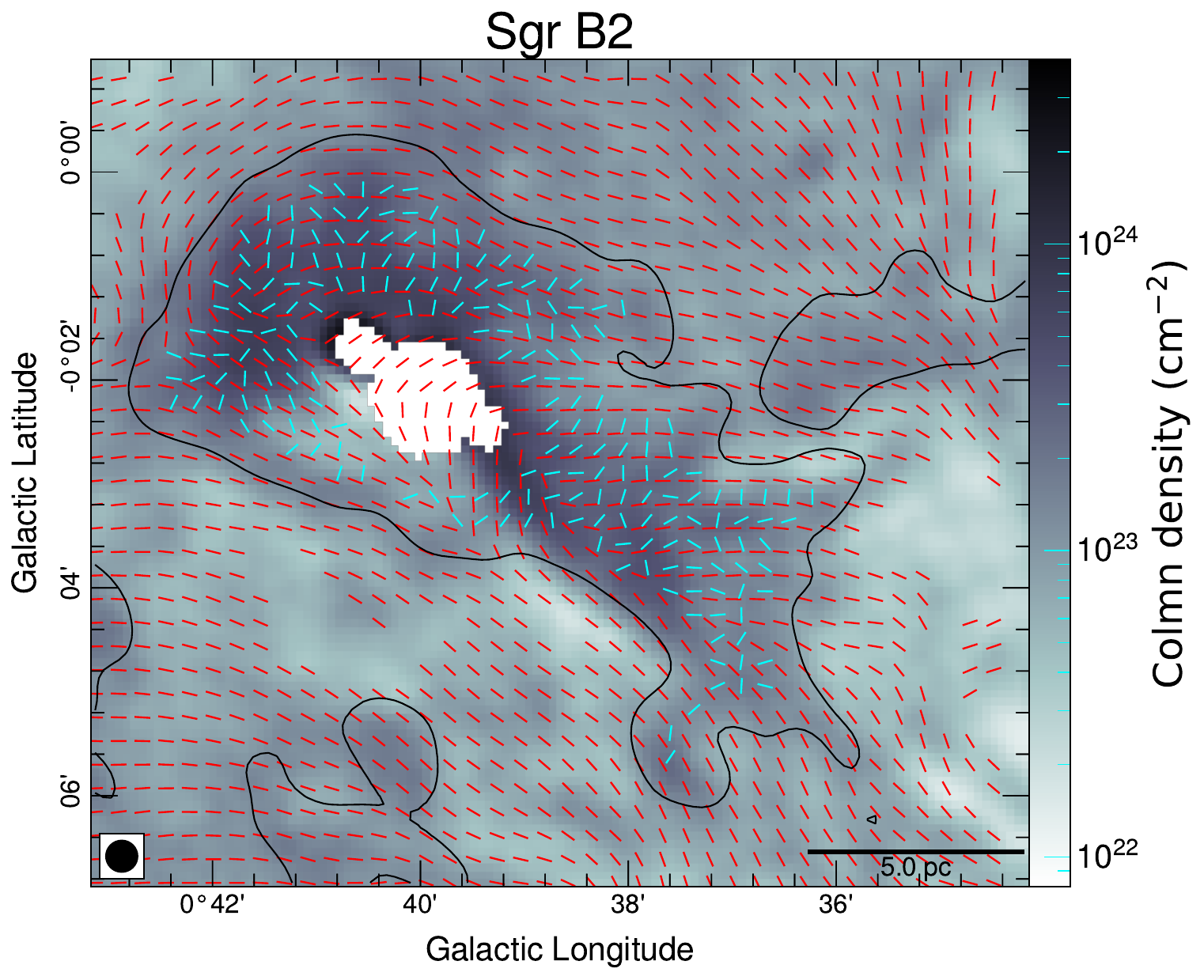}\\
    \caption{Comparison between the orientations of magnetic field and column density structure for Cloud E/F and Sgr B2. The panel format is the same as for Fig. \ref{fig:ro_plots}. For Sgr B2, the black contour marks an intensity threshold of 145 Jy/beam for 214 $\mu$m SOFIA/HAWC+ Stokes I emission.}
    \label{fig:ro_plots_more}
\end{figure*}

CMZ contains several molecular clouds like Sgr B2, Sgr C, $20~\mathrm{km~s^{-1}}$ cloud, $50~\mathrm{km~s^{-1}}$ cloud, G0.253+0.016 (the \emph{Brick}), Dust ridge clouds, showing different levels of star formation activity. Some of these clouds, like Sgr C and $50~\mathrm{km~s^{-1}}$ cloud, have higher star formation rates ($\sim10^{-2}M_\odot~\mathrm{yr^{-1}}$), in agreement with the empirical relation between the star formation and the amount of the dense gas \citep[][]{Kauffmann2017,Barnes2017,Walker2018,Lu2019}. With its high star formation rate and efficiency, Sgr B2 hosts one of the rare mini-starburst regions in our Galaxy \citep[][]{Gaume1995,Qin2011,Ginsburg2018,Pan2024}. In contrast, other clouds such as G0.253+0.016 are more quiescent \citep[][]{Walker2021,Lu2015}. In Fig. \ref{fig:HRO_MCs}, we present the distribution of relative orientations in these molecular clouds, with each density bin contains 180 pixels ($\sim$10 independent polarization measurements). We also verify that changing the number of independent measurements in each density bin by a factor of two does not significantly alter the trend on the relative orientation-column density relations. However, increasing the number of measurements per bin can smooth out alignment transitions, particularly in high-density bins where a parallel-to-perpendicular shift occurs. To preserve these alignment changes, we adopt 10 independent measurements per density bin. We use a intensity threshold of 108 Jy/beam (corresponding to a column density of $4\times10^{22}~\mathrm{cm^{-2}}$ with a dust temperature of 30 K) on 214 $\mu$m SOFIA/HAWC+ Stokes I emission to define the coverage of individual molecular clouds, expect for Sgr B2, as indicated by the black contour in Figs. \ref{fig:ro_plots} and \ref{fig:ro_plots_more}. The cyan segments show the orientations of density gradients in each molecular cloud. For Sgr B2, considering its high brightness, we applied a higher threshold of 145 Jy/beam (corresponding to a column density of $6\times10^{22}~\mathrm{cm^{-2}}$ with a dust temperature of 30 K). 

Some molecular clouds, such as Dust Ridge Clouds B and C, are relatively small and lack sufficient independent polarization detections ($\lesssim12$) to reliably trace magnetic field alignments across different density bins. As a result, we exclude these clouds from our analysis. Future higher-resolution observations will be necessary to study their magnetic field–column density relationships in detail. The $20~\mathrm{km~s^{-1}}$ and $50~\mathrm{km~s^{-1}}$ clouds (hereafter referred to 20MC and 50MC, respectively), on the other hand, are close to each other in the plane-of-sky and are both exceptionally bright, making it impossible to spatially distinguish them using a continuum intensity threshold of 108 Jy/beam. Instead, we separated them using kinematic information. The systematic velocities for 20MC and 50MC are 20 $\mathrm{km~s^{-1}}$ and 50 $\mathrm{km~s^{-1}}$, respectively. Based on \cite{Henshaw2016}, both clouds have velocity dispersions around 9 $\mathrm{km~s^{-1}}$, corresponding to a line full width at half maximum of about 21 $\mathrm{km~s^{-1}}$. We integrated the $\mathrm{HNCO~(4_{0,4}-3_{0,3})}$ data from the Mopra CMZ survey \citep[][]{Jones2012} over the velocity ranges of 6-26 $\mathrm{km~s^{-1}}$ for the 20MC and 38-58 $\mathrm{km~s^{-1}}$ for the 50MC, excluding the overlapping regions. The blue contours in Fig. \ref{fig:ro_plots} show the coverage of 20MC and 50 MC. The``Three Little Pigs" \citep[TLP,][]{Battersby2020} cloud complex consists of M0.145-0.086 (Straw), M0.106-0.082 (Sticks), and M0.068-0.075 (Stone). The spatial and kinematic separations between each other are relatively small, suggesting that they are likely associated. Therefore, we analyze TLP as a single cloud to study the relative orientation between the magnetic field and the column density structure.

We find various alignment behavior in different molecular clouds. In 20MC, 50 MC and G0.253+0.016, almost all density bins exhibit positive AM values, indicating a preferential parallel alignment between the magnetic field and the density structure in these regions. 
\begin{table*}[!ht]
\caption{Physical properties of the individual molecular clouds}
\begin{tabular}{lllllllllc}
\hline
\hline
Source & Mass\tablenotemark{a} & $\mathrm{R_{eff}}$\tablenotemark{b} & $n(H_2)$ & $\sigma_\mathrm{turb}$ & $B_{pos}$\tablenotemark{c}  & $\lambda$ & $\mathcal{M}_A$ & $\alpha_{k+B}$\tablenotemark{d} & RO\tablenotemark{e} \\
& $(10^4~\mathrm{M_\odot})$ & (pc) & $(10^3~\mathrm{cm}^{-3})$ & ($\mathrm{km~s^{-1}}$) & (mG) & & & & evolution \\
\hline
Sgr C & 6.7 & 3.3 & 6.1 & 6.9 & 0.28 & 1.8 & 2.0 & 1.8-2.9 & $\parallel$ to $\perp$/R \\
20 $\mathrm{km~s^{-1}}$ & 23.4 & 4.9 & 6.8 & 7.7 & 0.55 & 1.5 & 1.3 & 1.2-1.9 & $\parallel$ \\
50 $\mathrm{km~s^{-1}}$ & 8.4 & 3.7 & 5.4 & 10.2 & 0.98 & 0.5 & 0.8 & 3.8-6.0 & $\parallel$/R to $\parallel$ \\
Three Little Pigs & 9.1 & 3.8 & 5.6 & 8.9 & 0.28 & 1.9 & 2.5 & 2.4-4.0 & $\perp$/R to $\perp$ \\
G0.253+0.016 & 9.7 & 3.3 & 9.2 & 14.8 & 1.17 & 0.6 & 1.3 & 5.5-9.0 & $\parallel$ \\
Cloud D & 3.8 & 2.5 & 8.6 & 10.0 & 0.64 &  0.8 & 1.5 & 4.7-7.7 & R \\
Cloud E/F & 19.4 & 4.7 & 6.2 & 10.8 & 0.71 & 1.0 & 1.3 & 2.3-3.6 & $\perp$ to $\parallel$/R to $\perp$ \\
Sgr B2\tablenotemark{$\dagger$} & 325.0 & 6.1 & 48.1 & 9.3 & 0.45 & 20.3 & 6.1 & 0.1-0.2 & $\parallel$ to $\perp$\tablenotemark{*} \\
\hline
\end{tabular}
\label{tab:mc_proper}
\tablecomments{\tablenotemark{a} Gas mass for individual molecular clouds derived from the column density map. \tablenotemark{b} Effective radius for the molecular cloud. \tablenotemark{c} The total plane-of-sky magnetic field strength estimated by ADF method. \tablenotemark{d} Virial parameter with power-law index of density profile, $\rho\propto r^{-\beta}$, ranges from 0 to 2. \tablenotemark{e}The evolution of the relative orientation between the magnetic field and the column density structures within each cloud as a function of increasing column density. ``$\parallel$'' means preferential parallel orientation. ``$\perp$'' means preferential perpendicular orientation. ``R'' means no preferential orientation. \tablenotemark{*} means that Sgr B2 shows transition from parallel to perpendicular when we include the highest density regions which are saturated in 160 $\mu$m \emph{Herschel} data. \tablenotemark{f}Star formation rate for each molecular cloud referred from \cite{Hatchfield2024}. The star formation rate of Sgr B2 is from \cite{Ginsburg2018}. \tablenotemark{$\dagger$} The properties of Sgr B2 are from \cite{Pan2024} since the dense regions in Sgr B2 are masked out in the column density map.}
\end{table*}

Sgr B2 also shows positive AM values across most column density bins and a decreasing trend with increasing column density. However, some intermediate-density bins exhibit AM values close to zero, suggesting a lack of preferential alignment between the magnetic field and column density structures in those regions. In particular, the two massive dense cores (North, and Main) in Sgr B2 are saturated in the \emph{Herschel} 160 $\mu$m data, resulting in blank regions in the column density map. Previous studies \citep{Lis1991, Huttemeister1993, Pan2024} have shown that these regions have extremely high column densities, exceeding $10^{24}~\mathrm{cm^{-2}}$, leaving us with no information on the relative orientation of the magnetic field in the densest regions of Sgr B2. 

To address this, we derive the column density ($N_\mathrm{H_2}$) map for the densest regions using 214 $\mu$m continuum emission, assuming optically thin dust emission. The column density is calculated as:
\begin{equation}
    N_\mathrm{H_2}=\frac{gS_\nu D^2}{\mu m_H B_\nu(T_d)\kappa_\nu A},
\end{equation}
where $g=100$ is the gas-to-dust ratio, $S_\nu$ is the flux per pixel, $D=8.3$ kpc is the distance, $\mu=2.8$ is the mean molecular weight, $B_\nu(T_d)$ is the Planck function at the dust temperature $T_d$, $\kappa_\nu$ is the dust opacity, and $A$ is the physical area of each pixel. The dust opacity is assumed to follow $\kappa_\nu=\kappa_0(\nu/300~\mathrm{GHz})^\beta$, consistent with the approach used in our SED fitting. We adopt a dust emissivity index $\beta=2.0$, similar to the mean values observed in other dense molecular clouds such as Sgr C ($\bar{\beta}=2.2$) and G0.253+0.16 ($\bar{\beta}=1.9$). The dust temperature is fixed at 20 K for Sgr B2, as suggested by \cite{Pierce-Price2000} and \cite{Etxaluze2013}. Appendix \ref{app:hro_sgrb2} shows the derived column density map of Sgr B2, covering the densest regions. We find great consistency between the column densities obtained from SED fitting and those derived using the single-wavelength method. Appendix \ref{app:hro_sgrb2} also shows the HRO results for newly derived column density map, we find that the relative orientations between magnetic field and column density in Sgr B2 transit from parallel to perpendicular at regions with highest density ($N_\mathrm{H_2}\gtrsim8\times10^{23}~\mathrm{cm^{-2}}$). 

Sgr C shows positive AM values in low density regimes and decrease to the negative AM values as density increases, similar to what we observed in star forming regions in Galactic disk \citep[e.g.,][]{PlanckXXXV2016,Soler2017}. However, the critical density ($N_\mathrm{crit}$) where the transition happens is around $8\times10^{22}~\mathrm{cm^{-2}}$ which is much higher than the typical value ($\sim10^{21}-10^{22}~\mathrm{cm^{-2}}$) in the Galactic disk. Dust Ridge Cloud E/F (also known as Sgr B1 off) presents a more complex situation. The relative orientations between magnetic fields and column density structures shift from perpendicular to parallel and back to perpendicular as density increases. In the Three Little Pigs region, most density bins exhibit negative AM values, indicating a preferential perpendicular alignment between the magnetic field and the density structures. However, some intermediate-density bins have AM values close to zero, suggesting more complex or random relative orientations in those regions. In Dust Ridge Cloud D, all three density bins also show negative AM values, but their magnitudes are close to zero. This suggests that the magnetic field in this cloud may not exhibit a clear preferential orientation relative to the column density structures. Given that this region is not well resolved, higher-resolution observations are needed to more accurately determine the underlying alignment patterns. The relative alignments for each molecular cloud in the CMZ are summarized in Table \ref{tab:mc_proper}.

\cite{Pare2025} also examined the relative orientation between magnetic fields and column density structures in individual molecular clouds, categorizing the clouds into three density bins (low, intermediate, and high) for their analysis. While most clouds (e.g., 20 MC, TLP, G0.253+0.016) exhibited trends consistent with our findings, we find a difference in Sgr C and Cloud E/F when increasing the number of column density bins to better resolve high-density regions. Unlike the parallel alignment reported by \cite{Pare2025} across low, intermediate and high density bins, we find a transition from parallel to perpendicular relative orientations at the highest densities in these clouds. This is possibly due to more bins for high density regions in our analysis which are able to reveal more details of relative orientations in dense regions. In fact, the HRO plots for Sgr C and Cloud E/F in \cite{Pare2025} (see their Figs. 7 and 9) both showed one peak of relative orientation in perpendicular along with another peak in parallel for high density bin, indicating that there are still plenty of dense regions in these two molecular clouds showing perpendicular alignment.

\subsection{Comparison with MHD simulations of CMZ}\label{subsec:MHDsimu}
Numerical simulations \citep[e.g.,][]{Soler2013,Seifried2020,Girichidis2021} demonstrate that projection effects can significantly impact 2D relative orientation analyses. For example, a parallel or random alignment observed in 2D ($\mathrm{AM} > 0$) does not necessarily rule out a perpendicular alignment in 3D ($\mathrm{AM}< 0$) between magnetic fields and dense structures \citep[see Appendix C in][]{PlanckXXXV2016}. To check whether the preferential parallel alignment observed in the CMZ is genuine, we examined a 3D MHD simulation that approximate CMZ-like conditions (Tress et al., in prep).


The simulations are designed to study gas dynamics and star formation in the CMZ in the presence of magnetic fields and stellar feedback. They build upon those presented in \citet{Tress2020, Tress2024}, using the same initial conditions but incorporating additional physical processes and higher resolution. They are performed using the moving-mesh code AREPO \citep[][]{Springel2010,Weinberger2020} and include an external barred potential fine-tuned to the Milky Way \citep[][]{Hunter2024}, a time-dependent chemical network that keeps track of hydrogen and carbon chemistry \citep[][]{Glover2012}, a physically motivated model for the formation of new stars using star particles, supernova feedback and ionising radiation feedback from massive stars through the SWEEP method for “on the fly” radiative transfer \citep[][]{Peter2023}, and magnetic fields through the ideal MHD scheme implemented in AREPO \citep[][]{Pakmor2011,Pakmor2013}. The simulations follow the ISM evolution in the entire barred region of the simulated Milky Way and are therefore able to self-consistently follow the formation of magnetised molecular clouds and their embedded star formation from the large-scale flow. The adopted mass resolution is 20 $M_\odot$ which corresponds to spatial resolution $<$ 1 pc for ISM densities above $10^2~\mathrm{cm^{-3}}$. 

\begin{figure*}[!ht]
    \centering
    \includegraphics[width=0.95\textwidth]{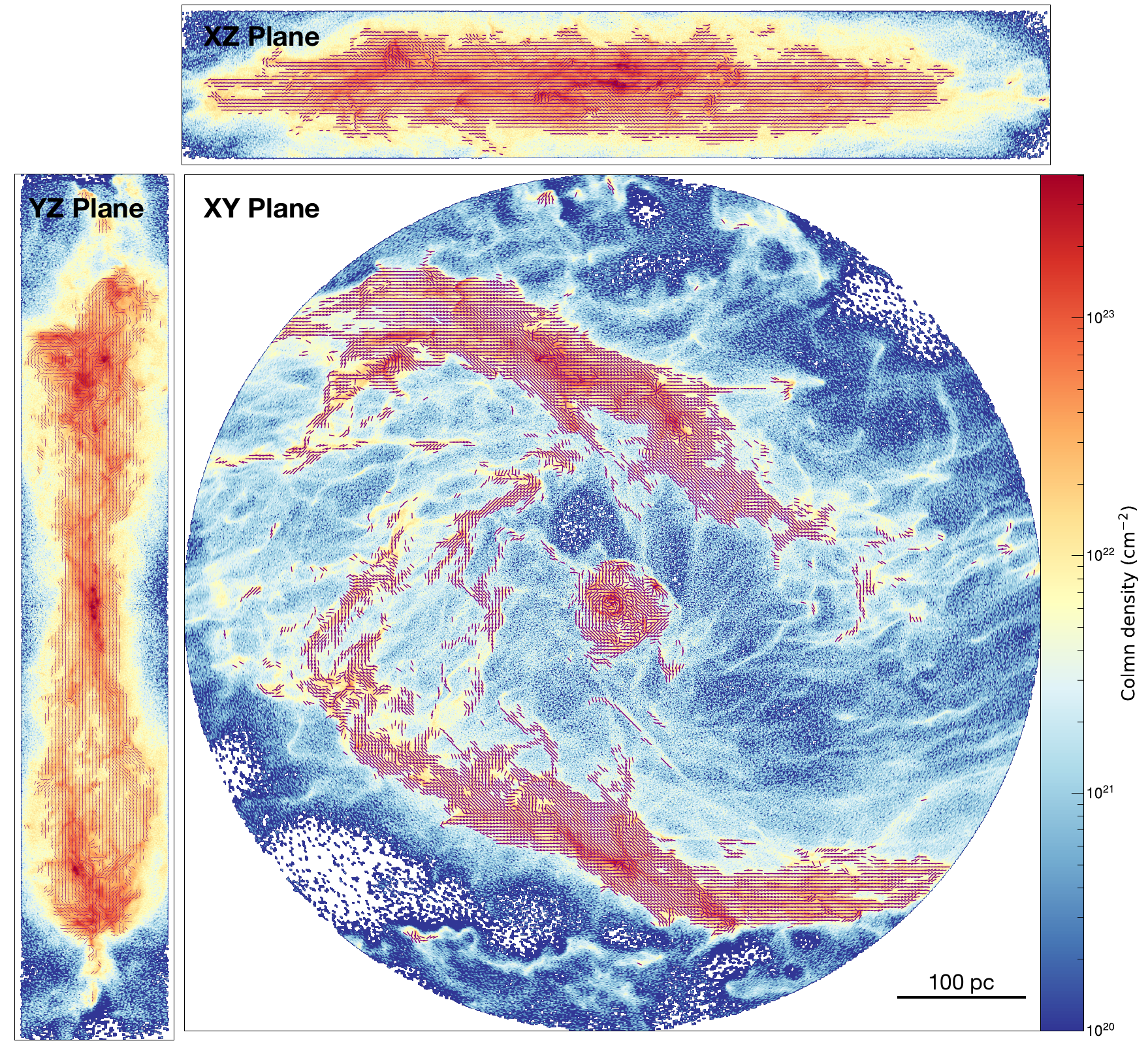}
    \caption{Visualization of the magnetic field and column density in the CMZ for the synthetic observations of the MHD simulation. The color-scale represents the gas column density. Purple segments represent the projected magnetic field orientation.}
    \label{fig:cmz_simu_xyz}
\end{figure*}

To compare MHD simulations with observations, we used the radiative transfer code POLArized RadIation Simulator \citep[POLARIS,][]{Reissl2016} to create synthetic dust polarization maps of the CMZ, assuming radiative torque (RAT) alignment to be the primary dust grain alignment mechanism in the CMZ as suggested in \cite{Butterfield2024}. The simulations produced by the AREPO code use unstructured grids with cell sizes that vary spatially. Because the coarsest resolution in the diffuse regions of the simulations is comparable to that of the observational data, we employ POLARIS to project the simulation onto a uniform grid with a spatial resolution of 0.6 pc, which matches the resolution of our observational data. Meanwhile, the AREPO MHD simulations provided the coordinates and masses of heating sources, and we derived the surface temperatures required for POLARIS inputs using the empirical mass-luminosity relation $L_*\propto M_*^{3.5}$ \citep{Kuiper1938} and mass-radius relation $R_*\propto M_*^{0.8}$ \citep[][]{Demircan1991}. In the synthetic observation, we adopts an external interstellar radiation field typical of the Galactic disk, with a strength of $\mathrm{G_0 = 1}$ \citep[][]{Mathis1983}. The gas-to-dust mass ratio is set to 100, and dust grains are modeled as a mixture of 62.5\% astronomical silicates and 37.5\% graphite \citep[][]{Mathis1977}, following a size distribution of $n_d(a) \propto a^{-3.5}$ for grain radii ($a$) ranging from 0.25 $\mathrm{\mu m}$ to 5 mm. The adopted settings for the interstellar radiation field, gas-to-dust ratio, and dust grain properties are representative of typical Galactic disk conditions, which may differ from those in the Galactic center. In this study, the synthetic observation based on these disk-like conditions serves as a simplified test case. Future work will incorporate CMZ-specific properties into the simulations to more accurately reflect the unique physical environment of the CMZ. \cite{Butterfield2024} and \cite{Pare2024} have shown that the FIREPLACE data likely trace the magnetic field local to the CMZ. Therefore, to investigate effect of the integration along LOS on the relative alignment, we only focus on the CMZ region in the simulation.

\begin{figure*}[!ht]
    \centering
    \includegraphics[width=0.48\textwidth]{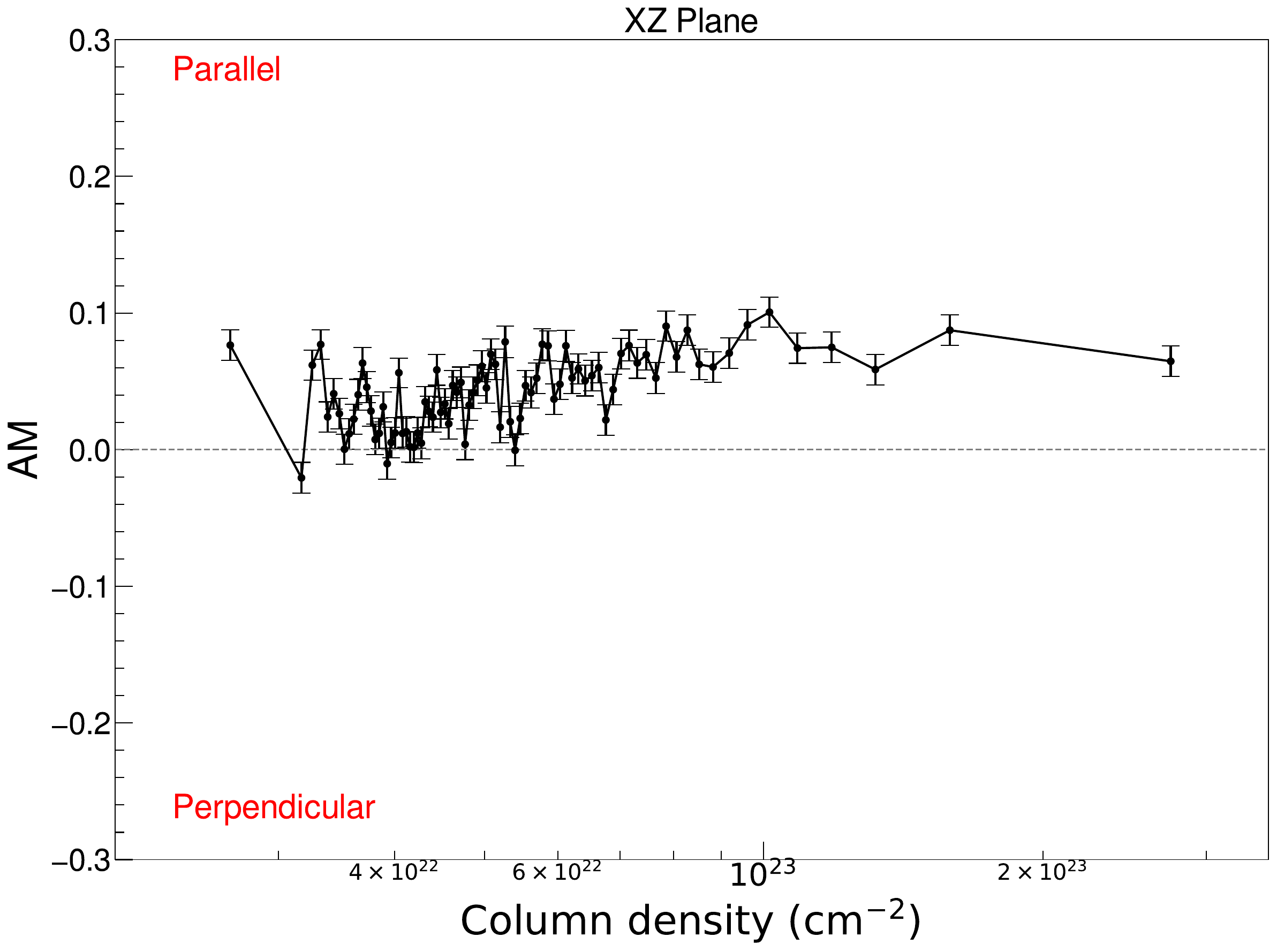}
    \includegraphics[width=0.48\textwidth]{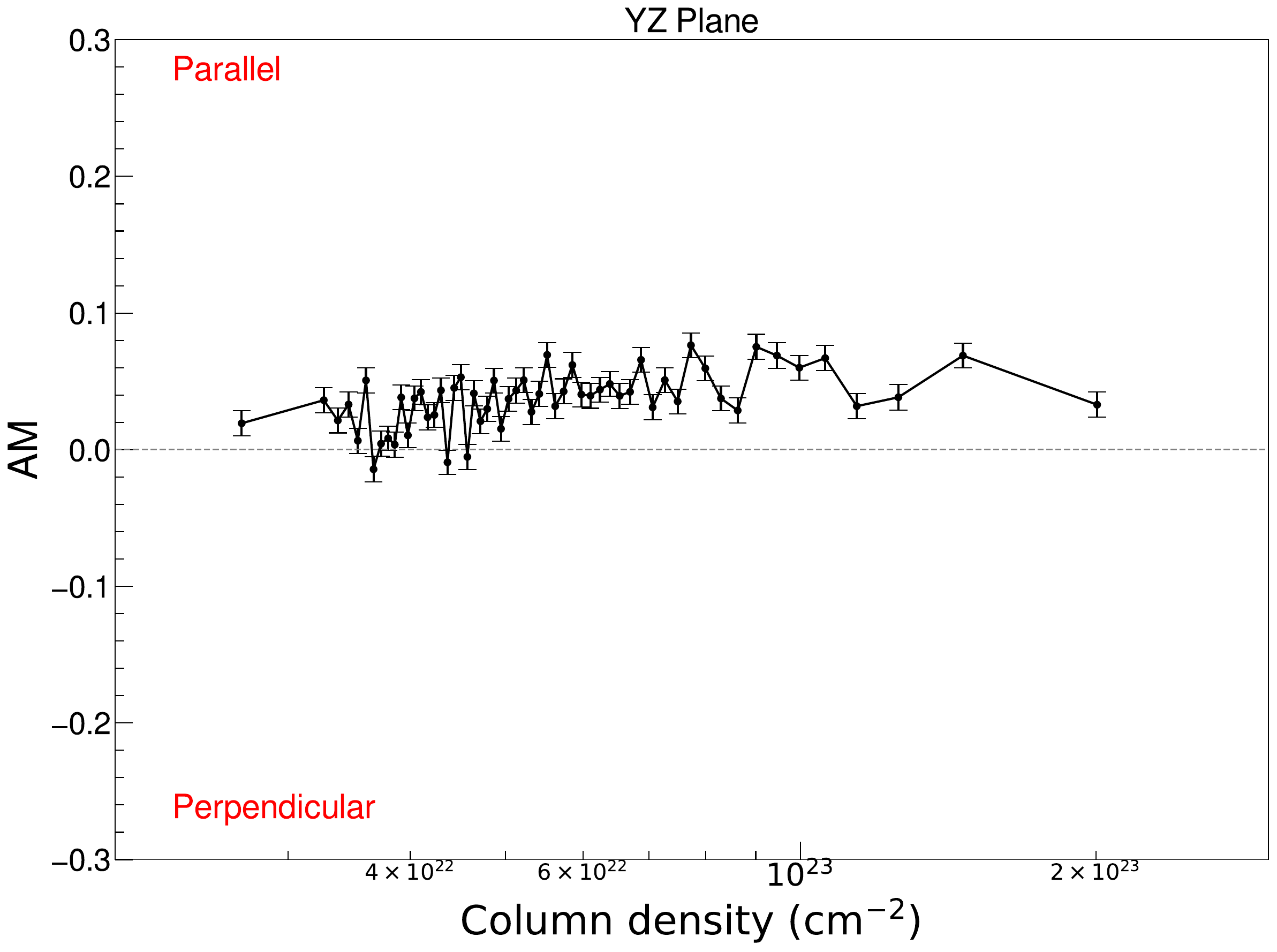}\\
    \includegraphics[width=0.48
    \textwidth]{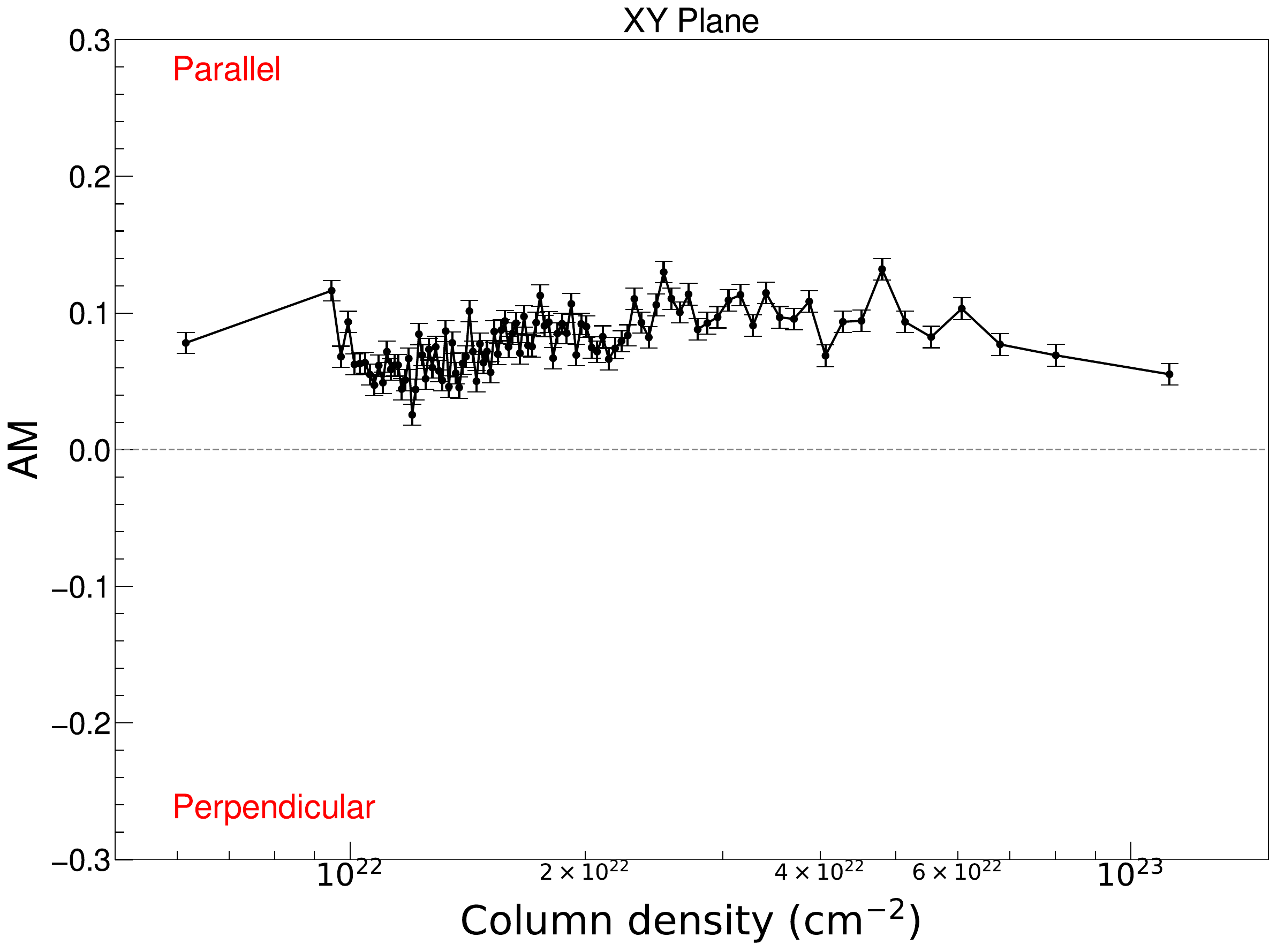}
    \includegraphics[width=0.48\textwidth]{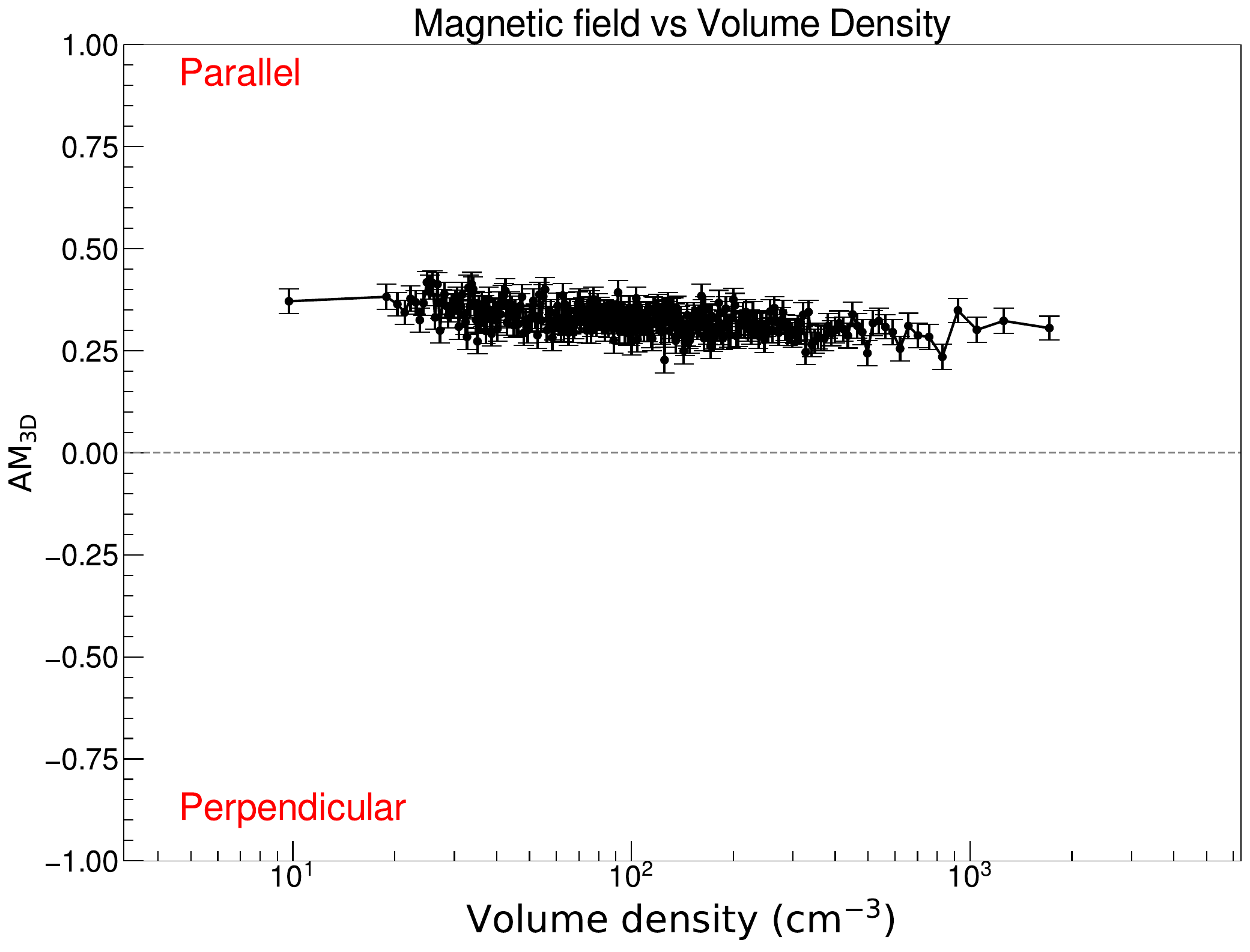}\\
    \caption{Alignment measures derived from MHD simulations of the CMZ. Top left: Projected 2D AM along the Y-axis (edge-on view). Top right: Projected 2D AM along the X-axis (edge-on view). Bottom left: Projected 2D AM along the Z-axis (face-on view). Bottom right: 3D alignment measures ($\mathrm{AM_{3D}}$) between magnetic fields and volume density structures across different density bins.}
    \label{fig:HRO_simu}
\end{figure*}

Fig. \ref{fig:cmz_simu_xyz} shows synthetic 214 $\mu$m observations of the simulated magnetic field overlaid on a column density map covering the entire CMZ, viewed along three orthogonal planes (xy, yz, and xz) at a distance of 8.1 kpc. Following the HRO analysis above, we derived the column density gradient and applied density masks for different viewing angles. For the edge-on views (xz and yz planes), we used a density threshold of $N_\mathrm{H_2} = 1.0 \times 10^{22}~\mathrm{cm^{-2}}$. For the face-on view (xy plane), a lower threshold of $N_\mathrm{H_2} = 3 \times 10^{21}~\mathrm{cm^{-2}}$ was used due to the reduced line-of-sight material integration. Applying HRO analysis to the synthetic polarization data across these directions (see Fig. \ref{fig:HRO_simu}), we found that edge-on views (xz and yz planes) show parallel alignment between magnetic fields and column density structures in high-density bins ($N_\mathrm{H_2} \gtrsim 5 \times 10^{22}~\mathrm{cm^{-2}}$) and no preferred alignment in low-density bins, which is consistent with what we observed in SOFIA data, while the face-on view (xy plane) shows parallel alignment across all density bins. 

To connect the alignment in the 2D polarization map with the actual conditions in the MHD simulation, we calculate the relative alignment between the magnetic field and the volume density ($n$) in 3D. This alignment is quantified by the angle $\psi = \measuredangle(\mathbf{B}, n)$, which measures the orientation between the 3D magnetic field and iso-density contours. To maintain consistency with the 2D case, we define $\mathrm{AM_{3D}} = \left\langle \cos 2\psi \right\rangle$, where $\mathrm{AM_{3D}} > 0$ indicates parallel alignment, $\mathrm{AM_{3D}} < 0$ indicates perpendicular alignment, and $\mathrm{AM_{3D}} \sim 0$ indicates no preferred alignment. The bottom right panel of Fig. \ref{fig:HRO_simu} shows positive $\mathrm{AM_{3D}}$ values across all density bins, indicating that the 3D magnetic fields are generally aligned with the volume density structures. In low-density regions, this alignment is also reflected in the face-on (XY plane) 2D projection. However, in the edge-on views (XZ and ZY planes), the projected 2D magnetic fields exhibit no clear preferential alignment at low column densities. A plausible explanation for this discrepancy is the presence of multiple diffuse gas components along the line of sight in these edge-on perspectives. This line-of-sight superposition can obscure the underlying 3D alignment, resulting in a randomization of the projected relative orientation due to averaging over many uncorrelated structures.

In general, we find preferentially parallel alignment in projected 2D and 3D synthetic observations across different densities, suggesting that the observed parallel alignments in SOFIA data are likely to be real instead of caused by a projection effect.
 
\section{Role of magnetic fields in CMZ clouds}\label{sec:BfieldRole}
Previous studies in the Galactic disk found a general trend in which relative orientations between gas column density structures and magnetic fields transition from parallel at low column densities to preferentially perpendicular at highest column densities \citep[e.g.,][]{PlanckXXXV2016,Soler2019,Chen2024}. The transition from parallel to perpendicular alignment usually occurs at $N_\mathrm{H_2}\sim10^{21}-10^{22}\mathrm{cm^{-2}}$. Numerical simulations \citep[e.g.,][]{Soler2013,Chen2016,Soler2017,Seifried2020} suggested that the change in relative orientation may be related to energy balance in the clouds and the transition occurs in regions in gravitational collapse. Therefore, due to the support provided by strong magnetic field and turbulence in the CMZ, the density where the transition happens in the CMZ may be much higher than that in the Galactic disk. To assess energy balance, we estimate the magnetic field strength and compare its energy with turbulent kinetic and gravitational energy.
\subsection{Magnetic field strength for individual clouds}
The Davis-Chandrasekhar-Fermi (DCF) method \citep{Davis1951, Chandrasekhar1953} is widely used to estimate the plane-of-sky magnetic field strength ($B_\mathrm{pos}$), assuming magnetic field perturbations are caused by turbulent motion. In this work, we apply a variant of the DCF method, the angular dispersion function (ADF) analysis \citep{Falceta-Goncavles2008, Houde2009, Hildebrand2009,Houde2016}, to quantify variations in magnetic field orientations. For a detailed review of the DCF and ADF methods, we refer readers to \cite{Liu2022}.

Following \cite{Hildebrand2009}, we can obtain the magnetic field strength by:
\begin{equation}
    B_0=\sqrt{\mu_0\rho}\ \sigma_\mathrm{turb}\left[\frac{\langle B^2_t \rangle}{\langle B^2_0 \rangle} \right]^{-1/2}
\end{equation}
\begin{equation}
B_\mathrm{pos}= Q_c B_0\sqrt{1+\left[\frac{\langle B^2_t \rangle}{\langle B^2_0 \rangle} \right]}
\end{equation}
where $\mu_0$ is the vacuum permeability, $\rho=\mu_\mathrm{H_2}m_\mathrm{H_2}n_\mathrm{H_2}$ is the average mass density of gas, $\sigma_\mathrm{turb}$ is the turbulent velocity dispersion of the cloud, $(\langle B^2_t \rangle/\langle B^2_0 \rangle)^{-1/2}$ is the turbulent-to-ordered magnetic field strength ratio, and $Q_c$ is the correction factor. Assuming a spherical geometry for each cloud, we can derive the average volume density ($n_\mathrm{H_2}$) of each cloud using an effective radius $R_\mathrm{eff}$ and gas mass ($M_\mathrm{gas}$) derived from the column density map:
\begin{equation}
    n_\mathrm{H_2}=3M_\mathrm{gas}/4\mu_\mathrm{H_2}m_\mathrm{H}\pi R^3_\mathrm{eff}   
\end{equation}
The effective radius of each molecular cloud, $R_\mathrm{eff} = \sqrt{A_\mathrm{cloud}/\pi} $, is derived from its enclosed area $A_\mathrm{cloud}$.
 $(\langle B^2_t \rangle/\langle B^2_0 \rangle)^{-1/2}$ can be derived from ADF expressed as \citep{Houde2009,Houde2016}:
\begin{multline}
    1-\langle\cos[\triangle\phi(\ell)]\rangle=\sum^\infty_{j=1}a_{2j}\ell^{2j}+\\ \frac{\langle B^2_t\rangle}{\langle B^2_0\rangle}\left[1-e^{-\ell^2/2(\delta^2+2W^2)}\right]
\end{multline}
where $\Delta\phi(\ell)$ is the angular dispersion of two polarization segments as a function of distance, $\ell$, between them, $\delta$ is the turbulence correlation length, and W is the standard deviation of the Gaussian beam. Here, we do not apply the line-of-sight (LOS) signal integration correction proposed by \cite{Houde2009} to determine the turbulent-to-ordered magnetic field ratio ($\sqrt{{\langle B^2_t\rangle}/{\langle B^2_0\rangle}}$). This is because the numerical study by \cite{Liu2021} indicates that this correction may not work well within the ADF framework. Instead, to account for the LOS signal integration effect, we adopt the numerically derived correction factor $Q_c=0.21$ from \cite{Liu2021} when we derive the plane-of-sky magnetic field strength. The ADF fitting results of the molecular clouds are shown in Appendix \ref{app:fitting_ADF}.

For turbulent velocity dispersion, we used $\mathrm{HNCO~(4_{0,4}-3_{0,3})}$ data from the Mopra CMZ survey \citep[][]{Jones2012}. Following Appendix B in \cite{Pan2024}, we shifted each pixel's spectra to the local intensity-weighted mean velocity and averaged them within the ADF analysis area to separate small-scale turbulence from large-scale bulk motion. A Gaussian function was then fitted to the averaged spectra to determine the velocity dispersion. Given that the dust temperature for most clouds is around 20 K, the thermal motion contribution is negligible. It is worth noting that some regions in the CMZ may exhibit multiple velocity components along the line of sight, potentially broadening the averaged line profiles and leading to an overestimation of the velocity dispersion. For example, \cite{Longmore2012} and \cite{Walker2015} reported smaller velocity dispersions for G0.253+0.016 and Cloud D using a multi-component
Gaussian-fitting of the $\mathrm{HN^{13}C (10-9)}$ emission from the Mopra CMZ survey \citep[][]{Jones2012}. However, the majority of the analyzed regions in our work show only a single dominant velocity component, and the derived velocity dispersions are consistent with the intensity-weighted mean values ($\langle\sigma\rangle$) reported by \citet{Henshaw2016}, in which they took multiple line-of-sight components into account. This consistency suggests that our measured dispersions provide reliable estimates of the turbulent motions within these clouds. Appendix \ref{app:turbvelo} shows the shifted line spectra. Table \ref{tab:mc_proper} lists the estimates of turbulent velocity dispersions.

The uncertainty in the estimated magnetic field strengths is difficult to quantify due to the inherent limitations of the DCF method. We adopt a typical uncertainty factor of 2, based on \cite{Liu2021}, who derived this by applying the DCF method to numerical simulations and comparing the estimates to input models. However, this uncertainty should be considered a lower limit, as real-world observations may introduce additional sources of error not accounted for in numerical simulations. The estimated $B_\mathrm{pos}$ values for the clouds are shown in Table \ref{tab:mc_proper}. The estimates of magnetic field strength of some molecular clous (e.g., Sgr C, 50 MC, G0.253+0.016, Cloud E/F) from FIREPLACE data are consistent with that from \cite{Lu2024} derived by JCMT data at comparable resolution.

\subsection{Comparison with turbulence and gravity}
With the derived magnetic field strength, we can compare the role of magnetic fields with other effects, such as turbulence and gravity.

To quantify the relative significance between magnetic field and turbulence of the individual cloud, we can use Alfv$\mathrm{\acute{e}}$nic Mach number:
\begin{equation}
\mathcal{M}_A=\sigma_\mathrm{turb,3D}/\upsilon_\mathrm{A,3D}
\end{equation}
where $\sigma_\mathrm{turb,3D}=\sqrt{3}~\sigma_\mathrm{turb}$ is an estimate for 3D turbulent velocity dispersion, assuming isotopic turbulence, and $\upsilon_\mathrm{A,3D}=B_\mathrm{3D}/\sqrt{\mu_0\rho}$ is the 3D Alfv$\mathrm{\acute{e}}$n velocity. The 3D magnetic field strength ($B_\mathrm{3D}$) can be estimated from the plane-of-sky field using a statistical relation. Assuming random inclination angles between the 3D and POS magnetic fields, \cite{Crutcher2004} proposed $B_\mathrm{3D} = 4B_\mathrm{pos}/\pi$. We found that nearly all molecular clouds in the CMZ have Alfv$\mathrm{\acute{e}}$nic Mach numbers greater than or approximately equal to 1 ($\mathcal{M}_A \gtrsim 1$), except the 50 MC cloud. Considering the uncertainties inherent in the DCF method, we conclude that turbulence is likely to play a role as significant as, or possibly more significant than, the magnetic field in most CMZ clouds.

With the estimated magnetic field strengths, we can also assess the balance between magnetic field and gravity in individual clouds by using the mass-to-flux ratio, $\lambda$, in units of critical value $1/(2\pi\sqrt{G})$ \citep[][]{Nakano1978,Crutcher2004}:
\begin{equation}
\lambda=\mu_\mathrm{H_2}m_\mathrm{H}\sqrt{\mu_0\pi G}\frac{N_\mathrm{H_2}}{B}\sim 7.6\times10^{-21}\frac{N_\mathrm{H_2}/\mathrm{cm}^{-2}}{B_\mathrm{3D}/\mu G}
\end{equation}
where $N_\mathrm{H_2}$ is the molecular hydrogen column density. In our sample, two clouds (50 MC and G0.253+0.016) exhibit relatively low mass-to-flux ratios ($\lambda \sim 0.5$), indicating that magnetic fields may play a dominant role over gravity in these regions. In contrast, the other clouds have mass-to-flux ratios greater than or approximately equal to 1 ($\lambda \gtrsim 1$), suggesting that gravity could be as significant as, or possibly more significant than, the magnetic field in those clouds.

The virial parameter is commonly used to evaluate the stability of a core against gravitational collapse. In the CMZ, where both magnetic fields and turbulence are strong, these forces both provide significant support against collapse \citep[e.g.,][]{Pillai2015,Myers2022}. Following \cite{Liu2020}, we estimate the virial parameter by incorporating magnetic and kinetic energy:
\begin{equation}
    \alpha_{k+B}=\frac{M_{k+B}}{M_\mathrm{gas}},
\end{equation}
where $M_{k+B}=\sqrt{M_B^2+(M_k/2)^2}$ is the critical virial mass, $M_\mathrm{gas}$ is the gas mass of the cloud, and $\beta$ is the power-law index of the cloud's density profile ($\rho\propto r^{-\beta}$). Here, $M_k$ and $M_B$ are defined as:
\begin{equation}
    M_k=\frac{3(5-2\beta)\sigma^2_\mathrm{tot}R_\mathrm{eff}}{(3-\beta)G},
\end{equation}
\begin{equation}
    M_B=\frac{\pi R_\mathrm{eff}^2 B_\mathrm{3D}}{\sqrt{\frac{3(3-\beta)}{2(5-2\beta)}\mu_0\pi G}},
\end{equation}
where $\beta$ ranges from 0 (uniform density) to 2 (centrally peaked). The effective radius of each molecular cloud, $R_\mathrm{eff} = \sqrt{A_\mathrm{cloud}/\pi} $, is derived from its enclosed area $A_\mathrm{cloud}$. Using this framework, we calculate the virial parameter for individual CMZ clouds. The results, listed in Table \ref{tab:mc_proper}, reveal that nearly all CMZ clouds are super-virial ($\alpha_{k+B}>1$), except Sgr B2. This suggests that most dense gas in the CMZ is stable against gravitational collapse, consistent with the observed quiescent star formation activity.

Interestingly, despite the diverse and complex evolution of alignment measures in individual clouds, a general trend emerges: clouds with relatively small virial parameters ($\alpha_\mathrm{k+B} \lesssim 3$) or high mass-to-flux ratios ($\lambda \gtrsim 1.0$), such as Sgr C, TLP, Cloud E/F, and Sgr B2, tend to show negative AM values in high-density regions. In contrast, clouds with larger virial parameters ($\alpha_\mathrm{k+B} \gtrsim 4$) or lower mass-to-flux ratios ($\lambda \lesssim 0.8$), such as 50 MC and G0.253+0.016, consistently exhibit positive AM values across all density bins. This trend suggests that the evolution of relative orientation between magnetic fields and density structures may be associated with the balance between magnetic, turbulent kinetic and gravitational energies as the MHD simulations suggested.

\subsection{Origin of the observed alignment in the CMZ}
Several MHD simulations \citep[e.g.,][]{Soler2013,Chen2016,Soler2017b,Seifried2020} explored the origin of the transition of relative orientations between magnetic field and column density structures from parallel to perpendicular with increasing densities and suggested that it is related to the energy balance between magnetic, kinetic, and gravitational energies. 

In diffuse, non-self-gravitating regions, velocity shear can stretch both the gas and magnetic field lines in the same direction, producing structures aligned with the field, as shown by \citet{Hennebelle2013} and \citet{PlanckXXXII2016}. Similarly, \citet{Xu2019} proposed that compressive MHD turbulence can induce mixing that aligns low-density structures in molecular clouds with the local magnetic field. As gas accumulates and becomes denser, a strong magnetic field can restrict gravitational collapse to occur preferentially along the field lines \citep{Mouschovias1976}, resulting in dense structures oriented perpendicular to the magnetic field. Meanwhile, the supersonic turbulence can generate dense structures in the molecular clouds by shock compression. When turbulence dominates over magnetic forces (super-Alfv\'{e}nic, $\mathcal{M}_A > 1$), the magnetic field is largely dragged with the gas due to flux freezing, leading to roughly isotropic compression and field-aligned structures. In contrast, the sub-Alfv\'{e}nic ($\mathcal{M}_A<1$) turbulence becomes anisotropic \citep[e.g.,][]{Sridhar1994, Xu2019}, and shocks tend to compress gas along the field lines, producing structures that are preferentially perpendicular to the magnetic field.

According to our HRO analysis, we find the relative orientations between magnetic fields and density structures in the CMZ are random in low density bins ($2\times10^{22}\lesssim N_\mathrm{H_2}\lesssim10^{23}~\mathrm{cm^{-2}}$) and parallel at high density bins ($N_\mathrm{H_2}\gtrsim10^{23}~\mathrm{cm^{-2}}$). As discussed in Section \ref{subsec:MHDsimu}, the synthetic observations of MHD simulations show that the random relative orientation in low density bins can be caused by the projection effects from line-of-sight integrations of unrelated diffuse components since the CMZ is observed in the edge-on perspective. For the parallel alignment observed in the high-density regime, two possible explanations exist: (1) The CMZ has low magnetization, allowing gas motion to drag the magnetic field lines, or (2) Most CMZ clouds remain stable against gravitational collapse despite their gas high densities, with strong shear aligning the matter and magnetic field lines. The magnetic field strength in the CMZ is substantially higher than in the Galactic disk, reaching values of \citep[$\sim0.1-1$ mG,][]{Ferriere2009,Longmore2013, Pan2024}. Notably, molecular clouds such as G0.253+0.016 \citep{Pillai2015} and 50 MC \citep{Lu2024} exhibit subcritical states, implying that magnetic fields play a dominant role over gravity. These observations suggest that low magnetization is unlikely to be a universal condition for all CMZ clouds. Additionally, nearly all clouds in our sample, except Sgr B2, are super-virial ($\alpha_{k+B} > 1$), meaning they are not self-gravitating. This supports the second explanation. We also found in the CMZ, dense regions such as Sgr C, Sgr B2, and Cloud E/F exhibit density structures that are perpendicular to the local magnetic field, while also showing relatively weak magnetic fields compared to turbulence ($\mathcal{M}_A\gtrsim1$). So, the shock compression by supersonic turbulence may not be the major factor that causes the dense structures perpendicular to the local magnetic field. We therefore propose that the parallel alignment in the CMZ arises because the majority of its gas remains stable against gravitational collapse, even at high densities, thanks to the additional support provided by strong turbulence and magnetic fields here. Furthermore, in some molecular clouds (e.g., Sgr C, Cloud E/F, and Sgr B2), the magnetic field becomes orthogonal to the column density structures only in regions of exceptionally high density ($\gtrsim10^{23}~\mathrm{cm^{-2}}$). This suggests that significantly more material is required in the CMZ compared to the Galactic disk to overcome support provided by strong turbulence and magnetic field and trigger gravitational collapse and star formation. This finding is consistent with the observed low star formation efficiency in the CMZ.

\section{Summary}\label{sec:summary}

We present a study of the relative orientation between magnetic fields in the Central Molecular Zone from SOFIA/HAWC+ data and column density structures derived by SED fitting of \emph{Herschel} and ATLASGAL data.

Our study reveals a random alignment in the low-density regime ($10^{22} < N_\mathrm{H_2} < 10^{23}\mathrm{cm^{-2}}$) and a trend toward parallel alignment at higher densities ($N_\mathrm{H_2} \gtrsim 10^{23}\mathrm{cm^{-2}}$), in contrast to the typical transition from parallel to perpendicular alignment observed in the Galactic disk with increasing column density. Numerical experiments using MHD simulations of the CMZ suggest parallel alignment in both projected 2D and 3D synthetic observations across all density regimes. This consistency between observations and simulations strongly suggests that the parallel alignment in the CMZ is intrinsic rather than an artifact of projection effects.

We also investigate individual molecular clouds in the CMZ, finding significant variations in the relative orientations between magnetic fields and density structures. In 20MC, 50MC, and G0.253+0.016, nearly all density bins exhibit parallel alignment. In contrast, TLP and Cloud D predominantly show perpendicular alignment. Sgr B2, after recovering its densest regions, displays a transition from parallel to perpendicular alignment, similar to Sgr C. Meanwhile, Cloud E/F undergoes a more complex evolution, shifting from perpendicular to parallel and back to perpendicular alignment as density increases.

Through analysis of the energy balance between magnetic, kinetic, and gravitational components, we that all CMZ clouds are super-virial despite their high densities—except for Sgr B2, which exhibits a significantly lower virial parameter and more active star formation. This indicates that most dense gas structures in the CMZ are stable against gravitational collapse, supported by strong turbulence and magnetic fields. Such stability explain the origin of parallel alignment observed throughout the CMZ region. 

Future studies incorporating higher resolution polarization data will help reveal the details of relative orientations between magnetic fields and column density structures within individual clouds, particularly for compact, dense structures (e.g., clumps and cores) that are more directly linked to star formation and small clouds (e.g., Dust Ridge Clouds B and C) which are excluded in our analysis due to a lack of sufficient data points for statistical analysis. Exploring the connection between star formation rates and the evolution of relative orientations in these regions will offer new insights into how magnetic field interacts with turbulence and gravity at smaller scales in the CMZ.

\begin{acknowledgements}
This work is based on observations made with the NASA/DLR Stratospheric Observatory for Infrared Astronomy (SOFIA). SOFIA was jointly operated by the Universities Space Research Association, Inc. (USRA), under NASA contract NNA17BF53C, and the Deutsches SOFIA Institut (DSI) under DLR contract 50 OK 2002 to the University of Stuttgart. X. P. is supported by the Smithsonian Astrophysical Observatory (SAO) Predoctoral Fellowship Program. Q. Z. gratefully acknowledges the support by the National Science Foundation under Award No. AST-2206512, and the Smithsonian Institute FY2024 Scholarly Studies Program. D.P. and D.C. acknowledge support from NASA ADAP award number 80NSSC25K7561. TGSP gratefully acknowledges support by the National Science Foundation under grant No. AST-2009842 and AST-2108989 and by NASA award \#09-0215 issued by USRA.
\end{acknowledgements}

\clearpage
\appendix
\section{Dust temperature map of the CMZ}\label{app:tdust}
\begin{figure}[!ht]
    \centering
    \includegraphics[width=0.8\linewidth]{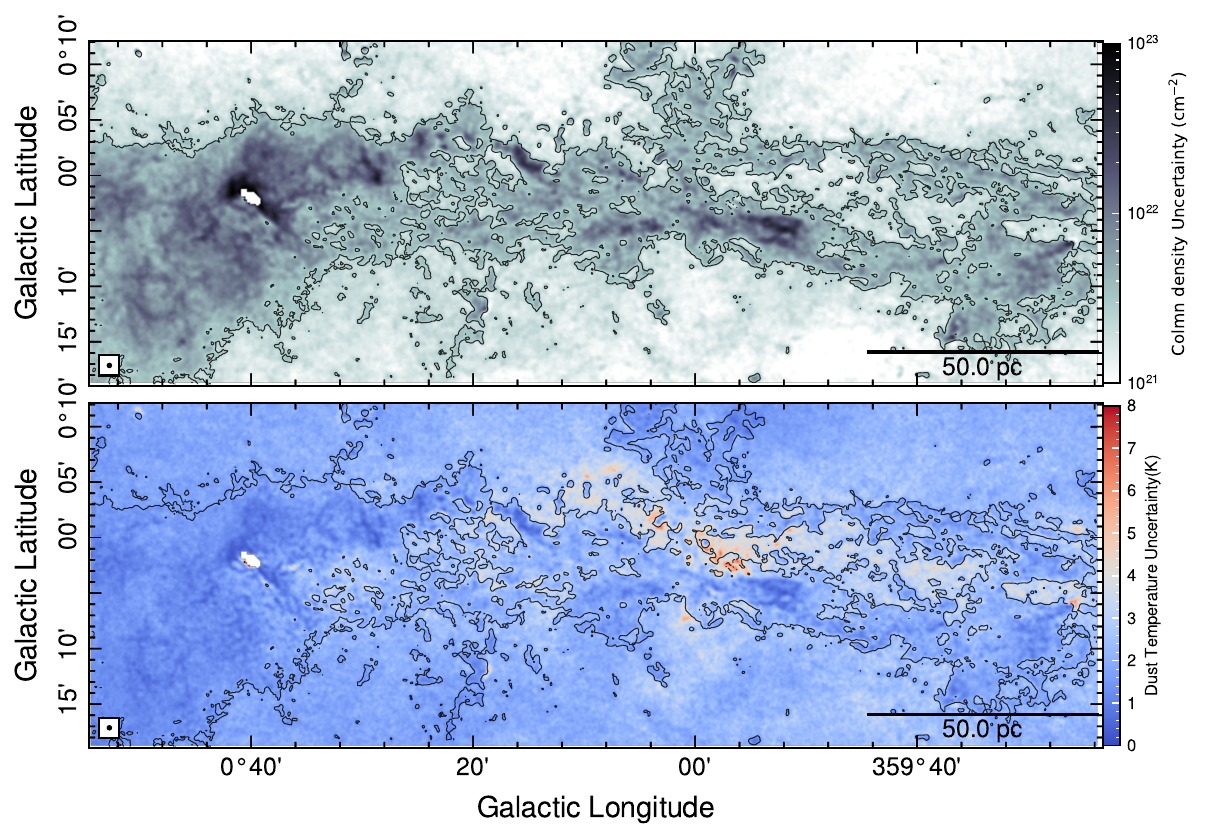}
    \caption{The distribution of column density and dust temperature uncertainty in the CMZ derived by SED fitting in Section \ref{subsec:colden}. The black contour marks a column density of $2.0\times10^{22}~\mathrm{cm^{-2}}$.}
    \label{fig:enter-label}
\end{figure}

\section{Relative orientation plot of Sgr B2}\label{app:hro_sgrb2}

\begin{figure}[!ht]
    \centering
    \includegraphics[width=0.35\textwidth]{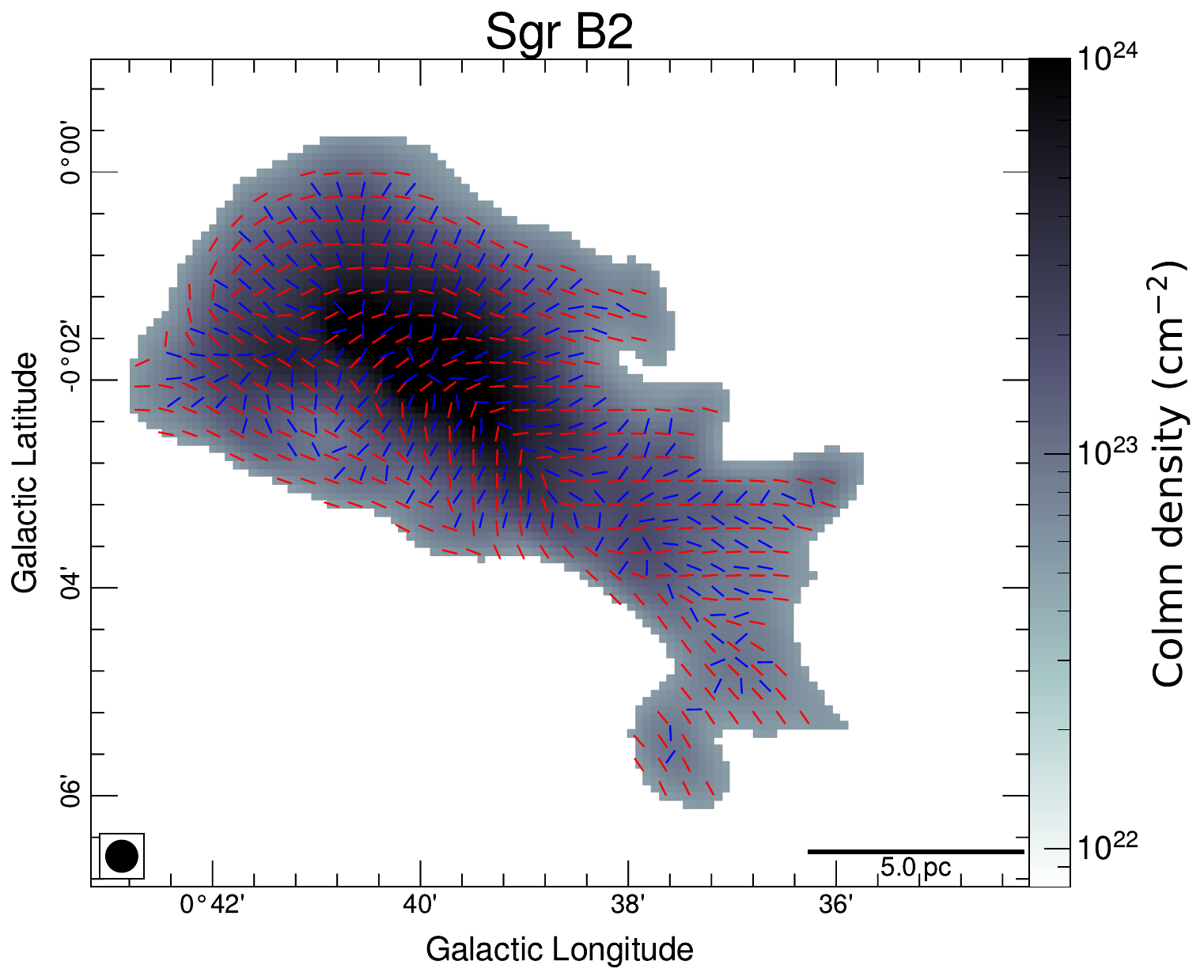}
    \includegraphics[width=0.45\linewidth]{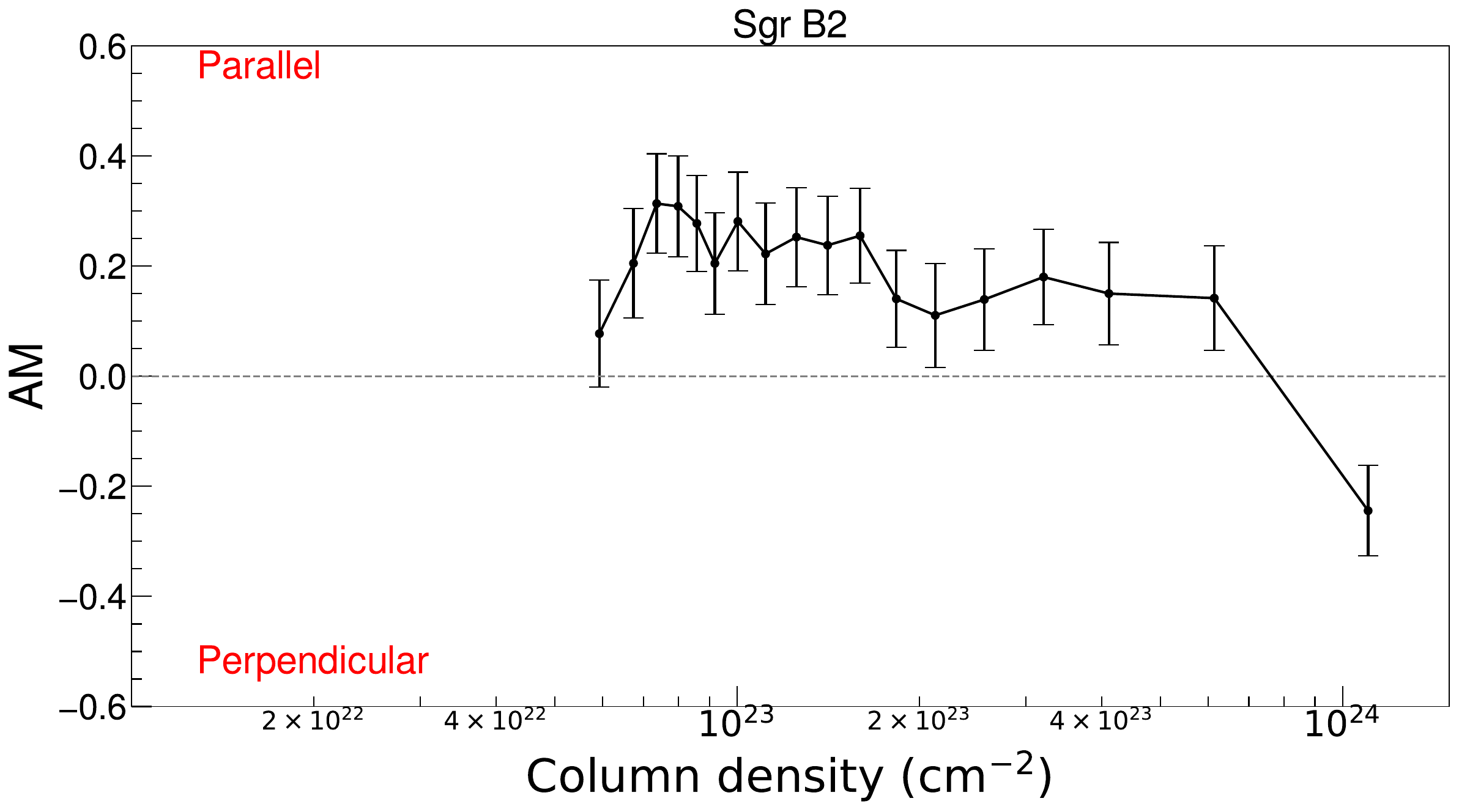}
    \caption{Relative orientations of Sgr B2 using 214 $\mu$m SOFIA/HAWC+ data. Left panel: Comparisons between the orientations of magnetic field and column density structure for Sgr B2. The column density map is derived by 214 $\mu$m SOFIA/HAWC+ Stokes I emission, recovering the densest regions in Sgr B2. The panel format is the same as for Fig \ref{fig:HRO_MCs}. Right panel: Alignment measurements for Sgr B2 with the column density derived by 214 $\mu$m SOFIA/HAWC+ Stokes I emission.}
    \label{fig:ro_plots_apx}
\end{figure}
\clearpage

\section{Fitting results for ADF method}\label{app:fitting_ADF}
\begin{figure}[!ht]
\centering
    \includegraphics[width=0.23\linewidth]{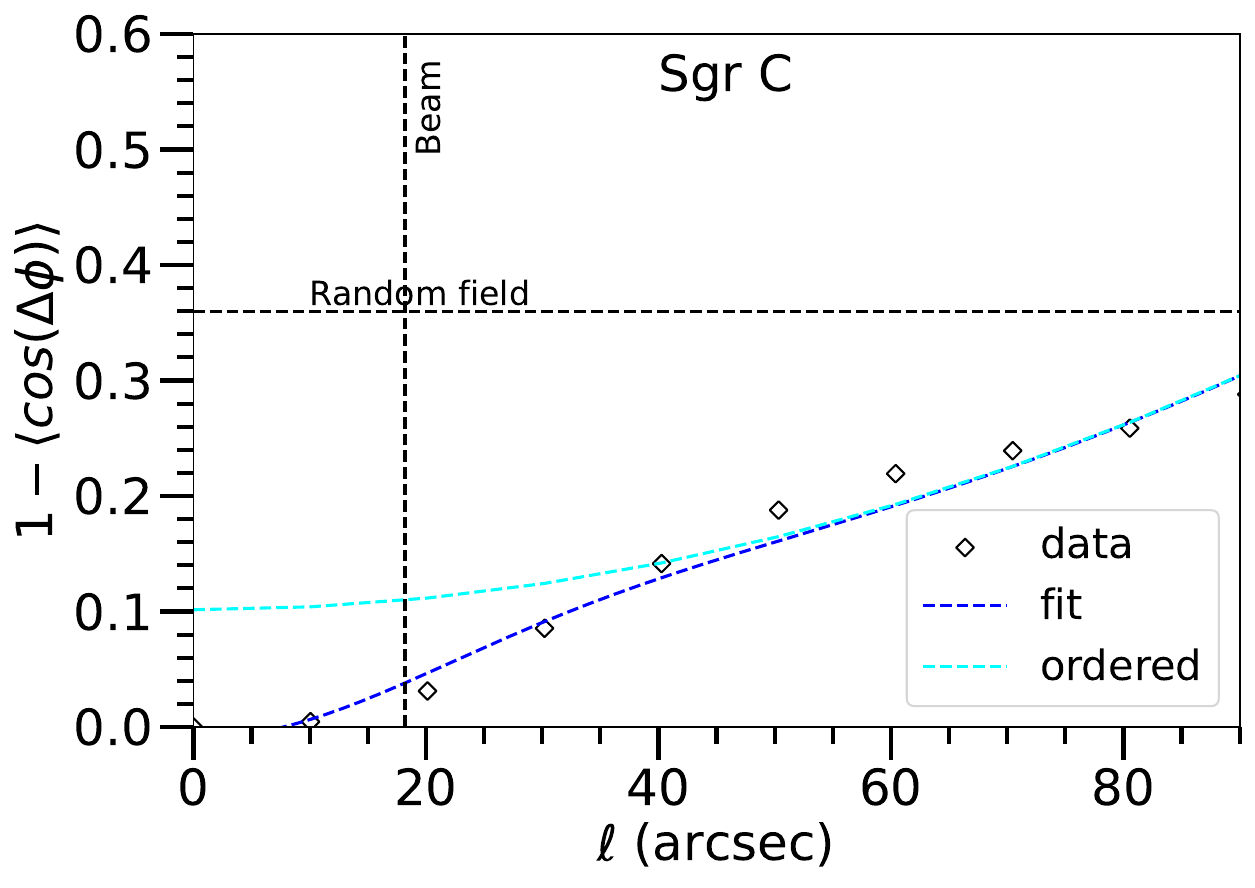}
    \includegraphics[width=0.23\linewidth]{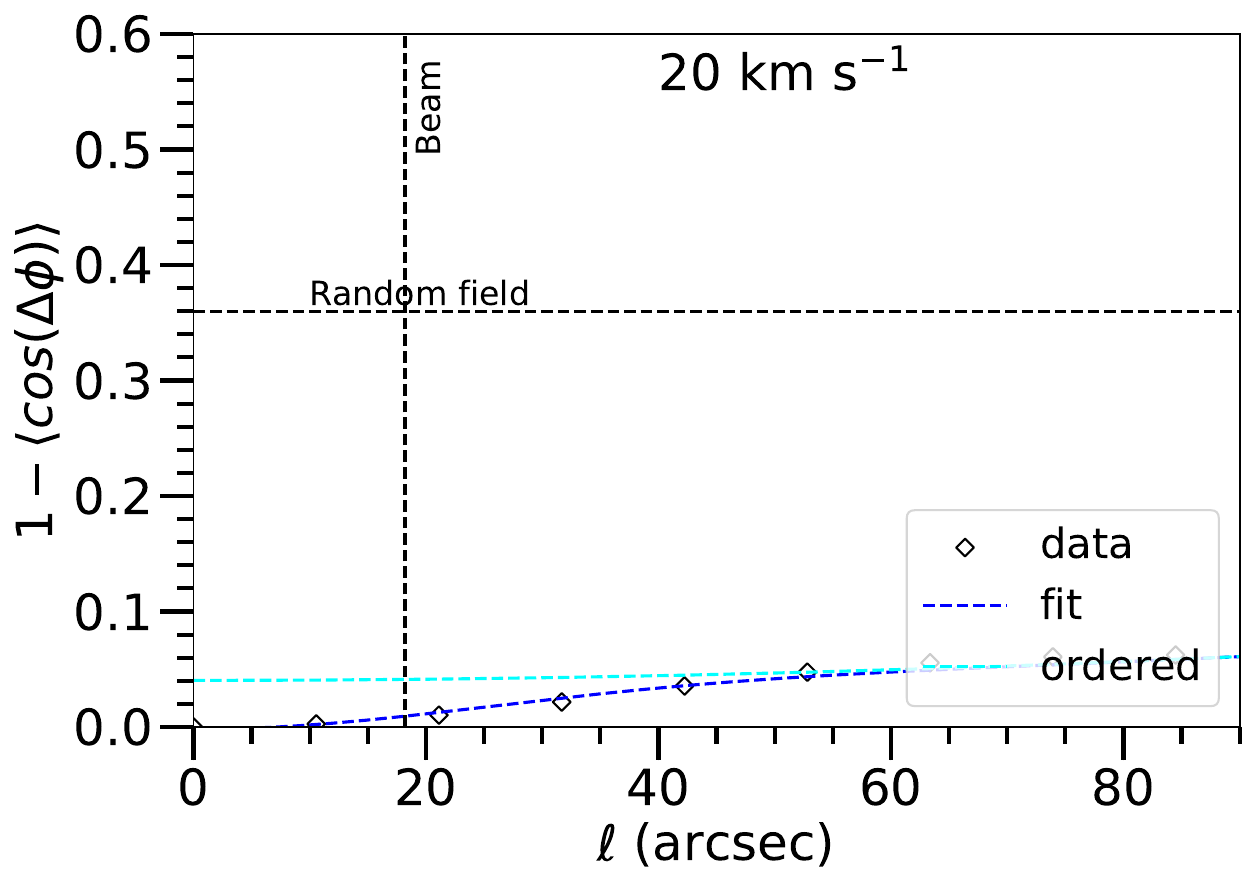}
    \includegraphics[width=0.23\linewidth]{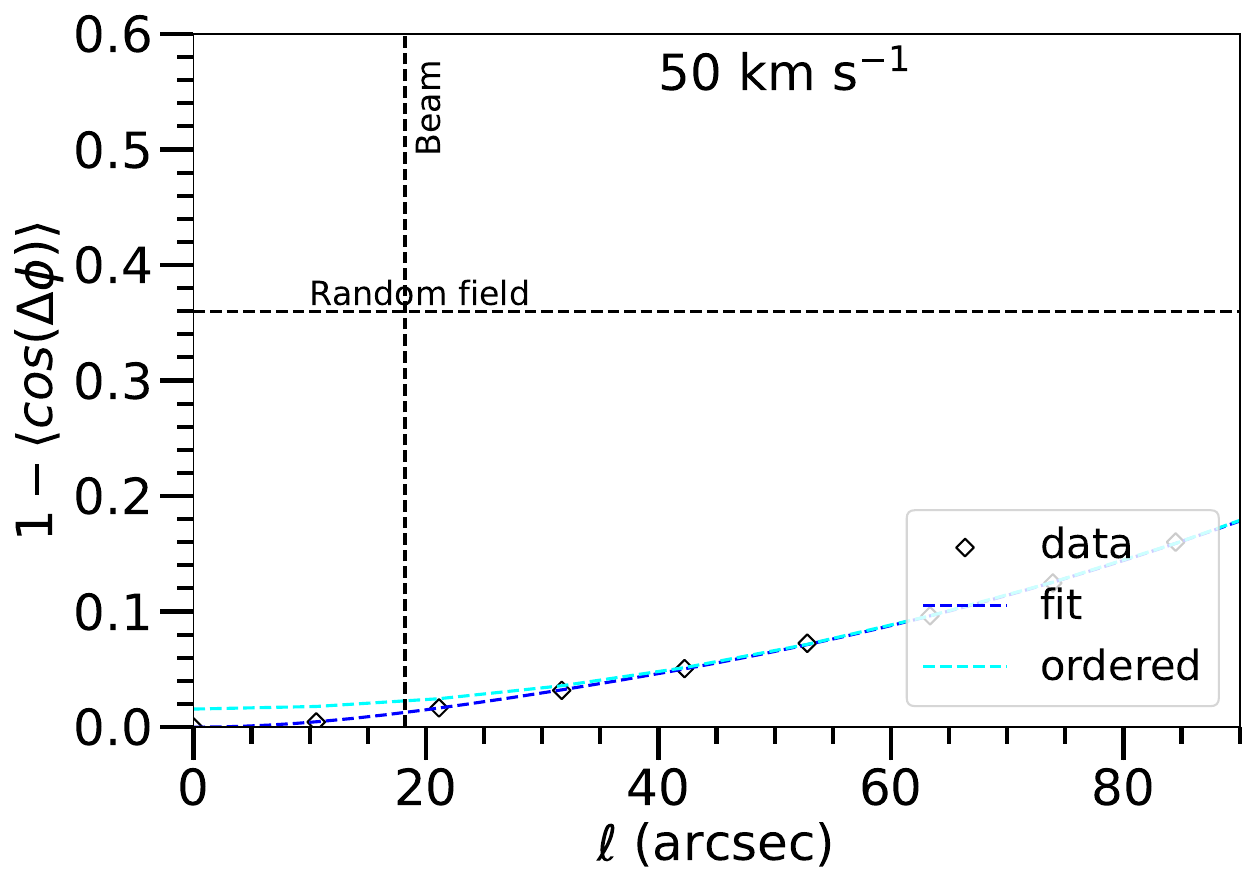}
    \includegraphics[width=0.23\linewidth]{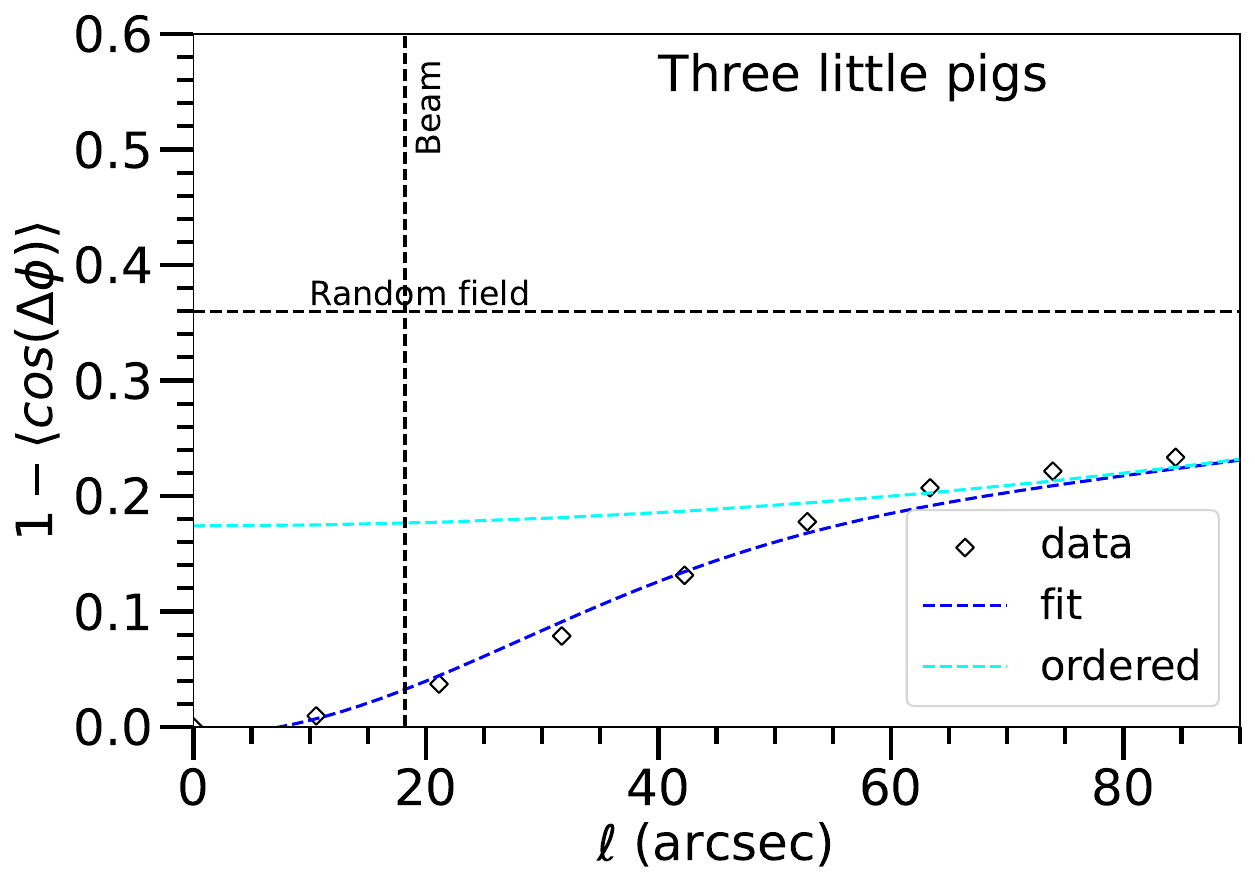}\\
    \includegraphics[width=0.23\linewidth]{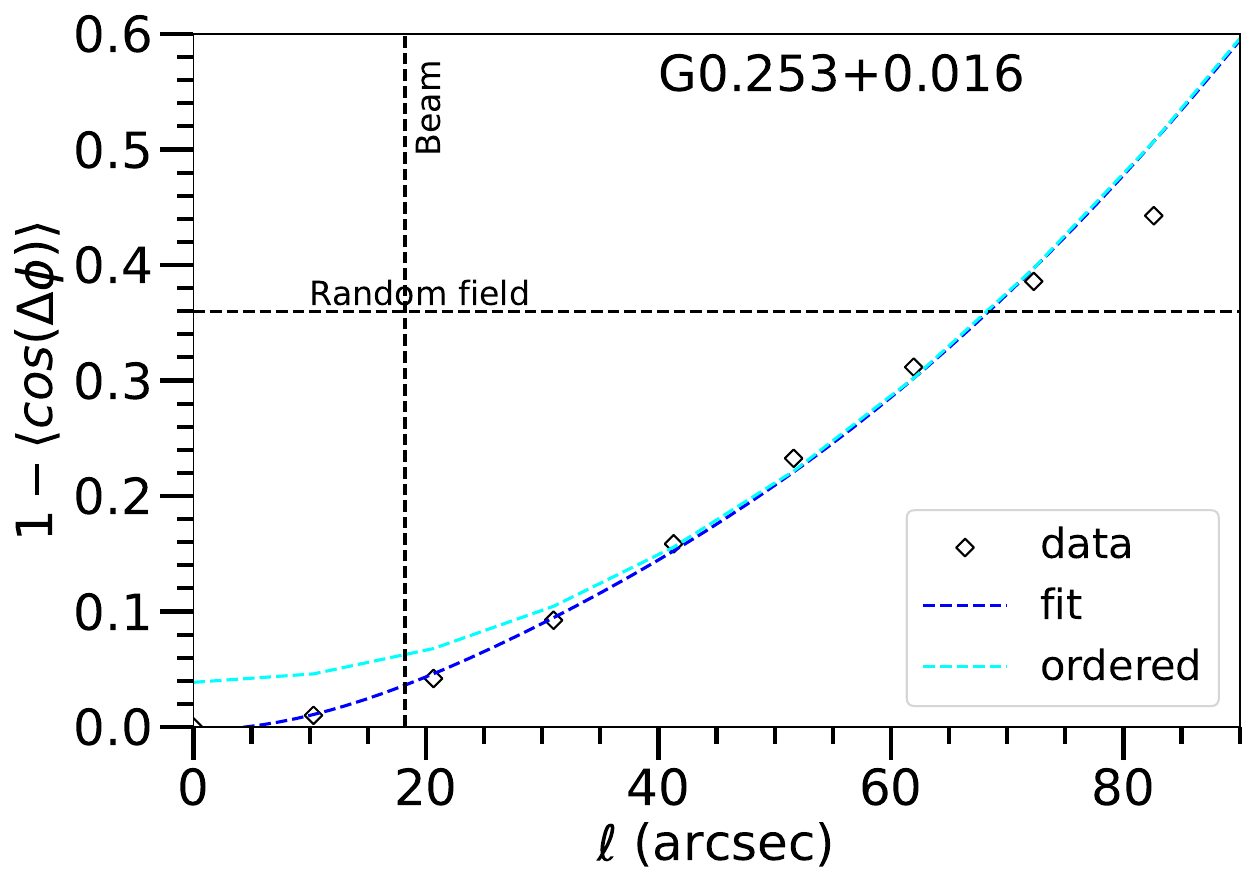}
    \includegraphics[width=0.23\linewidth]{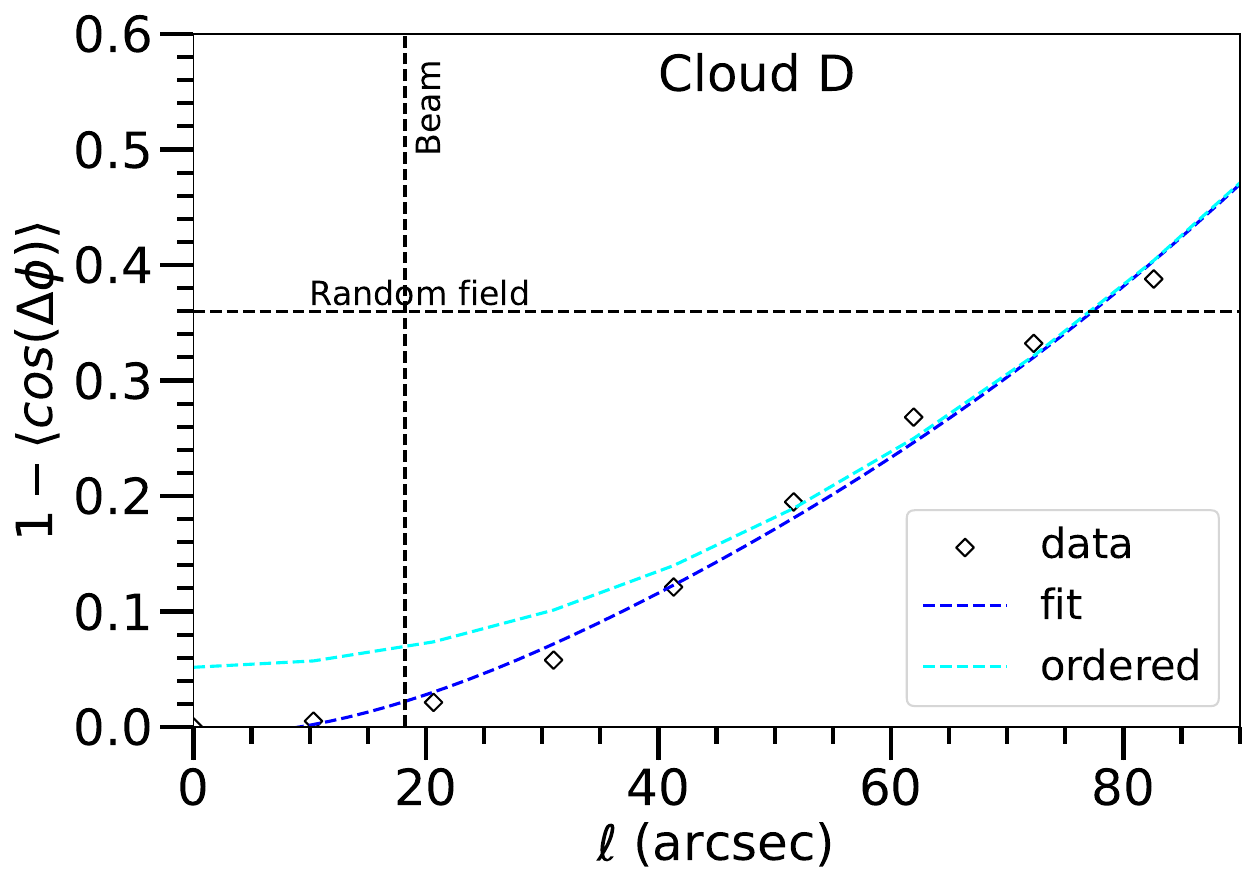}
    \includegraphics[width=0.23\linewidth]{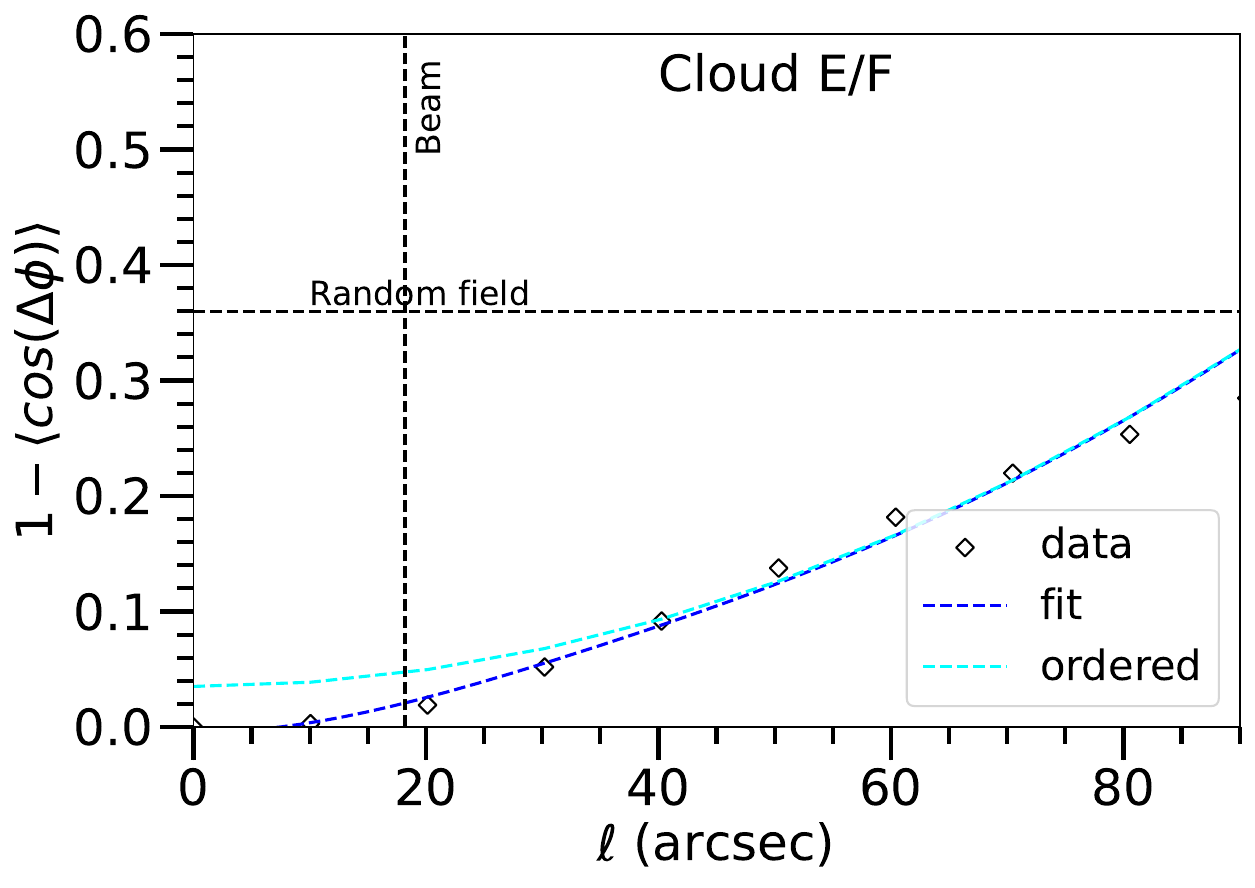}
\caption{Angular dispersion function for individual molecular clouds. Diamond symbols represent the angular dispersion segments. The blue dashed line indicates the fitted angular dispersion function, while the cyan dashed line represents the ordered component of the best-fit model. The vertical line marks the beam size.}
\label{fig:adf_plots}
\end{figure}

\section{Turbulent velocity dispersion}\label{app:turbvelo}
\begin{figure}[!ht]
    \centering
    \includegraphics[width=0.23\linewidth]{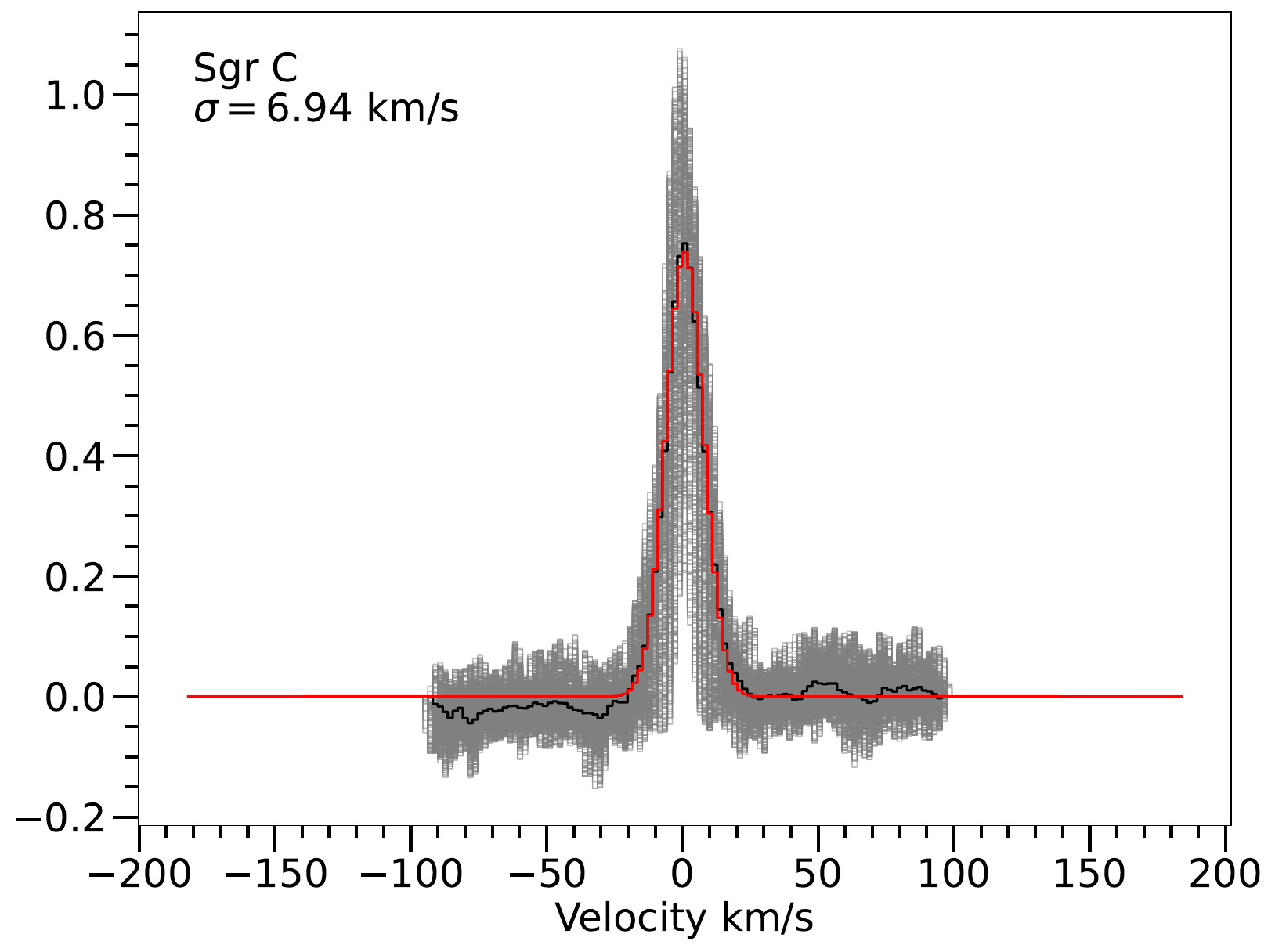}
    \includegraphics[width=0.23\linewidth]{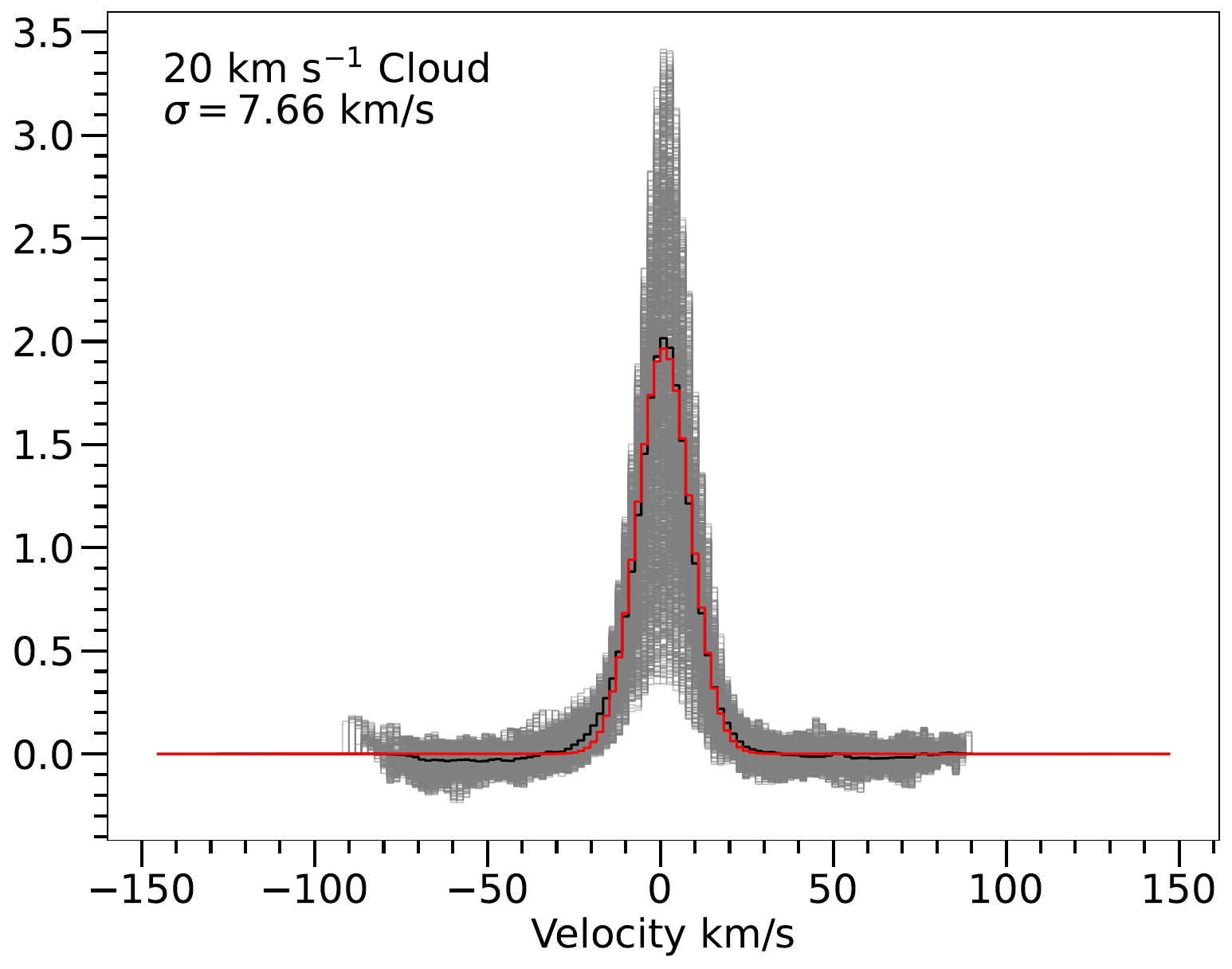}
    \includegraphics[width=0.23\linewidth]{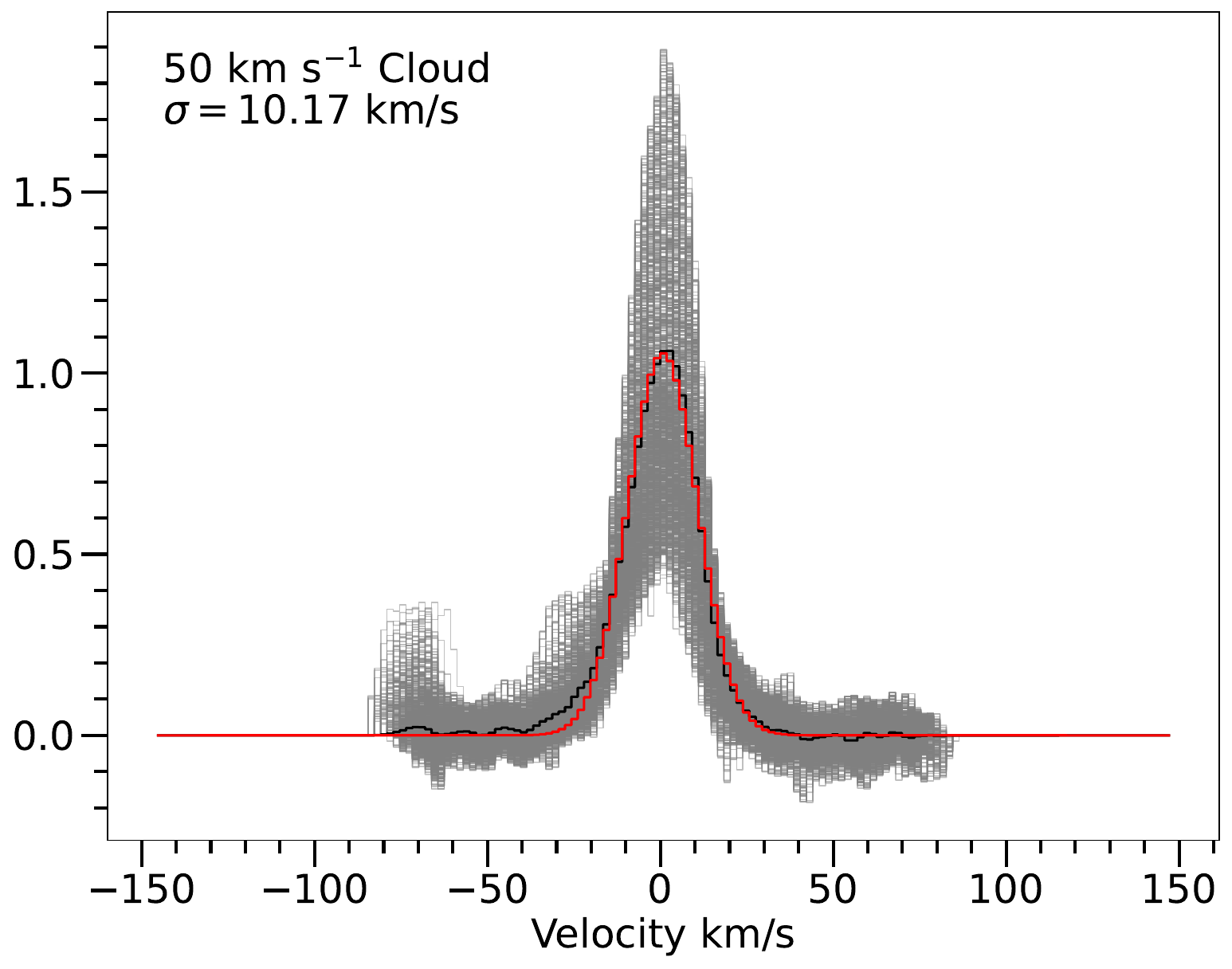}
    \includegraphics[width=0.23\linewidth]{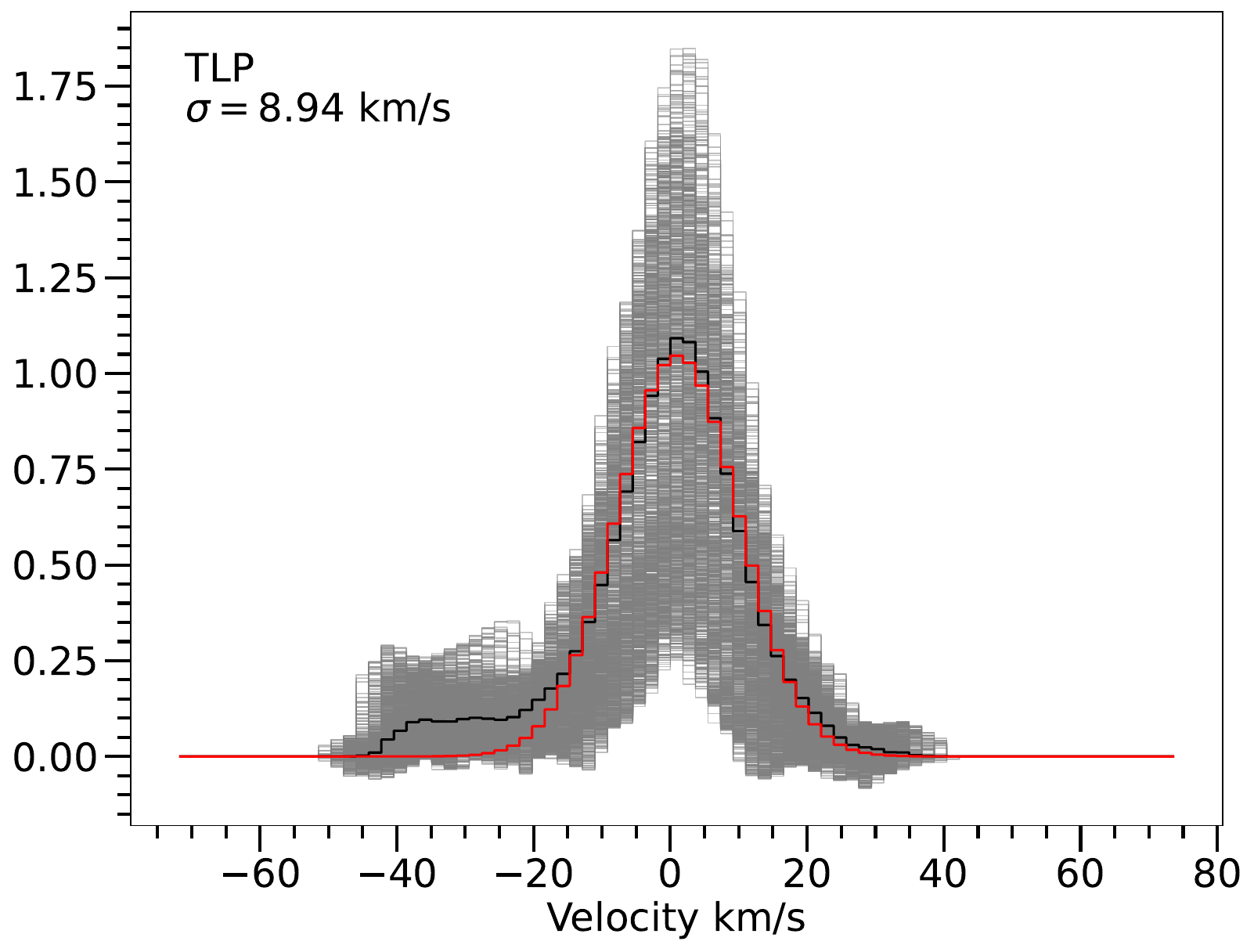}\\
    \includegraphics[width=0.23\linewidth]{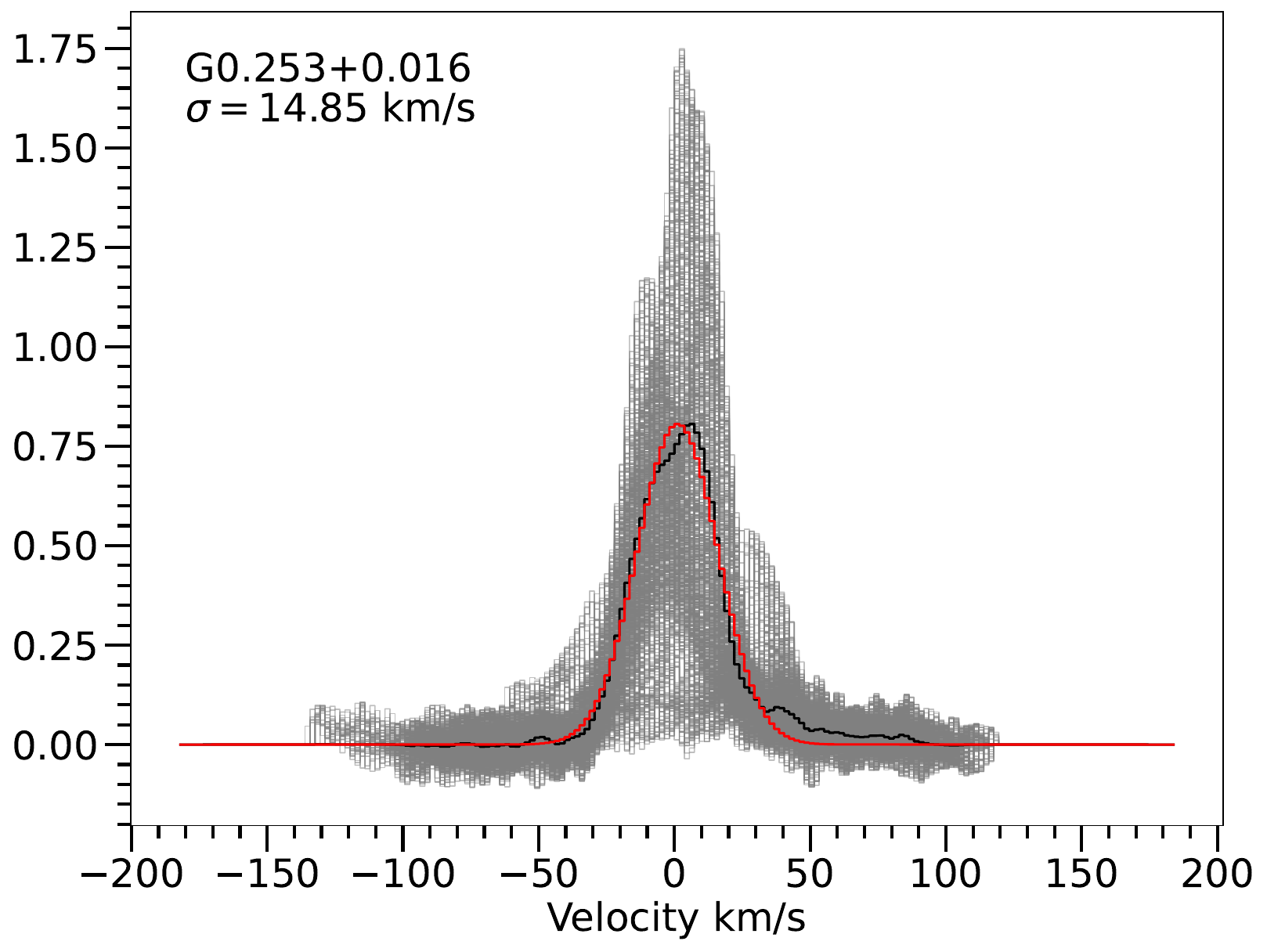}
    \includegraphics[width=0.23\linewidth]{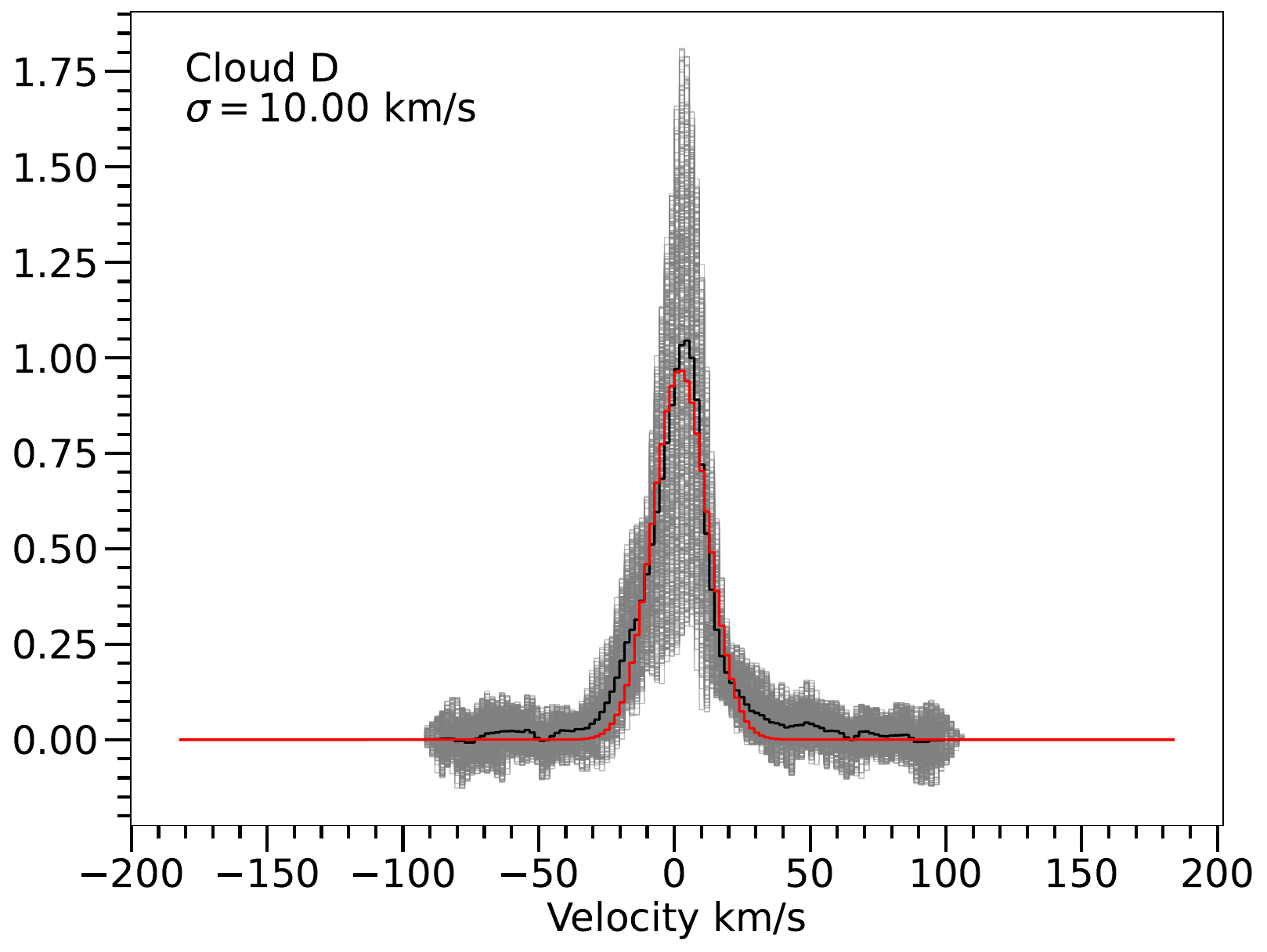}
    \includegraphics[width=0.23\linewidth]{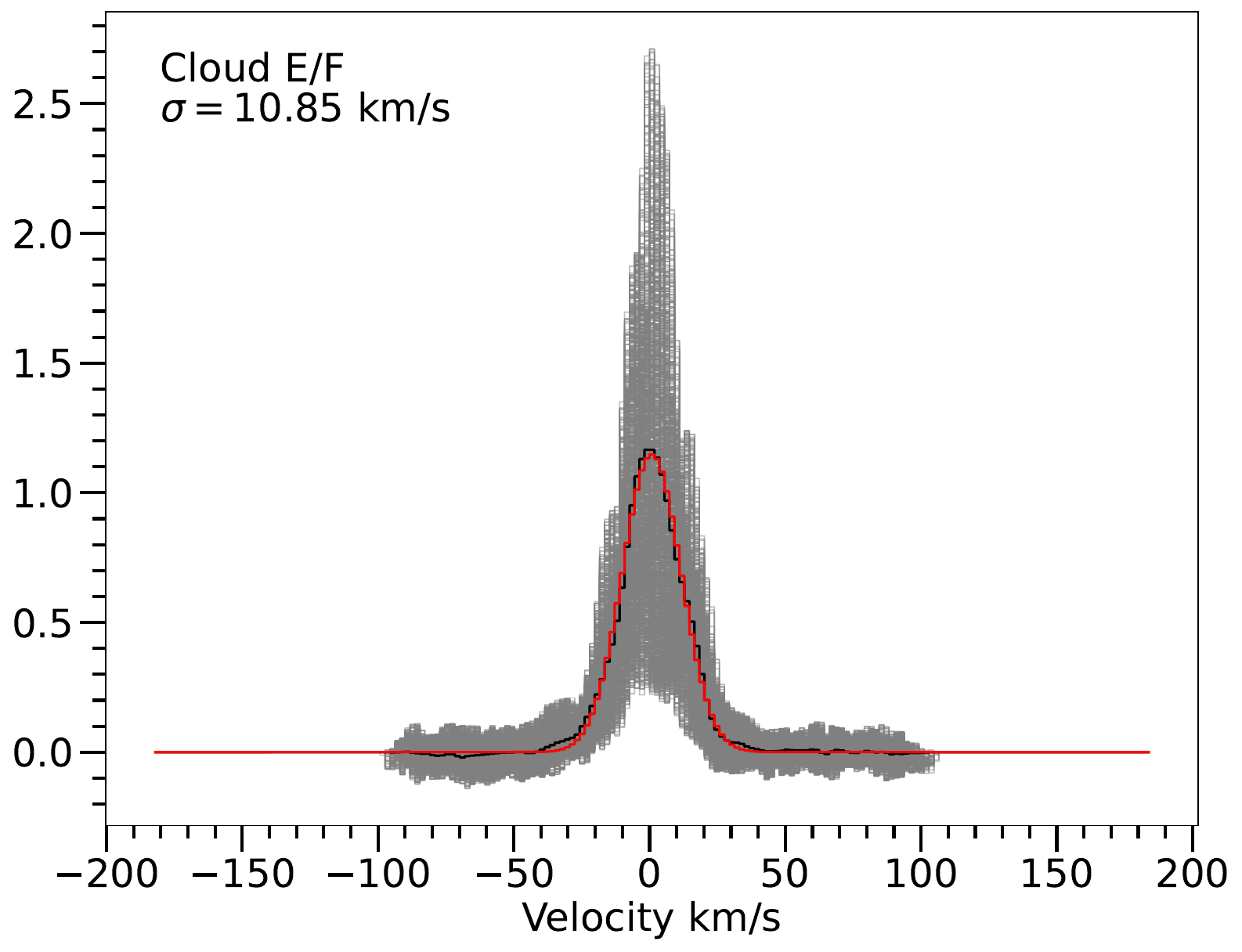}
    \caption{HNCO spectra of individual molecular clouds, corrected for local velocities derived from intensity-weighted velocity maps, are presented. Gray lines show the corrected HNCO spectra for all pixels within the coverage of each cloud. The black line indicates the average spectrum, weighted by the integrated intensity of each pixel, while the red line shows the result of the Gaussian fit. The fitted turbulent velocity dispersion is labeled in the upper left corner.}
\label{fig:hnco_plots}
\end{figure}
\clearpage

\bibliography{CMZ_SOFIA_HRO}{}

\begin{thebibliography}{}
\expandafter\ifx\csname natexlab\endcsname\relax\def\natexlab#1{#1}\fi
\providecommand{\url}[1]{\href{#1}{#1}}
\providecommand{\dodoi}[1]{doi:~\href{http://doi.org/#1}{\nolinkurl{#1}}}
\providecommand{\doeprint}[1]{\href{http://ascl.net/#1}{\nolinkurl{http://ascl.net/#1}}}
\providecommand{\doarXiv}[1]{\href{https://arxiv.org/abs/#1}{\nolinkurl{https://arxiv.org/abs/#1}}}

\bibitem[{{Barnes} {et~al.}(2017){Barnes}, {Longmore}, {Battersby}, {Bally}, {Kruijssen}, {Henshaw}, \& {Walker}}]{Barnes2017}
{Barnes}, A.~T., {Longmore}, S.~N., {Battersby}, C., {et~al.} 2017, \mnras, 469, 2263, \dodoi{10.1093/mnras/stx941}

\bibitem[{{Battersby} {et~al.}(2020){Battersby}, {Keto}, {Walker}, {Barnes}, {Callanan}, {Ginsburg}, {Hatchfield}, {Henshaw}, {Kauffmann}, {Kruijssen}, {Longmore}, {Lu}, {Mills}, {Pillai}, {Zhang}, {Bally}, {Butterfield}, {Contreras}, {Ho}, {Ott}, {Patel}, \& {Tolls}}]{Battersby2020}
{Battersby}, C., {Keto}, E., {Walker}, D., {et~al.} 2020, \apjs, 249, 35, \dodoi{10.3847/1538-4365/aba18e}

\bibitem[{{Battersby} {et~al.}(2024){Battersby}, {Walker}, {Barnes}, {Ginsburg}, {Lipman}, {Alboslani}, {Hatchfield}, {Bally}, {Glover}, {Henshaw}, {Immer}, {Klessen}, {Longmore}, {Mills}, {Molinari}, {Smith}, {Sormani}, {Tress}, \& {Zhang}}]{Battersby2024}
{Battersby}, C., {Walker}, D.~L., {Barnes}, A., {et~al.} 2024, arXiv e-prints, arXiv:2410.17334, \dodoi{10.48550/arXiv.2410.17334}

\bibitem[{{Butterfield} {et~al.}(2024){Butterfield}, {Chuss}, {Guerra}, {Morris}, {Par{\'e}}, {Wollack}, {Dowell}, {Hankins}, {Karpovich}, {Siah}, {Staguhn}, \& {Zweibel}}]{Butterfield2024}
{Butterfield}, N.~O., {Chuss}, D.~T., {Guerra}, J.~A., {et~al.} 2024, \apj, 963, 130, \dodoi{10.3847/1538-4357/ad12b9}

\bibitem[{{Chandrasekhar} \& {Fermi}(1953)}]{Chandrasekhar1953}
{Chandrasekhar}, S., \& {Fermi}, E. 1953, \apj, 118, 113, \dodoi{10.1086/145731}

\bibitem[{{Chen} {et~al.}(2016){Chen}, {King}, \& {Li}}]{Chen2016}
{Chen}, C.-Y., {King}, P.~K., \& {Li}, Z.-Y. 2016, \apj, 829, 84, \dodoi{10.3847/0004-637X/829/2/84}

\bibitem[{{Chen} {et~al.}(2024){Chen}, {Fissel}, {Sadavoy}, {Rosolowsky}, {Doi}, {Arzoumanian}, {Bastien}, {Coud{\'e}}, {di Francesco}, {Friesen}, {Furuya}, {Hwang}, {Inutsuka}, {Johnstone}, {Karoly}, {Kwon}, {Kwon}, {Le Gouellec}, {Liu}, {Mairs}, {Onaka}, {Pattle}, {Rawlings}, {Tahani}, {Tamura}, {Wang}, \& {BISTRO Team}}]{Chen2024}
{Chen}, M. C.-Y., {Fissel}, L.~M., {Sadavoy}, S.~I., {et~al.} 2024, \mnras, 533, 1938, \dodoi{10.1093/mnras/stae1829}

\bibitem[{{Crutcher} {et~al.}(2004){Crutcher}, {Nutter}, {Ward-Thompson}, \& {Kirk}}]{Crutcher2004}
{Crutcher}, R.~M., {Nutter}, D.~J., {Ward-Thompson}, D., \& {Kirk}, J.~M. 2004, \apj, 600, 279, \dodoi{10.1086/379705}

\bibitem[{{Csengeri} {et~al.}(2014){Csengeri}, {Urquhart}, {Schuller}, {Motte}, {Bontemps}, {Wyrowski}, {Menten}, {Bronfman}, {Beuther}, {Henning}, {Testi}, {Zavagno}, \& {Walmsley}}]{Csengeri2014}
{Csengeri}, T., {Urquhart}, J.~S., {Schuller}, F., {et~al.} 2014, \aap, 565, A75, \dodoi{10.1051/0004-6361/201322434}

\bibitem[{{Davis}(1951)}]{Davis1951}
{Davis}, L. 1951, Physical Review, 81, 890, \dodoi{10.1103/PhysRev.81.890.2}

\bibitem[{{Demircan} \& {Kahraman}(1991)}]{Demircan1991}
{Demircan}, O., \& {Kahraman}, G. 1991, \apss, 181, 313, \dodoi{10.1007/BF00639097}

\bibitem[{{Etxaluze} {et~al.}(2013){Etxaluze}, {Goicoechea}, {Cernicharo}, {Polehampton}, {Noriega-Crespo}, {Molinari}, {Swinyard}, {Wu}, \& {Bally}}]{Etxaluze2013}
{Etxaluze}, M., {Goicoechea}, J.~R., {Cernicharo}, J., {et~al.} 2013, \aap, 556, A137, \dodoi{10.1051/0004-6361/201321258}

\bibitem[{{Falceta-Gon{\c{c}}alves} {et~al.}(2008){Falceta-Gon{\c{c}}alves}, {Lazarian}, \& {Kowal}}]{Falceta-Goncavles2008}
{Falceta-Gon{\c{c}}alves}, D., {Lazarian}, A., \& {Kowal}, G. 2008, \apj, 679, 537, \dodoi{10.1086/587479}

\bibitem[{{Ferri{\`e}re}(2009)}]{Ferriere2009}
{Ferri{\`e}re}, K. 2009, \aap, 505, 1183, \dodoi{10.1051/0004-6361/200912617}

\bibitem[{{Fissel} {et~al.}(2019){Fissel}, {Ade}, {Angil{\`e}}, {Ashton}, {Benton}, {Chen}, {Cunningham}, {Devlin}, {Dober}, {Friesen}, {Fukui}, {Galitzki}, {Gandilo}, {Goodman}, {Green}, {Jones}, {Klein}, {King}, {Korotkov}, {Li}, {Lowe}, {Martin}, {Matthews}, {Moncelsi}, {Nakamura}, {Netterfield}, {Newmark}, {Novak}, {Pascale}, {Poidevin}, {Santos}, {Savini}, {Scott}, {Shariff}, {Soler}, {Thomas}, {Tucker}, {Tucker}, {Ward-Thompson}, \& {Zucker}}]{Fissel2019}
{Fissel}, L.~M., {Ade}, P. A.~R., {Angil{\`e}}, F.~E., {et~al.} 2019, \apj, 878, 110, \dodoi{10.3847/1538-4357/ab1eb0}

\bibitem[{{Gaume} {et~al.}(1995){Gaume}, {Claussen}, {de Pree}, {Goss}, \& {Mehringer}}]{Gaume1995}
{Gaume}, R.~A., {Claussen}, M.~J., {de Pree}, C.~G., {Goss}, W.~M., \& {Mehringer}, D.~M. 1995, \apj, 449, 663, \dodoi{10.1086/176087}

\bibitem[{{Ginsburg} {et~al.}(2018){Ginsburg}, {Bally}, {Barnes}, {Bastian}, {Battersby}, {Beuther}, {Brogan}, {Contreras}, {Corby}, {Darling}, {De Pree}, {Galv{\'a}n-Madrid}, {Garay}, {Henshaw}, {Hunter}, {Kruijssen}, {Longmore}, {Lu}, {Meng}, {Mills}, {Ott}, {Pineda}, {S{\'a}nchez-Monge}, {Schilke}, {Schmiedeke}, {Walker}, \& {Wilner}}]{Ginsburg2018}
{Ginsburg}, A., {Bally}, J., {Barnes}, A., {et~al.} 2018, \apj, 853, 171, \dodoi{10.3847/1538-4357/aaa6d4}

\bibitem[{{Girichidis}(2021)}]{Girichidis2021}
{Girichidis}, P. 2021, \mnras, 507, 5641, \dodoi{10.1093/mnras/stab2157}

\bibitem[{{Glover} \& {Clark}(2012)}]{Glover2012}
{Glover}, S. C.~O., \& {Clark}, P.~C. 2012, \mnras, 426, 377, \dodoi{10.1111/j.1365-2966.2012.21737.x}

\bibitem[{{Goldsmith} {et~al.}(1990){Goldsmith}, {Lis}, {Hills}, \& {Lasenby}}]{Goldsmith1990}
{Goldsmith}, P.~F., {Lis}, D.~C., {Hills}, R., \& {Lasenby}, J. 1990, \apj, 350, 186, \dodoi{10.1086/168372}

\bibitem[{{Guan} {et~al.}(2021){Guan}, {Clark}, {Hensley}, {Gallardo}, {Naess}, {Duell}, {Aiola}, {Atkins}, {Calabrese}, {Choi}, {Cothard}, {Devlin}, {Duivenvoorden}, {Dunkley}, {D{\"u}nner}, {Ferraro}, {Hasselfield}, {Hughes}, {Koopman}, {Kosowsky}, {Madhavacheril}, {McMahon}, {Nati}, {Niemack}, {Page}, {Salatino}, {Schaan}, {Sehgal}, {Sif{\'o}n}, {Staggs}, {Vavagiakis}, {Wollack}, \& {Xu}}]{Guan2021}
{Guan}, Y., {Clark}, S.~E., {Hensley}, B.~S., {et~al.} 2021, \apj, 920, 6, \dodoi{10.3847/1538-4357/ac133f}

\bibitem[{{Harper} {et~al.}(2018){Harper}, {Runyan}, {Dowell}, {Wirth}, {Amato}, {Ames}, {Amiri}, {Banks}, {Bartels}, {Benford}, {Berthoud}, {Buchanan}, {Casey}, {Chapman}, {Chuss}, {Cook}, {Derro}, {Dotson}, {Evans}, {Fixsen}, {Gatley}, {Guerra}, {Halpern}, {Hamilton}, {Hamlin}, {Hansen}, {Heimsath}, {Hermida}, {Hilton}, {Hirsch}, {Hollister}, {Hostetter}, {Irwin}, {Jhabvala}, {Jhabvala}, {Kastner}, {Kov{\'a}cs}, {Lin}, {Loewenstein}, {Looney}, {Lopez-Rodriguez}, {Maher}, {Michail}, {Miller}, {Moseley}, {Novak}, {Pernic}, {Rennick}, {Rhody}, {Sandberg}, {Sandford}, {Santos}, {Shafer}, {Sharp}, {Shirron}, {Siah}, {Silverberg}, {Sparr}, {Spotz}, {Staguhn}, {Toorian}, {Towey}, {Tuttle}, {Vaillancourt}, {Voellmer}, {Volpert}, {Wang}, \& {Wollack}}]{2018JAI.....740008H}
{Harper}, D.~A., {Runyan}, M.~C., {Dowell}, C.~D., {et~al.} 2018, Journal of Astronomical Instrumentation, 7, 1840008, \dodoi{10.1142/S2251171718400081}

\bibitem[{{Hatchfield} {et~al.}(2024){Hatchfield}, {Battersby}, {Barnes}, {Butterfield}, {Ginsburg}, {Henshaw}, {Longmore}, {Lu}, {Svoboda}, {Walker}, {Callanan}, {Mills}, {Ho}, {Kauffmann}, {Kruijssen}, {Ott}, {Pillai}, \& {Zhang}}]{Hatchfield2024}
{Hatchfield}, H.~P., {Battersby}, C., {Barnes}, A.~T., {et~al.} 2024, \apj, 962, 14, \dodoi{10.3847/1538-4357/ad10af}

\bibitem[{{Hennebelle}(2013)}]{Hennebelle2013}
{Hennebelle}, P. 2013, \aap, 556, A153, \dodoi{10.1051/0004-6361/201321292}

\bibitem[{{Henshaw} {et~al.}(2016){Henshaw}, {Longmore}, {Kruijssen}, {Davies}, {Bally}, {Barnes}, {Battersby}, {Burton}, {Cunningham}, {Dale}, {Ginsburg}, {Immer}, {Jones}, {Kendrew}, {Mills}, {Molinari}, {Moore}, {Ott}, {Pillai}, {Rathborne}, {Schilke}, {Schmiedeke}, {Testi}, {Walker}, {Walsh}, \& {Zhang}}]{Henshaw2016}
{Henshaw}, J.~D., {Longmore}, S.~N., {Kruijssen}, J.~M.~D., {et~al.} 2016, \mnras, 457, 2675, \dodoi{10.1093/mnras/stw121}

\bibitem[{{Heyer} \& {Dame}(2015)}]{Heyer2015}
{Heyer}, M., \& {Dame}, T.~M. 2015, \araa, 53, 583, \dodoi{10.1146/annurev-astro-082214-122324}

\bibitem[{{Hildebrand} {et~al.}(2009){Hildebrand}, {Kirby}, {Dotson}, {Houde}, \& {Vaillancourt}}]{Hildebrand2009}
{Hildebrand}, R.~H., {Kirby}, L., {Dotson}, J.~L., {Houde}, M., \& {Vaillancourt}, J.~E. 2009, \apj, 696, 567, \dodoi{10.1088/0004-637X/696/1/567}

\bibitem[{{Houde} {et~al.}(2016){Houde}, {Hull}, {Plambeck}, {Vaillancourt}, \& {Hildebrand}}]{Houde2016}
{Houde}, M., {Hull}, C. L.~H., {Plambeck}, R.~L., {Vaillancourt}, J.~E., \& {Hildebrand}, R.~H. 2016, \apj, 820, 38, \dodoi{10.3847/0004-637X/820/1/38}

\bibitem[{{Houde} {et~al.}(2009){Houde}, {Vaillancourt}, {Hildebrand}, {Chitsazzadeh}, \& {Kirby}}]{Houde2009}
{Houde}, M., {Vaillancourt}, J.~E., {Hildebrand}, R.~H., {Chitsazzadeh}, S., \& {Kirby}, L. 2009, \apj, 706, 1504, \dodoi{10.1088/0004-637X/706/2/1504}

\bibitem[{{Hu} {et~al.}(2022){Hu}, {Lazarian}, \& {Wang}}]{Hu2022}
{Hu}, Y., {Lazarian}, A., \& {Wang}, Q.~D. 2022, \mnras, 511, 829, \dodoi{10.1093/mnras/stac159}

\bibitem[{{Huettemeister} {et~al.}(1995){Huettemeister}, {Wilson}, {Mauersberger}, {Lemme}, {Dahmen}, \& {Henkel}}]{Huettemeister1995}
{Huettemeister}, S., {Wilson}, T.~L., {Mauersberger}, R., {et~al.} 1995, \aap, 294, 667

\bibitem[{{Hunter} {et~al.}(2024){Hunter}, {Sormani}, {Beckmann}, {Vasiliev}, {Glover}, {Klessen}, {Soler}, {Brucy}, {Girichidis}, {G{\"o}ller}, {Ohlin}, {Tress}, {Molinari}, {Gerhard}, {Benedettini}, {Smith}, {Hennebelle}, \& {Testi}}]{Hunter2024}
{Hunter}, G.~H., {Sormani}, M.~C., {Beckmann}, J.~P., {et~al.} 2024, \aap, 692, A216, \dodoi{10.1051/0004-6361/202450000}

\bibitem[{{Huttemeister} {et~al.}(1993){Huttemeister}, {Wilson}, {Henkel}, \& {Mauersberger}}]{Huttemeister1993}
{Huttemeister}, S., {Wilson}, T.~L., {Henkel}, C., \& {Mauersberger}, R. 1993, \aap, 276, 445

\bibitem[{{Jones} {et~al.}(2012){Jones}, {Burton}, {Cunningham}, {Requena-Torres}, {Menten}, {Schilke}, {Belloche}, {Leurini}, {Mart{\'\i}n-Pintado}, {Ott}, \& {Walsh}}]{Jones2012}
{Jones}, P.~A., {Burton}, M.~G., {Cunningham}, M.~R., {et~al.} 2012, \mnras, 419, 2961, \dodoi{10.1111/j.1365-2966.2011.19941.x}

\bibitem[{{Karoly} {et~al.}(2025){Karoly}, {Ward-Thompson}, {Pattle}, {Longmore}, {Di Francesco}, {Whitworth}, {Johnstone}, {Sadavoy}, {Koch}, {Yang}, {Furuya}, {Lu}, {Tamura}, {Debattista}, {Eden}, {Hwang}, {Poidevin}, {Najimudeen}, {Chen}, {Chung}, {Coude}, {Lin}, {Doi}, {Onaka}, {Fanciullo}, {Liu}, {Li}, {Bastien}, {Hasegawa}, {Kwon}, {Lai}, \& {Qiu}}]{Karoly2025}
{Karoly}, J., {Ward-Thompson}, D., {Pattle}, K., {et~al.} 2025, arXiv e-prints, arXiv:2502.11552, \dodoi{10.48550/arXiv.2502.11552}

\bibitem[{{Kauffmann} {et~al.}(2013){Kauffmann}, {Pillai}, \& {Zhang}}]{Kauffmann2013}
{Kauffmann}, J., {Pillai}, T., \& {Zhang}, Q. 2013, \apjl, 765, L35, \dodoi{10.1088/2041-8205/765/2/L35}

\bibitem[{{Kauffmann} {et~al.}(2017){Kauffmann}, {Pillai}, {Zhang}, {Menten}, {Goldsmith}, {Lu}, \& {Guzm{\'a}n}}]{Kauffmann2017}
{Kauffmann}, J., {Pillai}, T., {Zhang}, Q., {et~al.} 2017, \aap, 603, A89, \dodoi{10.1051/0004-6361/201628088}

\bibitem[{{Kendrew} {et~al.}(2013){Kendrew}, {Ginsburg}, {Johnston}, {Beuther}, {Bally}, {Cyganowski}, \& {Battersby}}]{Kendrew2013}
{Kendrew}, S., {Ginsburg}, A., {Johnston}, K., {et~al.} 2013, \apjl, 775, L50, \dodoi{10.1088/2041-8205/775/2/L50}

\bibitem[{{Kruijssen} {et~al.}(2014){Kruijssen}, {Longmore}, {Elmegreen}, {Murray}, {Bally}, {Testi}, \& {Kennicutt}}]{Kruijssen2014}
{Kruijssen}, J.~M.~D., {Longmore}, S.~N., {Elmegreen}, B.~G., {et~al.} 2014, \mnras, 440, 3370, \dodoi{10.1093/mnras/stu494}

\bibitem[{{Krumholz} \& {Kruijssen}(2015)}]{Krumholz2015}
{Krumholz}, M.~R., \& {Kruijssen}, J.~M.~D. 2015, \mnras, 453, 739, \dodoi{10.1093/mnras/stv1670}

\bibitem[{{Kuiper}(1938)}]{Kuiper1938}
{Kuiper}, G.~P. 1938, \apj, 88, 472, \dodoi{10.1086/143999}

\bibitem[{{Lada} {et~al.}(2010){Lada}, {Lombardi}, \& {Alves}}]{Lada2010}
{Lada}, C.~J., {Lombardi}, M., \& {Alves}, J.~F. 2010, \apj, 724, 687, \dodoi{10.1088/0004-637X/724/1/687}

\bibitem[{{Lazarian} \& {Yuen}(2018)}]{Lazarian2018}
{Lazarian}, A., \& {Yuen}, K.~H. 2018, \apj, 853, 96, \dodoi{10.3847/1538-4357/aaa241}

\bibitem[{{Lis} \& {Goldsmith}(1991)}]{Lis1991}
{Lis}, D.~C., \& {Goldsmith}, P.~F. 1991, \apj, 369, 157, \dodoi{10.1086/169746}

\bibitem[{{Liu} {et~al.}(2021){Liu}, {Zhang}, {Commer{\c{c}}on}, {Valdivia}, {Maury}, \& {Qiu}}]{Liu2021}
{Liu}, J., {Zhang}, Q., {Commer{\c{c}}on}, B., {et~al.} 2021, \apj, 919, 79, \dodoi{10.3847/1538-4357/ac0cec}

\bibitem[{{Liu} {et~al.}(2022){Liu}, {Zhang}, \& {Qiu}}]{Liu2022}
{Liu}, J., {Zhang}, Q., \& {Qiu}, K. 2022, Frontiers in Astronomy and Space Sciences, 9, 943556, \dodoi{10.3389/fspas.2022.943556}

\bibitem[{{Liu} {et~al.}(2020){Liu}, {Zhang}, {Qiu}, {Liu}, {Pillai}, {Girart}, {Li}, \& {Wang}}]{Liu2020}
{Liu}, J., {Zhang}, Q., {Qiu}, K., {et~al.} 2020, \apj, 895, 142, \dodoi{10.3847/1538-4357/ab9087}

\bibitem[{{Liu} {et~al.}(2023){Liu}, {Zhang}, {Koch}, {Liu}, {Li}, {Li}, {Girart}, {Chen}, {Ching}, {Ho}, {Lai}, {Qiu}, {Rao}, \& {Tang}}]{Liu2023}
{Liu}, J., {Zhang}, Q., {Koch}, P.~M., {et~al.} 2023, \apj, 945, 160, \dodoi{10.3847/1538-4357/acb540}

\bibitem[{{Longmore} {et~al.}(2012){Longmore}, {Rathborne}, {Bastian}, {Alves}, {Ascenso}, {Bally}, {Testi}, {Longmore}, {Battersby}, {Bressert}, {Purcell}, {Walsh}, {Jackson}, {Foster}, {Molinari}, {Meingast}, {Amorim}, {Lima}, {Marques}, {Moitinho}, {Pinhao}, {Rebordao}, \& {Santos}}]{Longmore2012}
{Longmore}, S.~N., {Rathborne}, J., {Bastian}, N., {et~al.} 2012, \apj, 746, 117, \dodoi{10.1088/0004-637X/746/2/117}

\bibitem[{{Longmore} {et~al.}(2013){Longmore}, {Bally}, {Testi}, {Purcell}, {Walsh}, {Bressert}, {Pestalozzi}, {Molinari}, {Ott}, {Cortese}, {Battersby}, {Murray}, {Lee}, {Kruijssen}, {Schisano}, \& {Elia}}]{Longmore2013}
{Longmore}, S.~N., {Bally}, J., {Testi}, L., {et~al.} 2013, \mnras, 429, 987, \dodoi{10.1093/mnras/sts376}

\bibitem[{{Lu} {et~al.}(2015){Lu}, {Zhang}, {Kauffmann}, {Pillai}, {Longmore}, {Kruijssen}, {Battersby}, \& {Gu}}]{Lu2015}
{Lu}, X., {Zhang}, Q., {Kauffmann}, J., {et~al.} 2015, \apjl, 814, L18, \dodoi{10.1088/2041-8205/814/2/L18}

\bibitem[{{Lu} {et~al.}(2019{\natexlab{a}}){Lu}, {Zhang}, {Kauffmann}, {Pillai}, {Ginsburg}, {Mills}, {Kruijssen}, {Longmore}, {Battersby}, {Liu}, \& {Gu}}]{Lu2019}
---. 2019{\natexlab{a}}, \apj, 872, 171, \dodoi{10.3847/1538-4357/ab017d}

\bibitem[{{Lu} {et~al.}(2019{\natexlab{b}}){Lu}, {Mills}, {Ginsburg}, {Walker}, {Barnes}, {Butterfield}, {Henshaw}, {Battersby}, {Kruijssen}, {Longmore}, {Zhang}, {Bally}, {Kauffmann}, {Ott}, {Rickert}, \& {Wang}}]{Lu2019b}
{Lu}, X., {Mills}, E. A.~C., {Ginsburg}, A., {et~al.} 2019{\natexlab{b}}, \apjs, 244, 35, \dodoi{10.3847/1538-4365/ab4258}

\bibitem[{{Lu} {et~al.}(2024){Lu}, {Liu}, {Pillai}, {Zhang}, {Liu}, {Gu}, {Hasegawa}, {Li}, {Tang}, {Hatchfield}, {Issac}, {Liu}, {Luo}, {Mai}, \& {Shen}}]{Lu2024}
{Lu}, X., {Liu}, J., {Pillai}, T., {et~al.} 2024, \apj, 962, 39, \dodoi{10.3847/1538-4357/ad1395}

\bibitem[{{Malinen} {et~al.}(2016){Malinen}, {Montier}, {Montillaud}, {Juvela}, {Ristorcelli}, {Clark}, {Bern{\'e}}, {Bernard}, {Pelkonen}, \& {Collins}}]{Malinen2016}
{Malinen}, J., {Montier}, L., {Montillaud}, J., {et~al.} 2016, \mnras, 460, 1934, \dodoi{10.1093/mnras/stw1061}

\bibitem[{{Mangilli} {et~al.}(2019){Mangilli}, {Aumont}, {Bernard}, {Buzzelli}, {de Gasperis}, {Durrive}, {Ferriere}, {Fo{\"e}nard}, {Hughes}, {Lacourt}, {Misawa}, {Montier}, {Mot}, {Ristorcelli}, {Roussel}, {Ade}, {Alina}, {de Bernardis}, {de Gouveia Dal Pino}, {Dubois}, {Engel}, {Guillet}, {Hargrave}, {Laureijs}, {Longval}, {Maffei}, {Magalhaes}, {Marty}, {Masi}, {Montel}, {Pajot}, {P{\'e}rot}, {Rodriguez}, {Salatino}, {Saccoccio}, {Savini}, {Stever}, {Tauber}, {Tibbs}, \& {Tucker}}]{Mangilli2019}
{Mangilli}, A., {Aumont}, J., {Bernard}, J.~P., {et~al.} 2019, \aap, 630, A74, \dodoi{10.1051/0004-6361/201935072}

\bibitem[{{Mathis} {et~al.}(1983){Mathis}, {Mezger}, \& {Panagia}}]{Mathis1983}
{Mathis}, J.~S., {Mezger}, P.~G., \& {Panagia}, N. 1983, \aap, 128, 212

\bibitem[{{Mathis} {et~al.}(1977){Mathis}, {Rumpl}, \& {Nordsieck}}]{Mathis1977}
{Mathis}, J.~S., {Rumpl}, W., \& {Nordsieck}, K.~H. 1977, \apj, 217, 425, \dodoi{10.1086/155591}

\bibitem[{{Mills} \& {Battersby}(2017)}]{Mills2017}
{Mills}, E. A.~C., \& {Battersby}, C. 2017, \apj, 835, 76, \dodoi{10.3847/1538-4357/835/1/76}

\bibitem[{{Molinari} {et~al.}(2010){Molinari}, {Swinyard}, {Bally}, {Barlow}, {Bernard}, {Martin}, {Moore}, {Noriega-Crespo}, {Plume}, {Testi}, {Zavagno}, {Abergel}, {Ali}, {Andr{\'e}}, {Baluteau}, {Benedettini}, {Bern{\'e}}, {Billot}, {Blommaert}, {Bontemps}, {Boulanger}, {Brand}, {Brunt}, {Burton}, {Campeggio}, {Carey}, {Caselli}, {Cesaroni}, {Cernicharo}, {Chakrabarti}, {Chrysostomou}, {Codella}, {Cohen}, {Compiegne}, {Davis}, {de Bernardis}, {de Gasperis}, {Di Francesco}, {di Giorgio}, {Elia}, {Faustini}, {Fischera}, {Fukui}, {Fuller}, {Ganga}, {Garcia-Lario}, {Giard}, {Giardino}, {Glenn}, {Goldsmith}, {Griffin}, {Hoare}, {Huang}, {Jiang}, {Joblin}, {Joncas}, {Juvela}, {Kirk}, {Lagache}, {Li}, {Lim}, {Lord}, {Lucas}, {Maiolo}, {Marengo}, {Marshall}, {Masi}, {Massi}, {Matsuura}, {Meny}, {Minier}, {Miville-Desch{\^e}nes}, {Montier}, {Motte}, {M{\"u}ller}, {Natoli}, {Neves}, {Olmi}, {Paladini}, {Paradis}, {Pestalozzi}, {Pezzuto}, {Piacentini}, {Pomar{\`e}s}, {Popescu}, {Reach}, {Richer}, {Ristorcelli},
  {Roy}, {Royer}, {Russeil}, {Saraceno}, {Sauvage}, {Schilke}, {Schneider-Bontemps}, {Schuller}, {Schultz}, {Shepherd}, {Sibthorpe}, {Smith}, {Smith}, {Spinoglio}, {Stamatellos}, {Strafella}, {Stringfellow}, {Sturm}, {Taylor}, {Thompson}, {Tuffs}, {Umana}, {Valenziano}, {Vavrek}, {Viti}, {Waelkens}, {Ward-Thompson}, {White}, {Wyrowski}, {Yorke}, \& {Zhang}}]{Molinari2010}
{Molinari}, S., {Swinyard}, B., {Bally}, J., {et~al.} 2010, \pasp, 122, 314, \dodoi{10.1086/651314}

\bibitem[{{Molinari} {et~al.}(2011){Molinari}, {Bally}, {Noriega-Crespo}, {Compi{\`e}gne}, {Bernard}, {Paradis}, {Martin}, {Testi}, {Barlow}, {Moore}, {Plume}, {Swinyard}, {Zavagno}, {Calzoletti}, {Di Giorgio}, {Elia}, {Faustini}, {Natoli}, {Pestalozzi}, {Pezzuto}, {Piacentini}, {Polenta}, {Polychroni}, {Schisano}, {Traficante}, {Veneziani}, {Battersby}, {Burton}, {Carey}, {Fukui}, {Li}, {Lord}, {Morgan}, {Motte}, {Schuller}, {Stringfellow}, {Tan}, {Thompson}, {Ward-Thompson}, {White}, \& {Umana}}]{Molinari2011}
{Molinari}, S., {Bally}, J., {Noriega-Crespo}, A., {et~al.} 2011, \apjl, 735, L33, \dodoi{10.1088/2041-8205/735/2/L33}

\bibitem[{{Morris} \& {Serabyn}(1996)}]{Morris1996}
{Morris}, M., \& {Serabyn}, E. 1996, \araa, 34, 645, \dodoi{10.1146/annurev.astro.34.1.645}

\bibitem[{{Mouschovias}(1976)}]{Mouschovias1976}
{Mouschovias}, T.~C. 1976, \apj, 206, 753, \dodoi{10.1086/154436}

\bibitem[{{Myers} {et~al.}(2022){Myers}, {Hatchfield}, \& {Battersby}}]{Myers2022}
{Myers}, P.~C., {Hatchfield}, H.~P., \& {Battersby}, C. 2022, \apj, 929, 34, \dodoi{10.3847/1538-4357/ac5906}

\bibitem[{{Nakano} \& {Nakamura}(1978)}]{Nakano1978}
{Nakano}, T., \& {Nakamura}, T. 1978, \pasj, 30, 671

\bibitem[{{Ossenkopf} \& {Henning}(1994)}]{Ossenkopf1994}
{Ossenkopf}, V., \& {Henning}, T. 1994, \aap, 291, 943

\bibitem[{{Pakmor} {et~al.}(2011){Pakmor}, {Bauer}, \& {Springel}}]{Pakmor2011}
{Pakmor}, R., {Bauer}, A., \& {Springel}, V. 2011, \mnras, 418, 1392, \dodoi{10.1111/j.1365-2966.2011.19591.x}

\bibitem[{{Pakmor} \& {Springel}(2013)}]{Pakmor2013}
{Pakmor}, R., \& {Springel}, V. 2013, \mnras, 432, 176, \dodoi{10.1093/mnras/stt428}

\bibitem[{{Pan} {et~al.}(2024){Pan}, {Zhang}, {Qiu}, {Rao}, {Zeng}, {Lu}, \& {Liu}}]{Pan2024}
{Pan}, X., {Zhang}, Q., {Qiu}, K., {et~al.} 2024, \apj, 972, 30, \dodoi{10.3847/1538-4357/ad5aea}

\bibitem[{{Par{\'e}} {et~al.}(2024){Par{\'e}}, {Butterfield}, {Chuss}, {Guerra}, {Iuliano}, {Karpovich}, {Morris}, \& {Wollack}}]{Pare2024}
{Par{\'e}}, D., {Butterfield}, N.~O., {Chuss}, D.~T., {et~al.} 2024, \apj, 969, 150, \dodoi{10.3847/1538-4357/ad4462}

\bibitem[{{Par{\'e}} {et~al.}(2025){Par{\'e}}, {Chuss}, {Karpovich}, {Butterfield}, {Iuliano}, {Pan}, {Wollack}, {Zhang}, {Morris}, {Nilsson}, \& {Zhao}}]{Pare2025}
{Par{\'e}}, D.~M., {Chuss}, D.~T., {Karpovich}, K., {et~al.} 2025, \apj, 978, 28, \dodoi{10.3847/1538-4357/ad9586}

\bibitem[{{Peter} {et~al.}(2023){Peter}, {Klessen}, {Kanschat}, {Glover}, \& {Bastian}}]{Peter2023}
{Peter}, T., {Klessen}, R.~S., {Kanschat}, G., {Glover}, S. C.~O., \& {Bastian}, P. 2023, \mnras, 519, 4263, \dodoi{10.1093/mnras/stac3034}

\bibitem[{{Pierce-Price} {et~al.}(2000){Pierce-Price}, {Richer}, {Greaves}, {Holland}, {Jenness}, {Lasenby}, {White}, {Matthews}, {Ward-Thompson}, {Dent}, {Zylka}, {Mezger}, {Hasegawa}, {Oka}, {Omont}, \& {Gilmore}}]{Pierce-Price2000}
{Pierce-Price}, D., {Richer}, J.~S., {Greaves}, J.~S., {et~al.} 2000, \apjl, 545, L121, \dodoi{10.1086/317884}

\bibitem[{{Pillai} {et~al.}(2015){Pillai}, {Kauffmann}, {Tan}, {Goldsmith}, {Carey}, \& {Menten}}]{Pillai2015}
{Pillai}, T., {Kauffmann}, J., {Tan}, J.~C., {et~al.} 2015, \apj, 799, 74, \dodoi{10.1088/0004-637X/799/1/74}

\bibitem[{{Planck Collaboration} {et~al.}(2014){Planck Collaboration}, {Ade}, {Aghanim}, {Alves}, {Armitage-Caplan}, {Arnaud}, {Ashdown}, {Atrio-Barandela}, {Aumont}, {Aussel}, {Baccigalupi}, {Banday}, {Barreiro}, {Barrena}, {Bartelmann}, {Bartlett}, {Bartolo}, {Basak}, {Battaner}, {Battye}, {Benabed}, {Beno{\^\i}t}, {Benoit-L{\'e}vy}, {Bernard}, {Bersanelli}, {Bertincourt}, {Bethermin}, {Bielewicz}, {Bikmaev}, {Blanchard}, {Bobin}, {Bock}, {B{\"o}hringer}, {Bonaldi}, {Bonavera}, {Bond}, {Borrill}, {Bouchet}, {Boulanger}, {Bourdin}, {Bowyer}, {Bridges}, {Brown}, {Bucher}, {Burenin}, {Burigana}, {Butler}, {Calabrese}, {Cappellini}, {Cardoso}, {Carr}, {Carvalho}, {Casale}, {Castex}, {Catalano}, {Challinor}, {Chamballu}, {Chary}, {Chen}, {Chiang}, {Chiang}, {Chon}, {Christensen}, {Churazov}, {Church}, {Clemens}, {Clements}, {Colombi}, {Colombo}, {Combet}, {Comis}, {Couchot}, {Coulais}, {Crill}, {Cruz}, {Curto}, {Cuttaia}, {Da Silva}, {Dahle}, {Danese}, {Davies}, {Davis}, {de Bernardis}, {de Rosa}, {de Zotti},
  {D{\'e}chelette}, {Delabrouille}, {Delouis}, {D{\'e}mocl{\`e}s}, {D{\'e}sert}, {Dick}, {Dickinson}, {Diego}, {Dolag}, {Dole}, {Donzelli}, {Dor{\'e}}, {Douspis}, {Ducout}, {Dunkley}, {Dupac}, {Efstathiou}, {Elsner}, {En{\ss}lin}, {Eriksen}, {Fabre}, {Falgarone}, {Falvella}, {Fantaye}, {Fergusson}, {Filliard}, {Finelli}, {Flores-Cacho}, {Foley}, {Forni}, {Fosalba}, {Frailis}, {Fraisse}, {Franceschi}, {Freschi}, {Fromenteau}, {Frommert}, {Gaier}, {Galeotta}, {Gallegos}, {Galli}, {Gandolfo}, {Ganga}, {Gauthier}, {G{\'e}nova-Santos}, {Ghosh}, {Giard}, {Giardino}, {Gilfanov}, {Girard}, {Giraud-H{\'e}raud}, {Gjerl{\o}w}, {Gonz{\'a}lez-Nuevo}, {G{\'o}rski}, {Gratton}, {Gregorio}, {Gruppuso}, {Gudmundsson}, {Haissinski}, {Hamann}, {Hansen}, {Hansen}, {Hanson}, {Harrison}, {Heavens}, {Helou}, {Hempel}, {Henrot-Versill{\'e}}, {Hern{\'a}ndez-Monteagudo}, {Herranz}, {Hildebrandt}, {Hivon}, {Ho}, {Hobson}, {Holmes}, {Hornstrup}, {Hou}, {Hovest}, {Huey}, {Huffenberger}, {Hurier}, {Ili{\'c}}, {Jaffe}, {Jaffe}, {Jasche},
  {Jewell}, {Jones}, {Juvela}, {Kalberla}, {Kangaslahti}, {Keih{\"a}nen}, {Kerp}, {Keskitalo}, {Khamitov}, {Kiiveri}, {Kim}, {Kisner}, {Kneissl}, {Knoche}, {Knox}, {Kunz}, {Kurki-Suonio}, {Lacasa}, {Lagache}, {L{\"a}hteenm{\"a}ki}, {Lamarre}, {Langer}, {Lasenby}, {Lattanzi}, {Laureijs}, {Lavabre}, {Lawrence}, {Le Jeune}, {Leach}, \& {Leahy}}]{Planck2013I}
{Planck Collaboration}, {Ade}, P.~A.~R., {Aghanim}, N., {et~al.} 2014, \aap, 571, A1, \dodoi{10.1051/0004-6361/201321529}

\bibitem[{{Planck Collaboration} {et~al.}(2016{\natexlab{a}}){Planck Collaboration}, {Ade}, {Aghanim}, {Alves}, {Arnaud}, {Arzoumanian}, {Ashdown}, {Aumont}, {Baccigalupi}, {Banday}, {Barreiro}, {Bartolo}, {Battaner}, {Benabed}, {Beno{\^\i}t}, {Benoit-L{\'e}vy}, {Bernard}, {Bersanelli}, {Bielewicz}, {Bock}, {Bonavera}, {Bond}, {Borrill}, {Bouchet}, {Boulanger}, {Bracco}, {Burigana}, {Calabrese}, {Cardoso}, {Catalano}, {Chiang}, {Christensen}, {Colombo}, {Combet}, {Couchot}, {Crill}, {Curto}, {Cuttaia}, {Danese}, {Davies}, {Davis}, {de Bernardis}, {de Rosa}, {de Zotti}, {Delabrouille}, {Dickinson}, {Diego}, {Dole}, {Donzelli}, {Dor{\'e}}, {Douspis}, {Ducout}, {Dupac}, {Efstathiou}, {Elsner}, {En{\ss}lin}, {Eriksen}, {Falceta-Gon{\c{c}}alves}, {Falgarone}, {Ferri{\`e}re}, {Finelli}, {Forni}, {Frailis}, {Fraisse}, {Franceschi}, {Frejsel}, {Galeotta}, {Galli}, {Ganga}, {Ghosh}, {Giard}, {Gjerl{\o}w}, {Gonz{\'a}lez-Nuevo}, {G{\'o}rski}, {Gregorio}, {Gruppuso}, {Gudmundsson}, {Guillet}, {Harrison}, {Helou},
  {Hennebelle}, {Henrot-Versill{\'e}}, {Hern{\'a}ndez-Monteagudo}, {Herranz}, {Hildebrandt}, {Hivon}, {Holmes}, {Hornstrup}, {Huffenberger}, {Hurier}, {Jaffe}, {Jaffe}, {Jones}, {Juvela}, {Keih{\"a}nen}, {Keskitalo}, {Kisner}, {Knoche}, {Kunz}, {Kurki-Suonio}, {Lagache}, {Lamarre}, {Lasenby}, {Lattanzi}, {Lawrence}, {Leonardi}, {Levrier}, {Liguori}, {Lilje}, {Linden-V{\o}rnle}, {L{\'o}pez-Caniego}, {Lubin}, {Mac{\'\i}as-P{\'e}rez}, {Maino}, {Mandolesi}, {Mangilli}, {Maris}, {Martin}, {Mart{\'\i}nez-Gonz{\'a}lez}, {Masi}, {Matarrese}, {Melchiorri}, {Mendes}, {Mennella}, {Migliaccio}, {Miville-Desch{\^e}nes}, {Moneti}, {Montier}, {Morgante}, {Mortlock}, {Munshi}, {Murphy}, {Naselsky}, {Nati}, {Netterfield}, {Noviello}, {Novikov}, {Novikov}, {Oppermann}, {Oxborrow}, {Pagano}, {Pajot}, {Paladini}, {Paoletti}, {Pasian}, {Perotto}, {Pettorino}, {Piacentini}, {Piat}, {Pierpaoli}, {Pietrobon}, {Plaszczynski}, {Pointecouteau}, {Polenta}, {Ponthieu}, {Pratt}, {Prunet}, {Puget}, {Rachen}, {Reinecke}, {Remazeilles},
  {Renault}, {Renzi}, {Ristorcelli}, {Rocha}, {Rossetti}, {Roudier}, {Rubi{\~n}o-Mart{\'\i}n}, {Rusholme}, {Sandri}, {Santos}, {Savelainen}, {Savini}, {Scott}, {Soler}, {Stolyarov}, {Sudiwala}, {Sutton}, {Suur-Uski}, {Sygnet}, {Tauber}, {Terenzi}, {Toffolatti}, {Tomasi}, {Tristram}, {Tucci}, {Umana}, {Valenziano}, {Valiviita}, {Van Tent}, {Vielva}, {Villa}, {Wade}, {Wandelt}, {Wehus}, {Ysard}, {Yvon}, \& {Zonca}}]{PlanckXXXV2016}
---. 2016{\natexlab{a}}, \aap, 586, A138, \dodoi{10.1051/0004-6361/201525896}

\bibitem[{{Planck Collaboration} {et~al.}(2016{\natexlab{b}}){Planck Collaboration}, {Adam}, {Ade}, {Aghanim}, {Alves}, {Arnaud}, {Arzoumanian}, {Ashdown}, {Aumont}, {Baccigalupi}, {Banday}, {Barreiro}, {Bartolo}, {Battaner}, {Benabed}, {Benoit-L{\'e}vy}, {Bernard}, {Bersanelli}, {Bielewicz}, {Bonaldi}, {Bonavera}, {Bond}, {Borrill}, {Bouchet}, {Boulanger}, {Bracco}, {Burigana}, {Butler}, {Calabrese}, {Cardoso}, {Catalano}, {Chamballu}, {Chiang}, {Christensen}, {Colombi}, {Colombo}, {Combet}, {Couchot}, {Crill}, {Curto}, {Cuttaia}, {Danese}, {Davies}, {Davis}, {de Bernardis}, {de Rosa}, {de Zotti}, {Delabrouille}, {Dickinson}, {Diego}, {Dole}, {Donzelli}, {Dor{\'e}}, {Douspis}, {Ducout}, {Dupac}, {Efstathiou}, {Elsner}, {En{\ss}lin}, {Eriksen}, {Falgarone}, {Ferri{\`e}re}, {Finelli}, {Forni}, {Frailis}, {Fraisse}, {Franceschi}, {Frejsel}, {Galeotta}, {Galli}, {Ganga}, {Ghosh}, {Giard}, {Gjerl{\o}w}, {Gonz{\'a}lez-Nuevo}, {G{\'o}rski}, {Gregorio}, {Gruppuso}, {Guillet}, {Hansen}, {Hanson}, {Harrison},
  {Henrot-Versill{\'e}}, {Hern{\'a}ndez-Monteagudo}, {Herranz}, {Hildebrandt}, {Hivon}, {Hobson}, {Holmes}, {Hovest}, {Huffenberger}, {Hurier}, {Jaffe}, {Jaffe}, {Jones}, {Juvela}, {Keih{\"a}nen}, {Keskitalo}, {Kisner}, {Kneissl}, {Knoche}, {Kunz}, {Kurki-Suonio}, {Lagache}, {Lamarre}, {Lasenby}, {Lattanzi}, {Lawrence}, {Leonardi}, {Levrier}, {Liguori}, {Lilje}, {Linden-V{\o}rnle}, {L{\'o}pez-Caniego}, {Lubin}, {Mac{\'\i}as-P{\'e}rez}, {Maffei}, {Maino}, {Mandolesi}, {Maris}, {Marshall}, {Martin}, {Mart{\'\i}nez-Gonz{\'a}lez}, {Masi}, {Matarrese}, {Mazzotta}, {Melchiorri}, {Mendes}, {Mennella}, {Migliaccio}, {Miville-Desch{\^e}nes}, {Moneti}, {Montier}, {Morgante}, {Mortlock}, {Munshi}, {Murphy}, {Naselsky}, {Natoli}, {N{\o}rgaard-Nielsen}, {Noviello}, {Novikov}, {Novikov}, {Oppermann}, {Oxborrow}, {Pagano}, {Pajot}, {Paoletti}, {Pasian}, {Perdereau}, {Perotto}, {Perrotta}, {Pettorino}, {Piacentini}, {Piat}, {Plaszczynski}, {Pointecouteau}, {Polenta}, {Ponthieu}, {Popa}, {Pratt}, {Prunet}, {Puget}, {Rachen},
  {Reach}, {Reinecke}, {Remazeilles}, {Renault}, {Ristorcelli}, {Rocha}, {Roudier}, {Rubi{\~n}o-Mart{\'\i}n}, {Rusholme}, {Sandri}, {Santos}, {Savini}, {Scott}, {Soler}, {Spencer}, {Stolyarov}, {Sudiwala}, {Sunyaev}, {Sutton}, {Suur-Uski}, {Sygnet}, {Tauber}, {Terenzi}, {Toffolatti}, {Tomasi}, {Tristram}, {Tucci}, {Umana}, {Valenziano}, {Valiviita}, {Van Tent}, {Vielva}, {Villa}, {Wade}, {Wandelt}, \& {Wehus}}]{PlanckXXXII2016}
{Planck Collaboration}, {Adam}, R., {Ade}, P.~A.~R., {et~al.} 2016{\natexlab{b}}, \aap, 586, A135, \dodoi{10.1051/0004-6361/201425044}

\bibitem[{{Qin} {et~al.}(2011){Qin}, {Schilke}, {Rolffs}, {Comito}, {Lis}, \& {Zhang}}]{Qin2011}
{Qin}, S.~L., {Schilke}, P., {Rolffs}, R., {et~al.} 2011, \aap, 530, L9, \dodoi{10.1051/0004-6361/201116928}

\bibitem[{{Reissl} {et~al.}(2016){Reissl}, {Wolf}, \& {Brauer}}]{Reissl2016}
{Reissl}, S., {Wolf}, S., \& {Brauer}, R. 2016, \aap, 593, A87, \dodoi{10.1051/0004-6361/201424930}

\bibitem[{{Schuller} {et~al.}(2009){Schuller}, {Menten}, {Contreras}, {Wyrowski}, {Schilke}, {Bronfman}, {Henning}, {Walmsley}, {Beuther}, {Bontemps}, {Cesaroni}, {Deharveng}, {Garay}, {Herpin}, {Lefloch}, {Linz}, {Mardones}, {Minier}, {Molinari}, {Motte}, {Nyman}, {Reveret}, {Risacher}, {Russeil}, {Schneider}, {Testi}, {Troost}, {Vasyunina}, {Wienen}, {Zavagno}, {Kovacs}, {Kreysa}, {Siringo}, \& {Wei{\ss}}}]{Schuller2009}
{Schuller}, F., {Menten}, K.~M., {Contreras}, Y., {et~al.} 2009, \aap, 504, 415, \dodoi{10.1051/0004-6361/200811568}

\bibitem[{{Seifried} {et~al.}(2020){Seifried}, {Walch}, {Weis}, {Reissl}, {Soler}, {Klessen}, \& {Joshi}}]{Seifried2020}
{Seifried}, D., {Walch}, S., {Weis}, M., {et~al.} 2020, \mnras, 497, 4196, \dodoi{10.1093/mnras/staa2231}

\bibitem[{{Soler}(2019)}]{Soler2019}
{Soler}, J.~D. 2019, \aap, 629, A96, \dodoi{10.1051/0004-6361/201935779}

\bibitem[{{Soler} \& {Hennebelle}(2017)}]{Soler2017b}
{Soler}, J.~D., \& {Hennebelle}, P. 2017, \aap, 607, A2, \dodoi{10.1051/0004-6361/201731049}

\bibitem[{{Soler} {et~al.}(2013){Soler}, {Hennebelle}, {Martin}, {Miville-Desch{\^e}nes}, {Netterfield}, \& {Fissel}}]{Soler2013}
{Soler}, J.~D., {Hennebelle}, P., {Martin}, P.~G., {et~al.} 2013, \apj, 774, 128, \dodoi{10.1088/0004-637X/774/2/128}

\bibitem[{{Soler} {et~al.}(2017){Soler}, {Ade}, {Angil{\`e}}, {Ashton}, {Benton}, {Devlin}, {Dober}, {Fissel}, {Fukui}, {Galitzki}, {Gandilo}, {Hennebelle}, {Klein}, {Li}, {Korotkov}, {Martin}, {Matthews}, {Moncelsi}, {Netterfield}, {Novak}, {Pascale}, {Poidevin}, {Santos}, {Savini}, {Scott}, {Shariff}, {Thomas}, {Tucker}, {Tucker}, \& {Ward-Thompson}}]{Soler2017}
{Soler}, J.~D., {Ade}, P.~A.~R., {Angil{\`e}}, F.~E., {et~al.} 2017, \aap, 603, A64, \dodoi{10.1051/0004-6361/201730608}

\bibitem[{{Springel}(2010)}]{Springel2010}
{Springel}, V. 2010, \mnras, 401, 791, \dodoi{10.1111/j.1365-2966.2009.15715.x}

\bibitem[{{Sridhar} \& {Goldreich}(1994)}]{Sridhar1994}
{Sridhar}, S., \& {Goldreich}, P. 1994, \apj, 432, 612, \dodoi{10.1086/174600}

\bibitem[{{Tang} {et~al.}(2021){Tang}, {Wang}, {Wilson}, {Heyer}, {Gutermuth}, {Schloerb}, {Yun}, {Bally}, {Loinard}, {Silich}, {Ch{\'a}vez}, {Haggard}, {Monta{\~n}a}, {S{\'a}nchez-Arg{\"u}elles}, {Zeballos}, {Zavala}, \& {Le{\'o}n-Tavares}}]{Tang2021}
{Tang}, Y., {Wang}, Q.~D., {Wilson}, G.~W., {et~al.} 2021, \mnras, 505, 2392, \dodoi{10.1093/mnras/stab1191}

\bibitem[{{Temi} {et~al.}(2018){Temi}, {Hoffman}, {Ennico}, \& {Le}}]{2018JAI.....740011T}
{Temi}, P., {Hoffman}, D., {Ennico}, K., \& {Le}, J. 2018, Journal of Astronomical Instrumentation, 7, 1840011, \dodoi{10.1142/S2251171718400111}

\bibitem[{{Tress} {et~al.}(2020){Tress}, {Sormani}, {Glover}, {Klessen}, {Battersby}, {Clark}, {Hatchfield}, \& {Smith}}]{Tress2020}
{Tress}, R.~G., {Sormani}, M.~C., {Glover}, S. C.~O., {et~al.} 2020, \mnras, 499, 4455, \dodoi{10.1093/mnras/staa3120}

\bibitem[{{Tress} {et~al.}(2024){Tress}, {Sormani}, {Girichidis}, {Glover}, {Klessen}, {Smith}, {Sobacchi}, {Armillotta}, {Barnes}, {Battersby}, {Bogue}, {Brucy}, {Colzi}, {Federrath}, {Garc{\'\i}a}, {Ginsburg}, {G{\"o}ller}, {Hatchfield}, {Henkel}, {Hennebelle}, {Henshaw}, {Hirschmann}, {Hu}, {Kauffmann}, {Kruijssen}, {Lazarian}, {Lipman}, {Longmore}, {Morris}, {Nogueras-Lara}, {Petkova}, {Pillai}, {Rivilla}, {S{\'a}nchez-Monge}, {Soler}, {Whitworth}, \& {Zhang}}]{Tress2024}
{Tress}, R.~G., {Sormani}, M.~C., {Girichidis}, P., {et~al.} 2024, \aap, 691, A303, \dodoi{10.1051/0004-6361/202450035}

\bibitem[{{Walker} {et~al.}(2015){Walker}, {Longmore}, {Bastian}, {Kruijssen}, {Rathborne}, {Jackson}, {Foster}, \& {Contreras}}]{Walker2015}
{Walker}, D.~L., {Longmore}, S.~N., {Bastian}, N., {et~al.} 2015, \mnras, 449, 715, \dodoi{10.1093/mnras/stv300}

\bibitem[{{Walker} {et~al.}(2018){Walker}, {Longmore}, {Zhang}, {Battersby}, {Keto}, {Kruijssen}, {Ginsburg}, {Lu}, {Henshaw}, {Kauffmann}, {Pillai}, {Mills}, {Walsh}, {Bally}, {Ho}, {Immer}, \& {Johnston}}]{Walker2018}
{Walker}, D.~L., {Longmore}, S.~N., {Zhang}, Q., {et~al.} 2018, \mnras, 474, 2373, \dodoi{10.1093/mnras/stx2898}

\bibitem[{{Walker} {et~al.}(2021){Walker}, {Longmore}, {Bally}, {Ginsburg}, {Kruijssen}, {Zhang}, {Henshaw}, {Lu}, {Alves}, {Barnes}, {Battersby}, {Beuther}, {Contreras}, {G{\'o}mez}, {Ho}, {Jackson}, {Kauffmann}, {Mills}, \& {Pillai}}]{Walker2021}
{Walker}, D.~L., {Longmore}, S.~N., {Bally}, J., {et~al.} 2021, \mnras, 503, 77, \dodoi{10.1093/mnras/stab415}

\bibitem[{{Weinberger} {et~al.}(2020){Weinberger}, {Springel}, \& {Pakmor}}]{Weinberger2020}
{Weinberger}, R., {Springel}, V., \& {Pakmor}, R. 2020, \apjs, 248, 32, \dodoi{10.3847/1538-4365/ab908c}

\bibitem[{{Xu} {et~al.}(2019){Xu}, {Ji}, \& {Lazarian}}]{Xu2019}
{Xu}, S., {Ji}, S., \& {Lazarian}, A. 2019, \apj, 878, 157, \dodoi{10.3847/1538-4357/ab21be}

\bibitem[{{Yang} {et~al.}(2025){Yang}, {Lai}, {Karoly}, {Pattle}, {Lu}, {Eden}, {Lin}, {Poidevin}, {Sharma}, {Hwang}, {Fanciullo}, {Tahani}, {Koch}, {Inutsuka}, {Le Gouellec}, {Duan}, {Wang}, {Fuller}, {Furuya}, {Gu}, {Hasegawa}, {Li}, {Liu}, {Akshaya}, {Najimudeen}, {Tram}, {Ward-Thompson}, {Arzoumanian}, {Di Francesco}, {Doi}, {Hoang}, {Kang}, {Kwon}, {Kwon}, {Lee}, {Liu}, {Onaka}, {Sadavoy}, {Tamura}, {Bastien}, {Berry}, {Coud{\'e}}, \& {Qiu}}]{Yang2025}
{Yang}, M.-Z., {Lai}, S.-P., {Karoly}, J., {et~al.} 2025, arXiv e-prints, arXiv:2503.05198, \dodoi{10.48550/arXiv.2503.05198}

\end{thebibliography}
\bibliographystyle{aasjournal}



\end{document}